\documentclass{ieeeaccess}
\usepackage{cite}
\usepackage{amsmath,amssymb,amsfonts}
\usepackage{algorithmic}
\usepackage{graphicx}
\usepackage{textcomp}

\def\BibTeX{{\rm B\kern-.05em{\sc i\kern-.025em b}\kern-.08em
    T\kern-.1667em\lower.7ex\hbox{E}\kern-.125emX}}

\usepackage{mathtools}
\usepackage{url}
\usepackage{xspace}
\usepackage{siunitx}
\usepackage{caption}
\usepackage{subcaption}
\usepackage{multirow}
\usepackage{multicol}
\usepackage{stackengine}
\usepackage{wrapfig}
\usepackage{textcase}
\usepackage{ifthen}
\usepackage{cases}
\usepackage{paralist}
\usepackage{booktabs}
\usepackage{makecell}
\usepackage{xspace}
\usepackage{array}
\usepackage{balance}
\usepackage[export]{adjustbox}
\usepackage[inline]{enumitem}
\usepackage{soul}
\usepackage{rotating}
\usepackage{stfloats}

\newcommand{\fakeparagraph}[1]{\vspace{1mm}\noindent\textbf{#1.}}
\newcommand{\vect}[1]{\boldsymbol{#1}}

\newcommand{\rssi}{\ensuremath{\mathit{RSSI}}\xspace}
\newcommand{\pdr}{\ensuremath{\mathit{PDR}}\xspace}
\newcommand{\ema}{\ensuremath{\mathit{EMA}}\xspace}
\newcommand{\dutycycle}{\textit{duty-cycle}\xspace}

\newcommand{\fpr}{\ensuremath{\mathit{FPR}}\xspace}

\newcommand{\wrt}{w.r.t.\ }

\newcommand{\home}{\textsc{home}\xspace}
\newcommand{\office}{\textsc{office}\xspace}
\newcommand{\first}{\textsc{first}\xspace}
\newcommand{\second}{\textsc{second}\xspace}
\newcommand{\third}{\textsc{third}\xspace}

\newcolumntype{L}[1]{>
{\raggedleft\arraybackslash}p{#1}}
\newcolumntype{C}[1]{>{\centering\arraybackslash}p{#1}}
\newcolumntype{R}[1]{>{\raggedright\arraybackslash}p{#1}}

\newcommand{\ie}{i.e.,\ }
\newcommand{\eg}{e.g.\ }
\newcommand{\vs}{vs.\ }
\newcommand{\free}{\textsc{free}\xspace}
\newcommand{\busy}{\textsc{busy}\xspace}

\newcommand{\thiat}{\ensuremath{TH_\mathit{IAT}}\xspace}
\newcommand{\thcount}{\ensuremath{TH_\mathit{count}}\xspace}    
\newcommand{\proposedmac}{LUCID\xspace}  

\newboolean{SUBMIT}
\newboolean{authnotes}

\setboolean{SUBMIT}{true} 

\ifthenelse{\boolean{SUBMIT}}
{
\setboolean{authnotes}{false}
}{
\setboolean{authnotes}{true}
}

\ifthenelse{\boolean{authnotes}}
{
\newcommand{\ram}[1]{\footnote{{\bf Ramona: #1}}}
\newcommand{\adis}[1]{\footnote{{\bf Indika: #1}}}
\newcommand{\dirk}[1]{\footnote{{\bf Dirk: #1}}}
\usepackage{draftwatermark}
\SetWatermarkLightness{0.85}
}{
\newcommand{\ram}[1]{}
\newcommand{\adis}[1]{}
\newcommand{\dirk}[1]{}
}

\begin{document}

\history{} 
\doi{} 

\title{\proposedmac: Receiver-aware Model-based Data Communication for Low-power Wireless Networks}

\author{
\uppercase{Indika S. A. Dhanapala\authorrefmark{1}, Ramona Marfievici\authorrefmark{2}, and Dirk Pesch\authorrefmark{3}}
}


\address[1]{Nimbus Research Centre, Munster Technological University, Cork, Ireland (e-mail: I.S.A.Dhanapala@mycit.ie)}
\address[2]{Digital Catapult, London, UK (e-mail: ramona.marfievici@digicatapult.org.uk)}
\address[3]{University College Cork, Cork, Ireland (e-mail: d.pesch@cs.ucc.ie)}
\tfootnote{This work was supported by the Irish Research Council in collaboration with the then United Technologies Research Centre, Cork, Ireland, and by Science Foundation Ireland (SFI) under grant 13/IA/1885.}

\markboth
{I. S. A. Dhanapala \headeretal: \proposedmac: Receiver-aware Model-based Data Communication for Low-power Wireless Networks}
{I. S. A. Dhanapala \headeretal: \proposedmac: Receiver-aware Model-based Data Communication for Low-power Wireless Networks}

\corresp{Corresponding author: Indika S. A. Dhanapala (e-mail: I.S.A.Dhanapala@mycit.ie).}

%

\begin{abstract}

In the last decade, the advancement of the Internet of Things (IoT) has caused unlicensed radio spectrum, especially the $2.4$~GHz ISM band, to be immensely crowded with smart wireless devices that are used in a wide range of application domains. Due to their diversity in radio resource use and channel access techniques, when collocated, these wireless devices create interference with each other, known as Cross-Technology Interference (CTI), which can lead to increased packet losses and energy consumption. CTI is a significant problem for low-power wireless networks, such as IEEE~$802.15.4$, as it decreases the overall dependability of the wireless network.  

To improve the performance of low-power wireless networks under CTI conditions, we propose a data-driven proactive receiver-aware MAC protocol, \proposedmac, based on interference estimation and white space prediction. We leverage statistical analysis of real-world traces from two indoor environments characterised by varying channel conditions to develop CTI prediction methods. 
The CTI models that generate accurate predictions of interference behaviour are an intrinsic part of our solution. 
\proposedmac is thoroughly evaluated in realistic simulations and we show that depending on the application data rate and the network size, our solution achieves higher dependability, $1.2$\% increase in \textit{packet delivery ratio} and $0.02$\% decrease in \dutycycle under bursty indoor interference than state of the art alternative methods. 

\end{abstract}



\titlepgskip=-15pt

\maketitle

\section{Introduction}
\label{sec:introduction}

Wireless communication systems operating in unlicensed radio spectrum, such as the $2.4$~GHz ISM band, often suffer from Cross Technology Interference (CTI), which is the overlapping of transmissions from heterogeneous wireless systems in time and frequency. This interference occurs due to the broadcast nature of wireless transmissions of collocated devices of different technologies, such as IEEE~$802.11$ (WiFi), IEEE~$802.15.1$ (Bluetooth), or IEEE~$802.15.4$, which cannot coordinate their transmissions. CTI creates packet losses, increases channel contention, and ultimately under-utilises the scarce frequency spectrum~\cite{Liang2010, Hithnawi2014}. 

These problems are exacerbated for IEEE~$802.15.4$ based low-power wireless networks due to their lower transmission power levels compared to competing technologies such as IEEE~$802.11$. In the presence of interference, low-power wireless nodes need to adapt to these interference patterns and schedule their transmissions in order to avoid interference, maximise the reliability of their communication and minimise energy consumption. One way to achieve this is for nodes to acquire a detailed model of the surrounding interference through interference power measurements. Utilising this, nodes can parameterise \textit{interference estimation} and \textit{white space prediction} models to schedule their channel access or tune their communication protocols to avoid interference. 

In this paper, we exploit a large set of real-world data traces to create models for both estimating the interference and predicting white spaces. 
Note that we define white space as the length in time in which an IEEE~$802.15.4$ packet and its ACK can be transmitted without preemption. 
The traces were collected in two different indoor environments: \office and \home. The former is an office building, while the latter is a student dormitory. We characterised and analysed the traces using two measures, the mean interference Inter Arrival Time (IAT) and the number of interference signals in a slot of fixed duration. The analysis revealed:
\begin{inparaenum}[$i)$] 
\item interference traces of arbitrary distribution, and
\item the presence of peak and off-peak patterns. 
\end{inparaenum}
The first observation motivated us to evaluate the potential of a Gaussian Mixture Model (GMM) to estimate the interference, and the second led us to use two interleaved models to account for the observed patterns. The estimated interference generated by the GMM is used as input for predicting white spaces 
to allow low-power wireless nodes to better schedule their transmissions. 

The accuracy of our GMM-based interference estimation model is evaluated \wrt ground truth traces. Our results show that the accuracy we obtain with our approach, over $94.7\%$ in all tested cases, 
is significantly higher than the state of the art Pareto model~\cite{Huang2010} and our previous approach based on a second-order Markov-Modulated Poisson Process (MMPP(2)) model~\cite{IndikaLCN}. 
Moreover, the accuracy of the white space prediction is $97.7$\% and $89.5$\% in the two tested environments, the \office and the \home. Furthermore, we propose a proactive model-based receiver-aware MAC protocol, \proposedmac,
for low-power wireless networks, which is based on the interference estimation and the white space prediction mechanisms. 
\proposedmac is a data-driven MAC protocol which requires interference models for its operation. Moreover, as the same interference model is used on both sides of the communication channel, \proposedmac's white space prediction mechanism is able not only to predict transmission opportunities at the sender but also to synchronise the sender and the receiver to find a rendezvous point. The added functionality of the white space prediction mechanism considerably alleviated the design of \proposedmac to address CTI. 
\proposedmac achieves higher dependability, a $1.2$\% increase in \textit{packet delivery ratio}, and a $0.02$\% decrease in \dutycycle than the state of the art CRYSTAL~\cite{crystal} under bursty indoor interference. 

It is worth noting that according to the literature dependability of systems comprises of multiple attributes, such as availability, reliability, safety, confidentiality, integrity, and maintainability, and is defined as follows: ``dependability of a system is the ability to avoid service failures that are more frequent and more severe than is acceptable to the user(s)''~\cite{Avizienis2004, Avizienis2004_2}. In this work, following the aforementioned definition, a wireless network is considered to be highly dependable when it delivers high communication reliability and high energy efficiency.
Furthermore, IEEE~$802.11$ is a wireless Ethernet standard that provides interoperability guidelines for vendors that produce wireless devices for local area networking based on the aforementioned standard. WiFi is a wireless technology that is based on the IEEE~$802.11$ standard. However, commonly IEEE~$802.11$ and WiFi are used interchangeably and this work will follow the same.

The remainder of the paper is structured as follows. In Section~\ref{sec:related_work}, interference detection and classification methods, interference models, and dependable data collection systems are reviewed. An overview of our MAC protocol is presented in Section~\ref{sec:overview}. Section~\ref{sec:pre_deployment} and Section~\ref{sec:solution} illustrate the proposed solution in more detail and Section~\ref{sec:perf_eval} demonstrates its performance. We discuss the limitations of our models, explore possible future work, and conclude the paper with brief but important remarks in Section~\ref{sec:discussion}.

\section{Related Work}
\label{sec:related_work}

\fakeparagraph{Interference Detection and Classification} Several works aim to measure interference and understand its impact on low-power wireless networks, and classify interference sources~\cite{Noda2011,Iyer2015,TIIM,sonic,crosszig,Musaloiu-E2008b}. 
Musaloiu and Terzis~\cite{Musaloiu-E2008b}, use \rssi based features to quantify the interference on all IEEE~$802.15.4$ channels to select the least interfered one. 
Noda et al.~\cite{Noda2011} compute the ratio of channel idle and busy time for assessing channel quality in the presence of interference. 
SpeckSense~\cite{Iyer2015} classifies \rssi bursts to characterise the channel as periodic, bursty or a combination of both. SoNIC~\cite{sonic} uses information from corrupted packets for interference source classification. While these works succeed in detecting interference and identifying the source of interference, it is not clear how the techniques are useful for autonomous interference mitigation due to the diversity of interference sources, such as IEEE~$802.11$/WiFi, Bluetooth, and radio emissions from microwave oven.  
TIIM~\cite{TIIM} goes a step further and extracts features from corrupted packets to quantify the interference conditions instead of identifying the interferer. Thus the interference condition can be mapped to a specific mitigation technique. Nonetheless, TIIM only recommends 
countermeasures that can be applied to prevailing interference conditions but does not provide their implementation. 
CrossZig~\cite{crosszig}, the follow-up work to TIIM, contains an implementation of an adaptive packet recovery and FEC coding scheme to address the problem. 
According to real-time interference level assessment based on a measure of Packet Delivery Ratio (\pdr), ART~\cite{ART} proposes a probabilistic mechanism to adaptively enable CSMA only under severe interference and otherwise uses their flexible multi-channel access mechanism called FAVOR, which facilitates to fine-tune the trade-off between throughput and \pdr. 
Grimaldi et al.~\cite{Grimaldi2019} use manifold supervised-learning classifiers for real-time identification of multiple sources of interference, such as WiFi, IEEE~$802.15.4$, Bluetooth, and Bluetooth Low Energy (BLE), by extracting envelope and spectral features of the underlying interference signals. Their technique can identify statistics of concurrent interference in adverse conditions. 
All these solutions, however, are reactive, depending on the prevailing channel conditions, power hungry, and do not aim to predict the white spaces through modelling, which is the goal in this paper.

\fakeparagraph{Interference Modelling} Creating lightweight models of interference is not a trivial task. Several researchers have proposed models for channel occupancy~\cite{Stabellini2010,Huang2010,boanojag,Geirhofer2008,Lagana2012} and for emulating interference caused by WiFi and Bluetooth~\cite{jamlab}. 
A two-state semi-Markov model for channel occupancy is defined in~\cite{Stabellini2010}, and exploited by each node to identify the least interfered channel and to switch accordingly. In comparison, we do not limit interference caused only by WiFi, but identify the white spaces for a specific channel through modelling interference in the time domain. 
For modelling WiFi interference, Geirhofer et al.~\cite{Geirhofer2008} propose a semi-Markov model and its continuous-time Markov chain, while Lagan\`a et al.~\cite{Lagana2012} enhance this model with the ability to distinguish detected and undetected WiFi activities. This model considers the limited detection range of sensor nodes and uses likelihood maximisation and neural networks for estimating model parameters. 
Boano et al.~\cite{Boano:2010:MSM:2127940.2127963,boanojag} define a two-state semi-Markov model for channel occupancy and noise measurements are used to measure the duration of the \free and \busy instants, and compute their CDFs. Based on the longest \busy period, the MAC protocol parameters are derived to meet the application requirements. 
JamLab~\cite{jamlab} models and regenerates WiFi/Bluetooth/microwave interference patterns using sensor nodes, considering both saturated (always \busy) and unsaturated traffic scenarios. A Markov chain model is used for saturated traffic and a probability mass function of empirical data for the non-saturated one. In contrast, our goal is not to emulate interference traffic but to estimate it, and for this we use a Gaussian Mixture Model (GMM) to capture the ambient interference conditions. 
The work in~\cite{Huang2010} is closely related to ours, focusing on a model-based prediction of the length of the immediate white space when a ZigBee frame is ready to be transmitted in the presence of WiFi interference. Depending on the length of the white space, the MAC frame is split in order to minimise collision probability. Nevertheless, continuous sampling of the operating channel is required as the model's parameters are calibrated whenever there is a frame to be transmitted. Moreover, their prediction is short-term in contrast to ours, which is long-term and provides more information as to when to transmit. 

\fakeparagraph{Cognitive Radio Solutions} Cognitive Radio (CR) is a technology envisaged to solve problems in wireless networks emerging due to scarce frequency spectrum and its inefficient allocation/usage~\cite{Akyildiz2009}. CR-enabled devices are able to change their transmitter parameters based on interaction with the environment in which they operate~\cite{fcc2003docket}. Several works have proposed to enhance the communication in wireless networks using CR~\cite{Cordeiro2007, Su2007, Kondareddy2008, COMAC, sca_mac, Chen2013, Hossain2018}. 

C-MAC~\cite{Cordeiro2007} exploits a superframe based distributed multi-channel MAC protocol to tackle the dynamics of resource availability due to primary user activities. Here, the coordination amongst the nodes about channel usage is accomplished with a dynamically assigned Common Control Channel (CCC). 
Su and Zhang~\cite{Su2007} use two transceivers, one for conveying control data over a dedicated CCC while the other is used for data communication.
The authors use different sensing policies for finding available idle channels and a time-slotted mechanism for coordination between nodes. The node that detects an idle channel informs the other nodes via the CCC with the use of beacons in mini-slots. 

The use of a CCC leads to problems such as single point of failure and channel saturation with an increasing number of users. 
SYN-MAC~\cite{Kondareddy2008} avoids the CCC and uses a hybrid MAC protocol wherein the exchange of control signals is done in a time-slotted fashion while data transmission is based on random access. SYN-MAC shows better connectivity and higher throughput than CCC based protocols in a congested network. 
COMAC~\cite{COMAC} uses a contention-based handshaking mechanism for the exchange of control information. The protocol shares information regarding locally available channels with the receiver for selecting the set of data channels based on dynamically adjusted signal to interference and noise ratio. 

SCA-MAC~\cite{sca_mac} exploits the statistics of spectrum usage for decision making on channel access. To this end, for each channel, a list of the last $1000$ channel idle duration is maintained. 
CR-CSMA/CA~\cite{Chen2013} is another multi-channel MAC protocol that extends the traditional RTS/CTS to a three-way-handshake mechanism PTS (Prepare To Send)/RTS/CTS for channel access coordination. 

CR-RDV~\cite{Hossain2018} is a CSMA/CA-based distributed CR rendezvous MAC protocol to overcome the channel contention and rendezvous problem (rendezvous collision) in wireless networks, which occurs when multiple devices achieve rendezvous on the same channel. The data channel is selected based on the receiver preferences. To this end, the protocol maintains a list of backup channels to be used during service interruptions and the channel list is integrated into RTS/CTS packets. 

Most of these CR solutions require multi-channel spectrum sensing and the use of a global or a local CCC for the exchange of control information. To satisfy those requirements, wireless devices need to spend a considerable amount of energy. Thus, these solution need more research before adopting them in low-power wireless networks. 

\fakeparagraph{Low-power Data Collection} Wireless sensor networks are used to monitor, record, and disseminate physical conditions, such as temperature and humidity, in their operating environments. Once deployed, these functions are executed without human interactions, thus the reliability and the lifetime of the wireless network are crucial factors. Several works have emerged over the last decade that aim to improve the dependability of low-power wireless networks, \eg \cite{Dozer2007, CompetitionRedFixHopCH2017, CompetitionSparkle2016, CompetitionBigBangBus2018, CompetitionRedNodeBus2019, CompetitionCrystal2018}. 

A network stack for periodic data-gathering is proposed in Dozer~\cite{Dozer2007}, wherein the scheduling of the transmission is distributed using tree-based routing. The energy consumption is minimised with the proper coordination of MAC, topology control, and routing. 
RedFixHop~\cite{CompetitionRedFixHopCH2017} exploits flooding and constructive interference for hardware-triggered simultaneous transmissions for high communication reliability, which is further increased by channel hopping. Even though they achieve high communication reliability, the re-transmissions based on time redundancy increases the energy consumption of the wireless network. Moreover, due to the limitation in the commodity hardware, the payload is restricted to $1$~Byte only. 
By combining flooding, topology control, and transmission power control, Sparkle~\cite{CompetitionSparkle2016} decreases the energy consumption of low-power wireless networks with a slight improvement in reliability. 
Different to RedFixHop~\cite{CompetitionRedFixHopCH2017}, BigBangBus~\cite{CompetitionBigBangBus2018} uses software-triggered transmissions to increase the payload of the transmissions, while exploiting flooding, capture effect, and frequency diversity. Nonetheless, the improvement of the payload size is $2$~Bytes. 
RedNodeBus~\cite{CompetitionRedNodeBus2019} overcomes the restrictions of very short packets by using long preambles, which eases tight synchronisation and boosts the possibility of capture effect. 
CRYSTAL Clear~\cite{CompetitionCrystal2018} combines synchronous transmissions with channel hopping and noise detection in which nodes go to sleep if there is strong noise in the operating channel to save energy. The channel is marked noisy when the clear channel assessment reads above a predefined threshold. CRYSTAL tackles harsh interference by escaping it with channel hopping, while the noise detection schedules extra transmissions in a decentralised way for fighting interference. The latter, however, may keep nodes unnecessarily active, which deteriorates the dependability of the low-power wireless network. 

With the design and implementation of Glossy~\cite{glossy}, almost all communication protocols/systems tend to exploit synchronous transmissions, constructive interference, and capture effect for designing dependable solutions. 
Nonetheless, our work differs from the current trend of designing dependable low-power wireless networks, as our approach is data-driven and is based on interference models and a prediction mechanism to find rendezvous points for the communication between nodes. Moreover, our approach will pave the way toward new research directions in the wireless research community in general. 

\section{ \proposedmac Overview}
\label{sec:overview}

Cross-technology interference (CTI) in a shared communication medium can be a significant problem for coexistence of collocated dissimilar wireless networks, especially for low-power, resource-constrained networks such as IEEE~$802.15.4$ based networks. There are many reactive techniques to address the coexistence problem wherein the medium is checked before transmissions. 
However, there is still a possibility that the packets may be corrupted during transmission due to concurrent transmissions from collocated wireless devices. 
Because of this reason, the reliability of communication decreases while the energy consumption of devices increases because of re-transmissions to maintain reliability, leading to a low-dependable network. 

Therefore, we propose \proposedmac, a proactive model-based receiver-aware MAC protocol, to address the coexisting problem in a shared wireless medium, such as the $2.4$~GHz ISM band. \proposedmac can estimate interference patterns and predict transmission opportunities for IEEE~$802.15.4$ based WSNs to have high dependability.  


\proposedmac consists of two phases: \textbf{deployment} and \textbf{model-based data communication}. Since the proposed solution is a model-based technique, the models in use need to be trained before they can be used. A prerequisite is to collect interference traces that will be used for computing model parameters. This is done in a \textbf{deployment} phase before commencement of the data communication phase. The nodes collect interference traces in order to assess interference characteristics in the target environment and the traces are used to compute the interference model parameters.
Section~\ref{sec:pre_deployment} illustrates the methodology that we used to collect interference traces, how we characterised interference, and how the model parameters were computed. 

At the end of the deployment phase, the \textbf{model-based data communication} can start its operation. Here, nodes go first through an initialisation phase for
\begin{inparaenum}[$a)$]
\item time synchronisation, 
\item acquiring routing information, and 
\item model exchanges.
\end{inparaenum}
 
In our proposed approach, the interference model is used to predict when nodes can transmit without interference being present, so called white spaces. As such, nodes will wake up based on a local prediction of white spaces to receive a senders transmission based on the predicted white space. Therefore, \proposedmac uses a time slot based mechanism with short time-slots for medium access that reflect white spaces. This requires a tight time synchronisation mechanism in the sensor network to make sure time-slots are aligned across neighbour nodes. The periodic time synchronisation allows all nodes in the network to update their clock \wrt the clock of the network coordinator. 

The knowledge of a node's next hop neighbour is important for the correct operation of \proposedmac. This knowledge is acquired from the routing protocol running on the node, and \proposedmac relies on the underlying protocols that perform neighbour discovery and routing. 

During the network initialisation phase, nodes also exchange their interference models that were computed during the deployment phase with each other. Nodes broadcast their interference models in a round-robin fashion to their neighbours.

After exchanging models, nodes go into radio duty cycling to save energy. A node maintains two radio states related to communication: active and sleeping. A node goes into the former state, in which the radio transceiver is on, whenever there is a data/control packet to be transmitted/received, and sleeps otherwise. Receiver-aware communication starts after the model exchange timeout which is a configurable parameter to ensure nodes in the network can exchange their models fully. Nodes are now ready for data communication. 

In the following, we consider that a typical periodic data collection application runs on the wireless sensor network. 
Whenever a packet is ready to be transmitted, a node wakes up and utilises a free slot predicted by the interference model of the next hop, which  was shared by the neighbour during network initialisation. For successful communication, both ends of the communication channel should synchronise their radio state. It is worth noting that each node goes into its active state, \ie switches on its radio, to receive packets from its neighbours, following the predictions of its own interference model. In the circumstances where the own interference model does not predict a free slot, the node considers the very next slot as free. Therefore, the rendezvous is readily achieved when the sender utilises the predictions from the neighbour's interference model. The packet transmission is done irrespective of the presence of interference at the sender. 

Because the interference models need to be adapted to the dynamic interference conditions, while the application is running, the network coordinator periodically scrutinises the network-wide Packet Delivery Ratio (\pdr). If the moving average of \pdr crosses below a predefined threshold, the network coordinator triggers a command to 
select new models. Dissemination of the command is done by flooding a 
model selection control packet throughout the network. Upon receiving the 
model selection control packet, a node switches to the second interference model which was computed during the deployment phase, \ie from peak to off-peak model, and vice versa. Section~\ref{sec:solution} describes \proposedmac in more detail.

\section{Deployment}
\label{sec:pre_deployment}

\begin{figure*}
\centering
\includegraphics[trim={0cm, 0cm, 0cm, 0cm}, clip, width=0.75\textwidth, keepaspectratio]{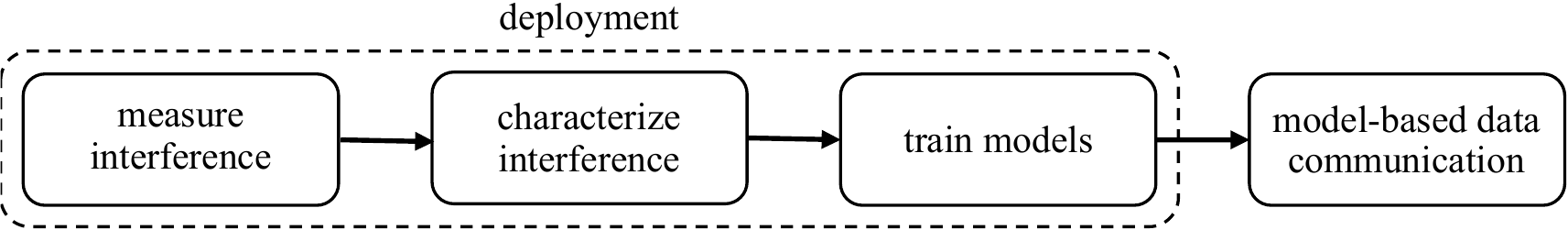}
\caption{Overview of the deployment phase.}
\label{fig:overview_pre_deployment}
\end{figure*}

In this section, we present the key aspects of the deployment phase in detail. Once deployed, an IoT network, such as an IEEE~$802.15.4$ low-power wireless network, works autonomously to execute the application tasks that it is programmed to do. However, in \proposedmac, before going operational, the nodes in the network must acquire an accurate understanding of the interference in the surrounding radio environment. During this phase, interference traces are collected in order to assess interference characteristics in the deployed environment. This process helps the nodes to perform three key functions depicted in Fig.~\ref{fig:overview_pre_deployment}: measure and characterise interference, and train interference models. 

\subsection{Measure Interference}
\label{sec:interference_traces}

Before commencing data communication, all the nodes in the wireless network measure interference. This is the fundamental task a node must perform to understand the radio interference. To achieve this, a node listens on its operating IEEE~$802.15.4$ channel and classifies the receiving signals as interference or not based on the two hardware interrupts of the transceiver: Clear Channel Assessment (CCA) and Start Frame Delimiter (SFD).

Sampling the radio environment for interference characterisation and parametrising the interference models costs nodes a significant energy budget. Therefore, it is sensible to use a separate network of nodes for collecting interference traces during the deployment phase. The models trained from the collected traces can then be used by the primary sensor network. 

\fakeparagraph{Location} 
We conducted interference measurements in two typical indoor environments: \office and \home. The former is an office building, and the latter is a shared apartment in a student dormitory exhibiting more bursty traffic. Both environments consist of multiple sources of radio signals, such as WiFi, Bluetooth, and Microwave ovens, operating in the $2.4$~GHz ISM band. 

\fakeparagraph{Hardware/software platforms} 
We used the TMote Sky hardware platform \cite{tmotesky}, equipped with the ChipCon2420 (CC$2420$) radio module, which is compliant with the IEEE~$802.15.4$ standard, for interference measurements. The nodes were placed on walls along the hallway in both the \office and the \home at $1.75$~m height and were connected to a USB port of a PC. Leveraging the experience from~\cite{Huang2010}, hardware interrupts from the Clear Channel Assessment (CCA) and the Start of Frame Delimiter (SFD) pins of the CC$2420$ radio module are used for fast interference detection. When the radio module receives a signal above the CCA threshold (\ie $-77$~dBm), the CCA pin goes low to indicate a busy channel; the SFD pin goes high if the receiving signal is an IEEE~$802.15.4$ packet. Therefore, we can identify the presence of interference by capturing time instances where both CCA and SFD pins are low. When the interference is present, a packet is sent from the node to the connected PC via USB for time-stamping. We ensure that the time-stamping delay is constant as the node always transmits one character for marking interference.  

\fakeparagraph{Measurement execution} 
We conducted three measurement campaigns for gathering interference traces: \first, \second and \third. The \first and the \second were performed in the \office. During the \first, three nodes were deployed at the same location between two WiFi APs (Access Points) detecting the interference on IEEE~$802.15.4$ channels $13$, $18$ and $23$. For the \second, the three nodes and the APs were interleaved and used channel $18$ only. These choices paved the way to explore interference traces from IEEE~$802.15.4$ channels overlapping with different WiFi channels and different characteristics of WiFi traffic. The collection of traces was executed for $24$~hours, during a working day of the week, from $1$:$00$~PM and $4$:$00$~PM in the \office and the \home respectively. For the \third campaign, a single node was used to collect traces on channel $18$ from the \office and the \home. As the goal was to assess the interference for the long-term, the measurement campaign was run for two weeks, during September $12$-$26$, $2017$. 


\subsection{Characterise Interference}
\label{sec:interference_characterization}

\fakeparagraph{Methodology}
We recall that our goal is to predict white spaces 
(un-interfered slots)
for transmission by nodes of the low-power wireless network in the presence of interference. For the interference characterisation, we divided the time axis into slots of a fixed duration of $100$~ms. We empirically determined that a longer slot lengths lead to better prediction accuracy but reduce the throughput of the application. We found $100$~ms to be a good trade-off for both. More information on the impact on the slot length toward the performance of low-power wireless networks is presented in Section~\ref{sec:perf_eval}. 

To mark a time-slot as \free, it should contain at least $8.512$~ms interference-free duration, which is the time required for a $133$ bytes IEEE~$802.15.4$ frame and its ACK to be transmitted without preemption. 
Since we use time-slots, the lower and upper bounds on the length of white spaces are $8.512$~ms and $100$~ms, respectively. We characterise the traces in terms of the mean interference Inter-Arrival Time (IAT) and the number of arrival signals per slot statistics. Although mean IAT is the most directly informative statistical property of the interference trace, if used alone to characterise the traffic within a slot leads to an increase in the false discovery rate of bursts of interference signals (\busy periods). Therefore, we use two thresholds, \thiat and \thcount, respectively for mean IAT and the number of interference signal arrivals per slot, to decide the status of the channel as follows:
\[
   Channel = 
\begin{cases}
    \busy,\text{ if } IAT\leq \thiat \text{ \& } \\ \mkern63mu count\geq TH\textsubscript{count}\\
    \free,  \text{ otherwise }
\end{cases}
\]
We refer readers to our previous work~\cite{our_dcoss_paper_short} for more information on this. 

\begin{table}
\centering
\captionsetup{width=.47\textwidth}
\caption{NCLR comparison of the interference traces from \first and \second.}
\setlength\extrarowheight{1pt}
\label{tab:NCLR}
\resizebox{\columnwidth}{!}{%
\begin{tabular}{|c|c|c|c||c|c|c|c|} \hline
\multicolumn{4}{|c||}{\textbf{FIRST}}                    & \multicolumn{4}{c|}{\textbf{SECOND}}                 \\ \hline
\textbf{Channel} & \textbf{13} & \textbf{18} & \textbf{23} & \textbf{Location} & \textbf{1} & \textbf{2} & \textbf{3} \\ \hline
\textbf{13} & 0           & 0.78        & 0.22        & \textbf{1}   & 0          & 0.12       & 0.88       \\
\textbf{18} & 0.78        & 0           & 1        & \textbf{2}   & 0.12       & 0          & 1          \\
\textbf{23} & 0.22        & 1           & 0           & \textbf{3}   & 0.88       & 1          & 0    \\ \hline
\end{tabular}
}
\end{table}

\begin{figure*}[ht!]
\centering
\begin{subfigure}[t]{0.225\textwidth}
\includegraphics[trim={0.5cm, 0.4cm, 4.6cm, 0.4cm}, clip, height=0.8\textwidth, keepaspectratio, right]{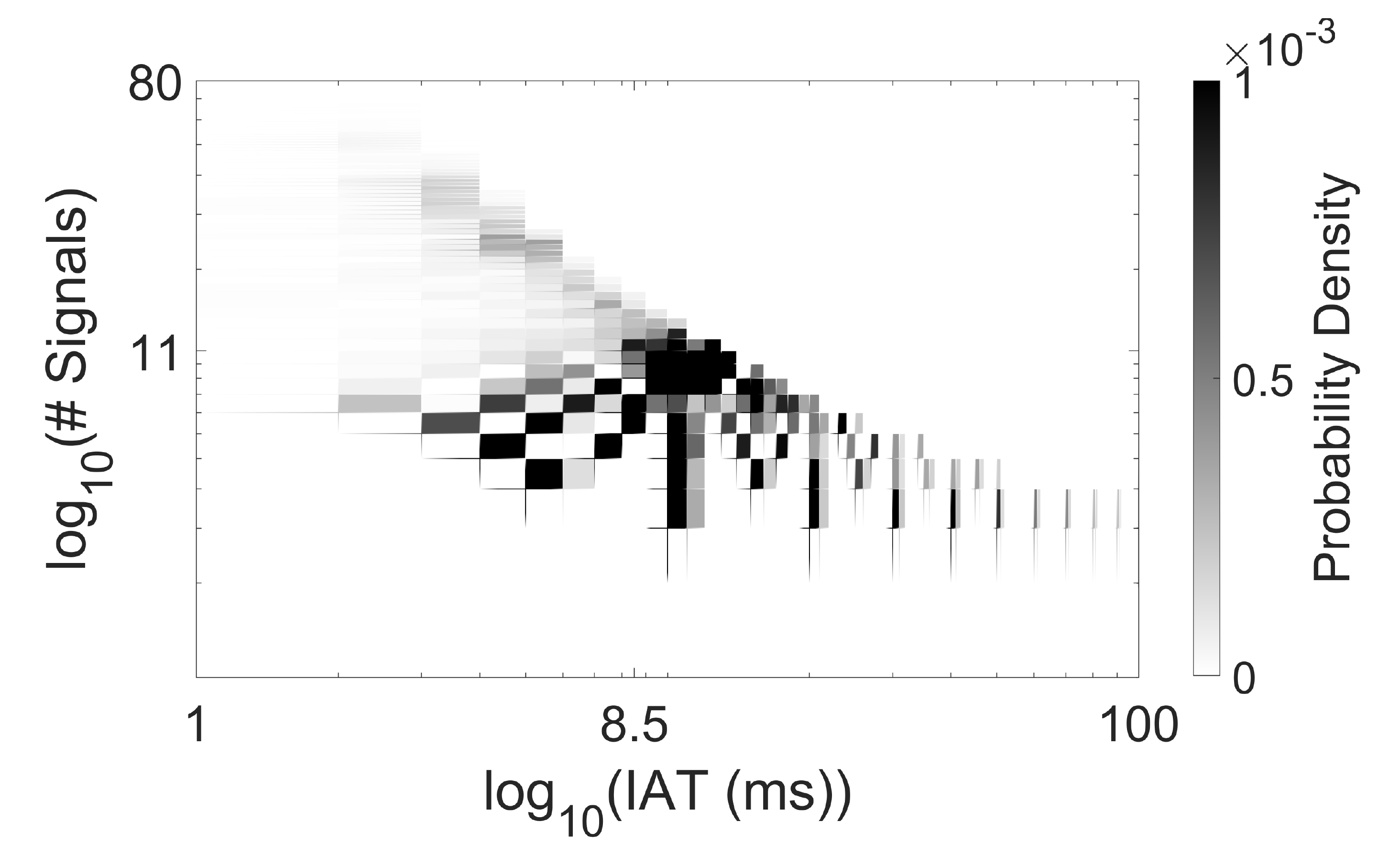} 
\label{fig:trace_scenario1_ch13}	
\end{subfigure}
\begin{subfigure}[t]{0.225\textwidth}
\includegraphics[trim={4.0cm, 0.4cm, 4.6cm, 0.4cm}, clip, height=0.8\textwidth, keepaspectratio, center]{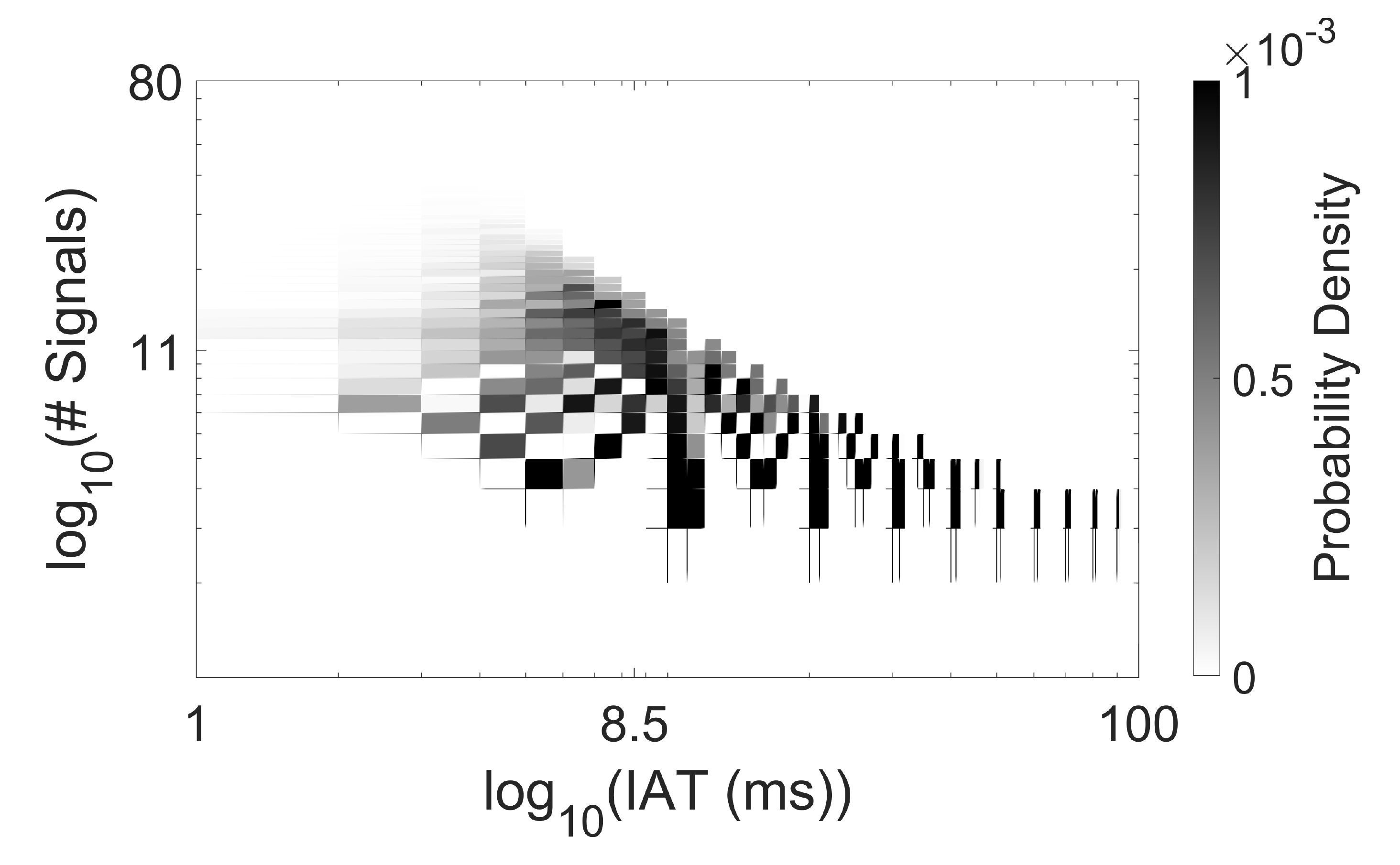}
\label{fig:trace_scenario1_ch18}	
\end{subfigure}
\begin{subfigure}[t]{0.225\textwidth}
\includegraphics[trim={4.0cm, 0.4cm, 0.8cm, 0.4cm}, clip, height=0.8\textwidth, keepaspectratio, left]{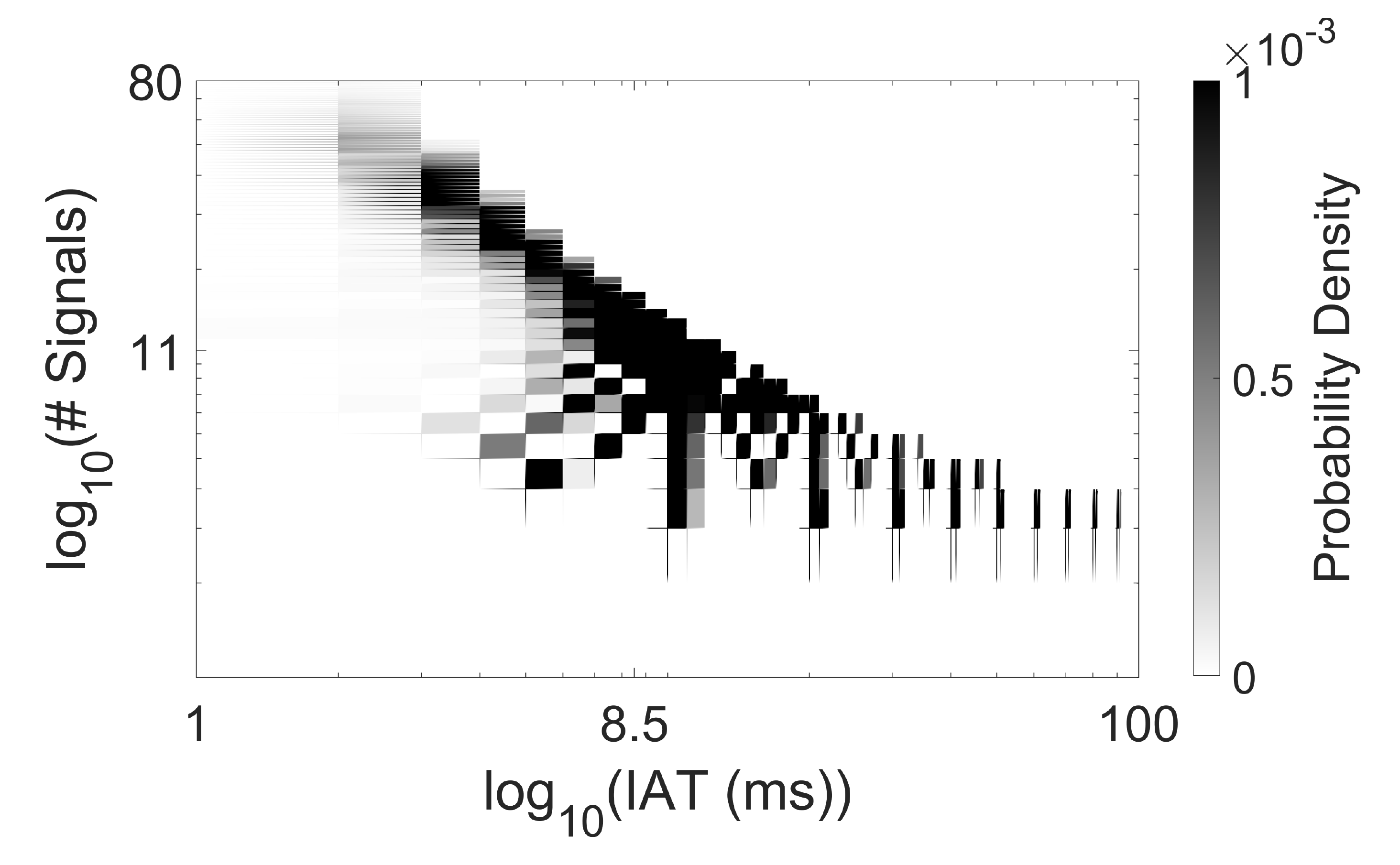}
\label{fig:trace_scenario1_ch23}	
\end{subfigure}
\caption{Probability distribution function of traces from \first on channel 13 (left), 18 (center) and 23 (right).} 
\par\bigskip
\label{fig:traces_scenario1}
\begin{subfigure}[t]{0.225\textwidth}
\includegraphics[trim={0.5cm, 0.4cm, 4.6cm, 0.4cm}, clip, height=0.8\textwidth, keepaspectratio, right]{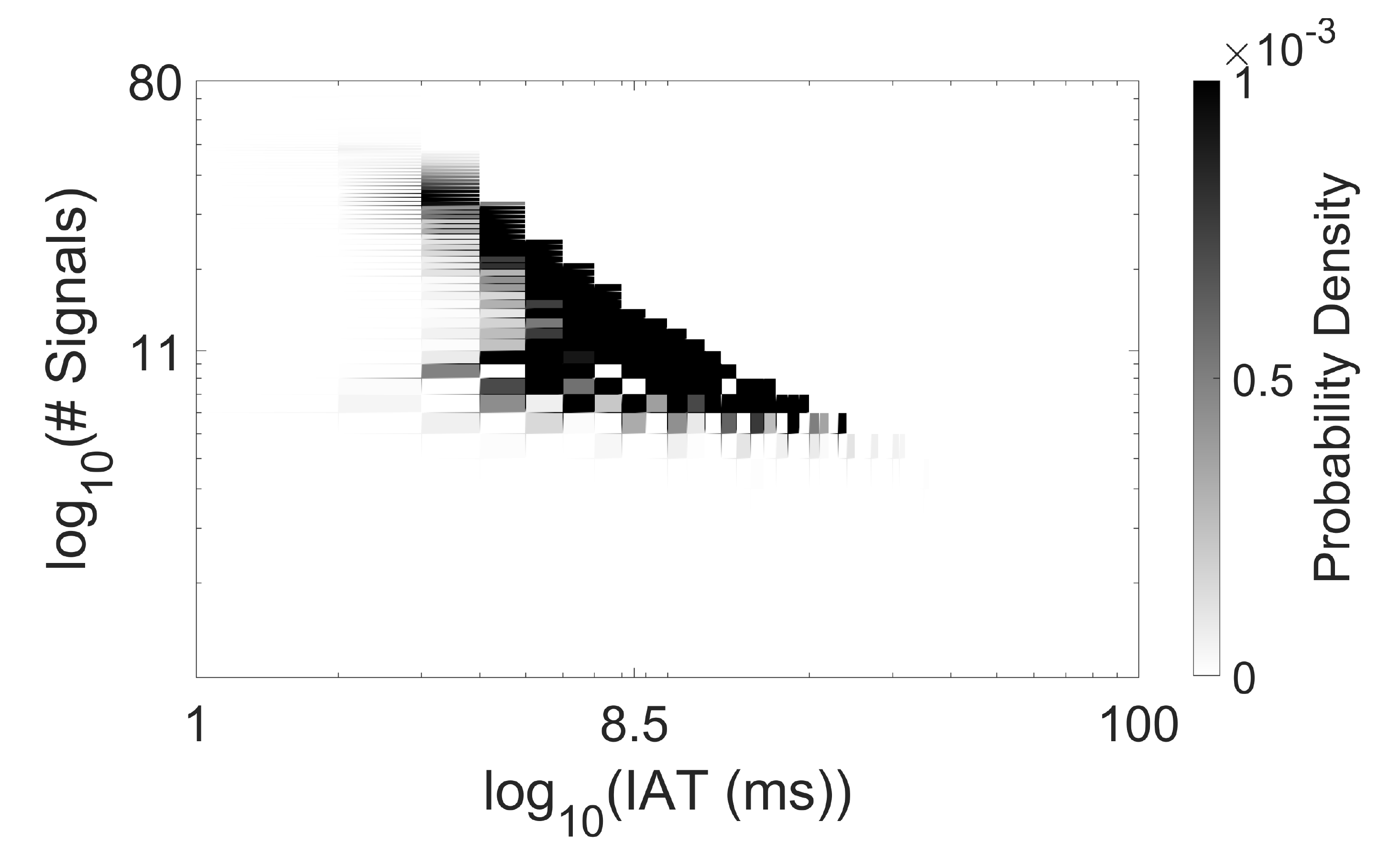}
\label{fig:trace_scenario2_loc1}	
\end{subfigure}
\begin{subfigure}[t]{0.225\textwidth}
\includegraphics[trim={4.0cm, 0.4cm, 4.6cm, 0.4cm}, clip, height=0.8\textwidth, keepaspectratio, center]{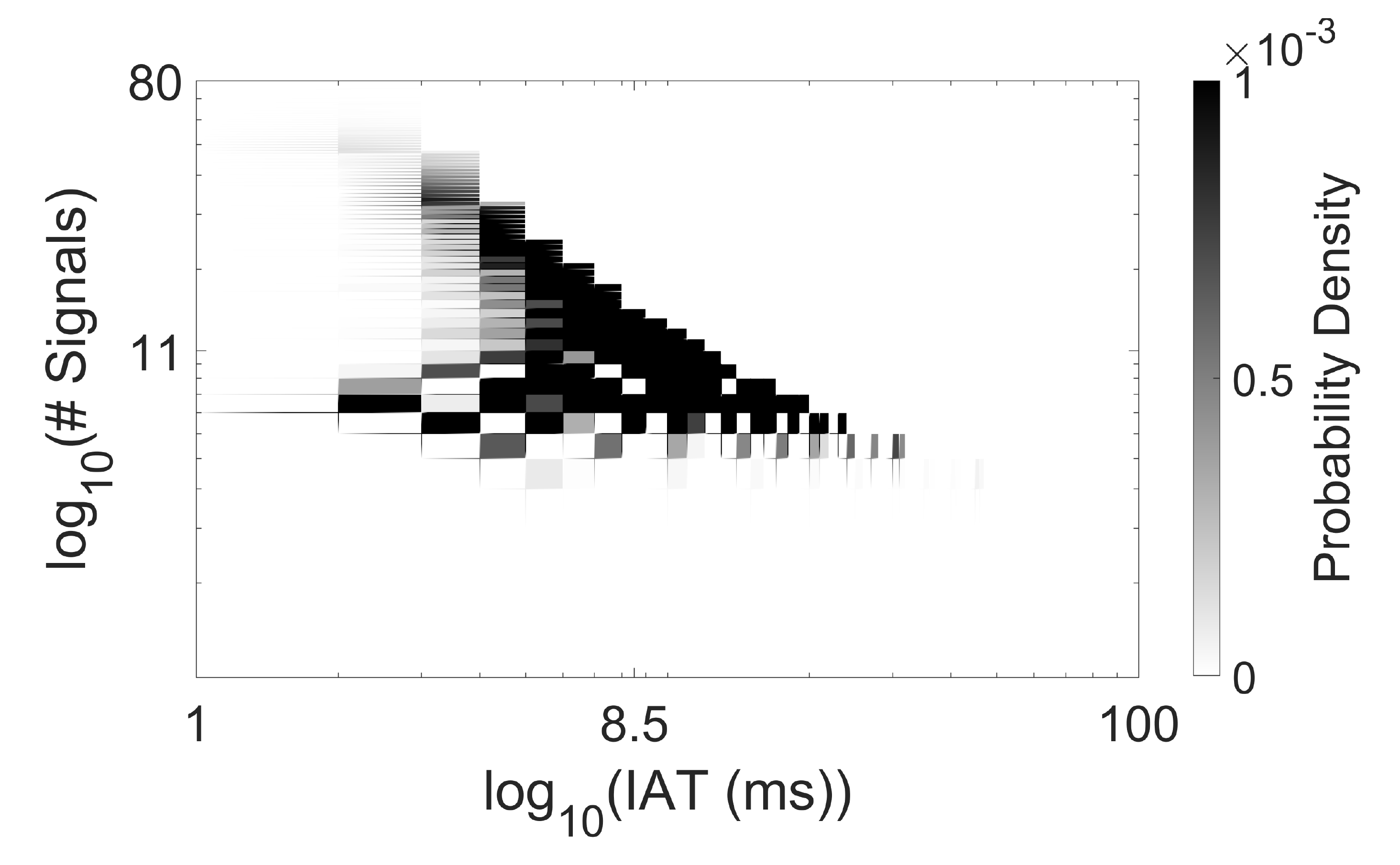}
\label{fig:trace_scenario2_loc2}	
\end{subfigure}
\begin{subfigure}[t]{0.225\textwidth}
\includegraphics[trim={4.0cm, 0.4cm, 0.8cm, 0.4cm}, clip, height=0.8\textwidth, keepaspectratio, left]{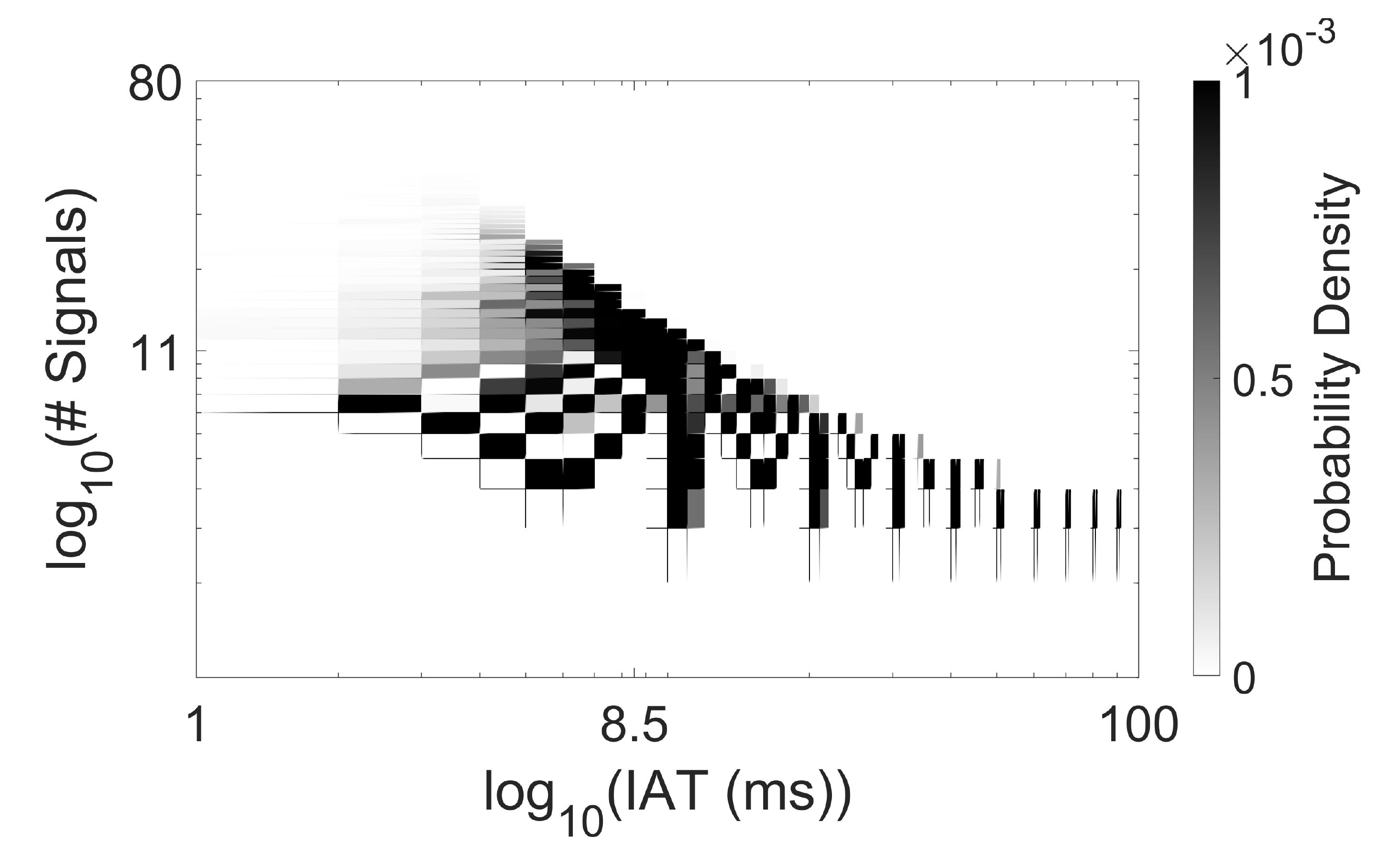}
\label{fig:trace_scenario2_loc3}		
\end{subfigure}
\caption{Probability distribution function of traces from \second at location 1 (left), 2 (center), and 3 (right).} 
\par\bigskip
\label{fig:traces_scenario2}
\begin{subfigure}[t]{0.225\textwidth}
\includegraphics[trim={0.5cm, 0.4cm, 4.6cm, 0.4cm}, clip, height=0.8\textwidth, keepaspectratio, right]{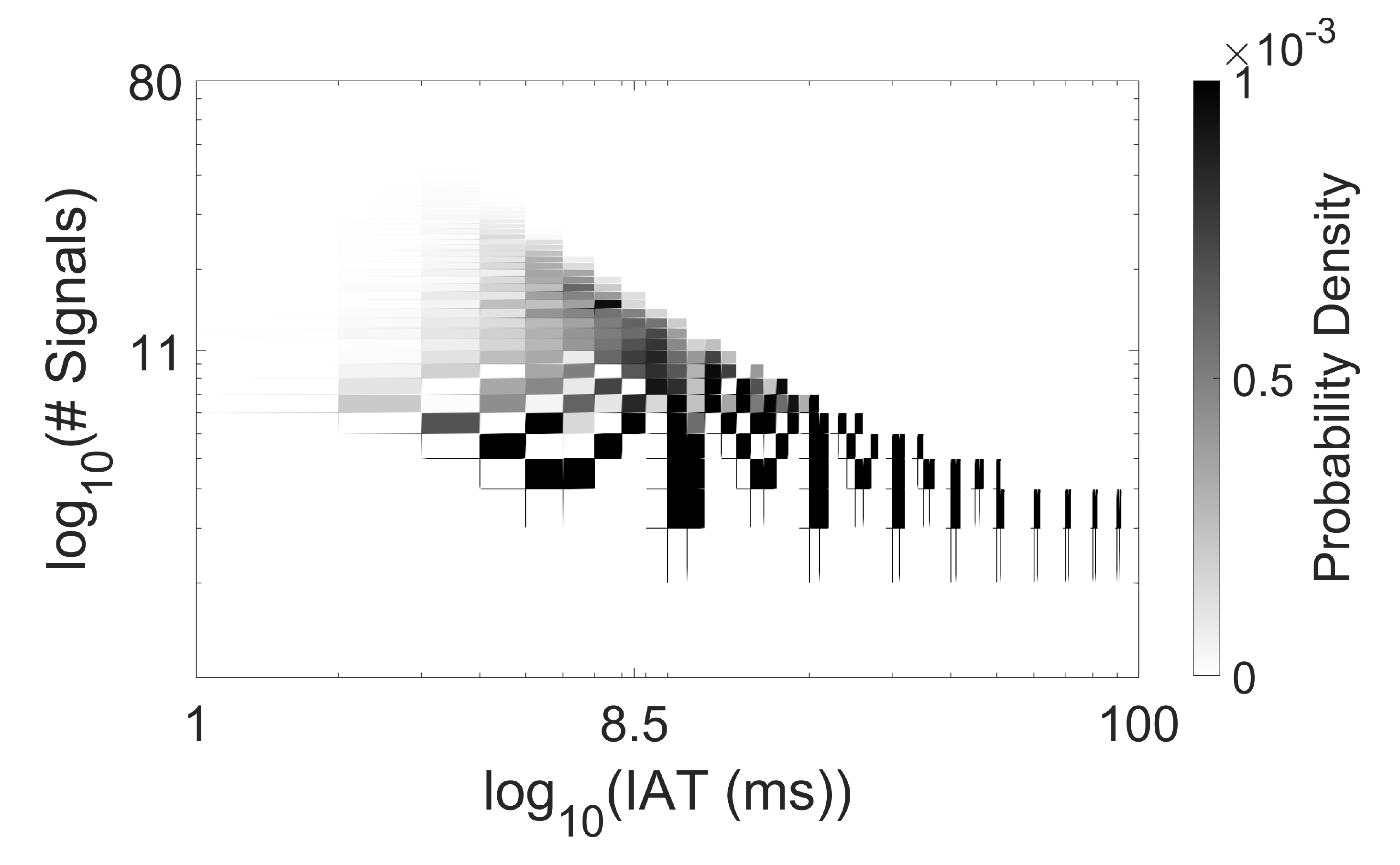}
\caption{First week \office.} 
\label{fig:trace_office_1}	
\end{subfigure}
\begin{subfigure}[t]{0.225\textwidth}
\includegraphics[trim={4.0cm, 0.4cm, 4.6cm, 0.4cm}, clip, height=0.8\textwidth, keepaspectratio, right]{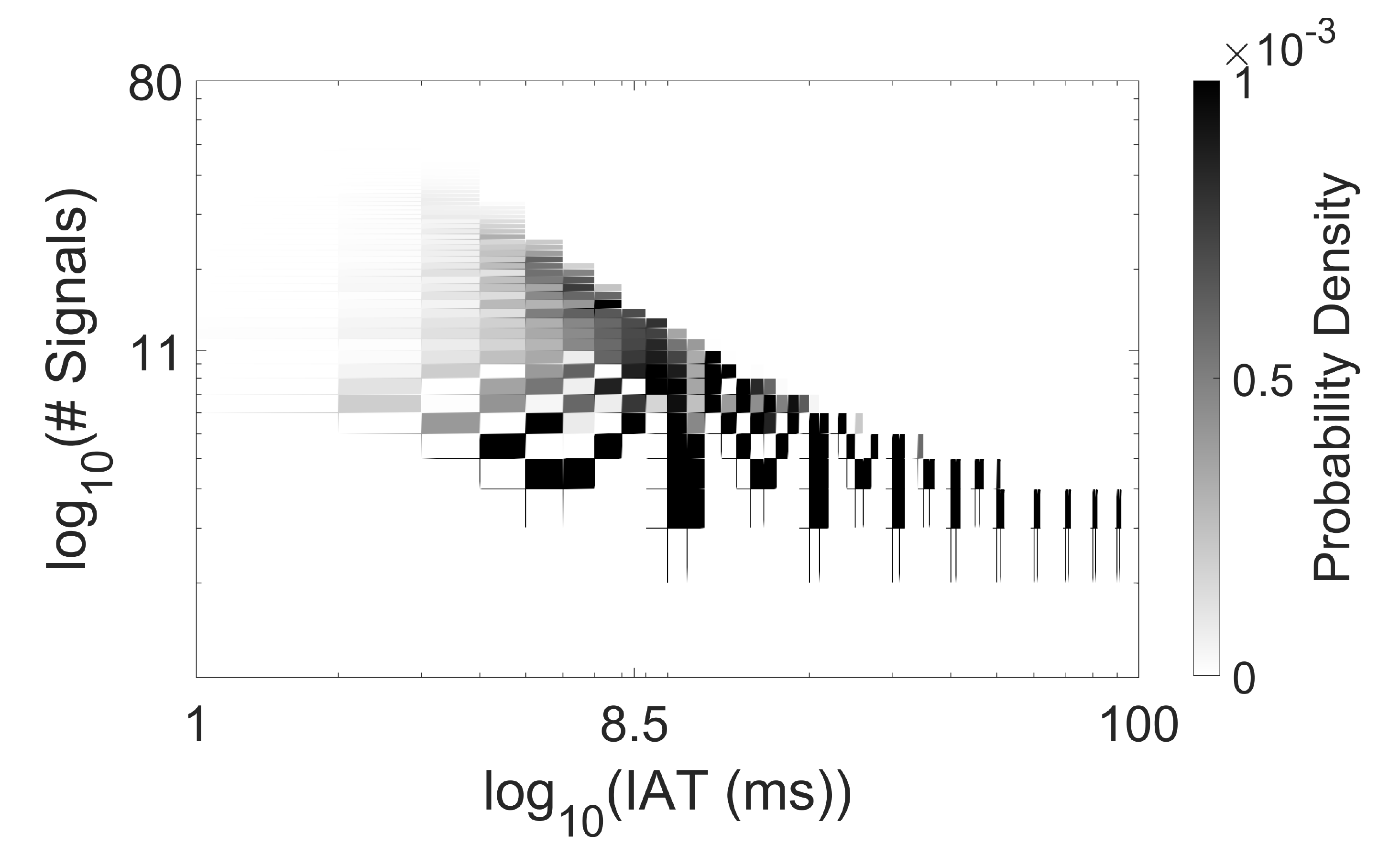}
\caption{Second week \office.} 
\label{fig:trace_office_2}	
\end{subfigure}
\hspace{0.5mm}
\begin{subfigure}[t]{0.225\textwidth}
\includegraphics[trim={4.0cm, 0.4cm, 4.6cm, 0.4cm}, clip, height=0.8\textwidth, keepaspectratio, left]{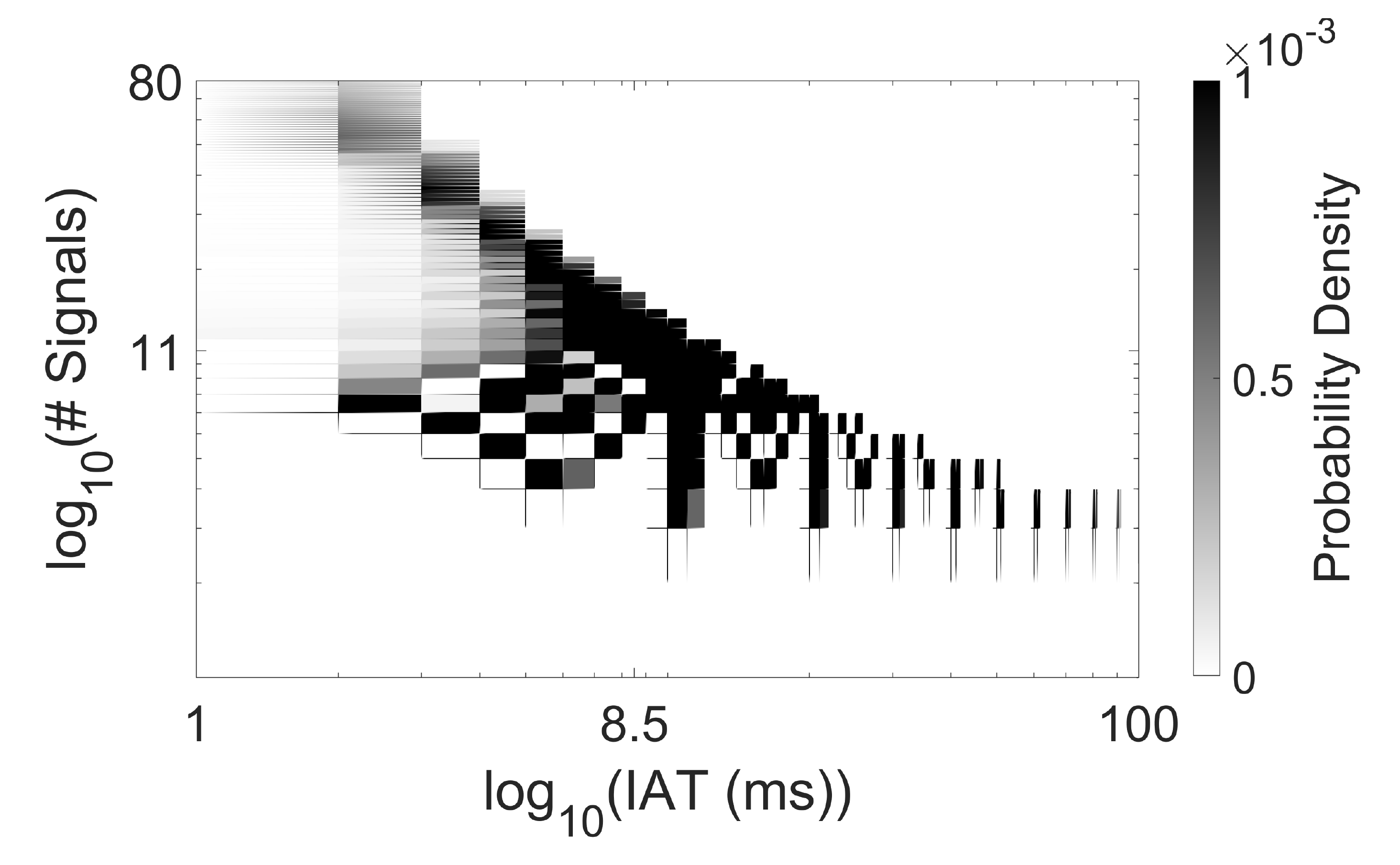}
\caption{First week \home.} 
\label{fig:trace_home_1}	
\end{subfigure}
\begin{subfigure}[t]{0.225\textwidth}
\includegraphics[trim={4.0cm, 0.4cm, 0.8cm, 0.4cm}, clip, height=0.8\textwidth, keepaspectratio, left]{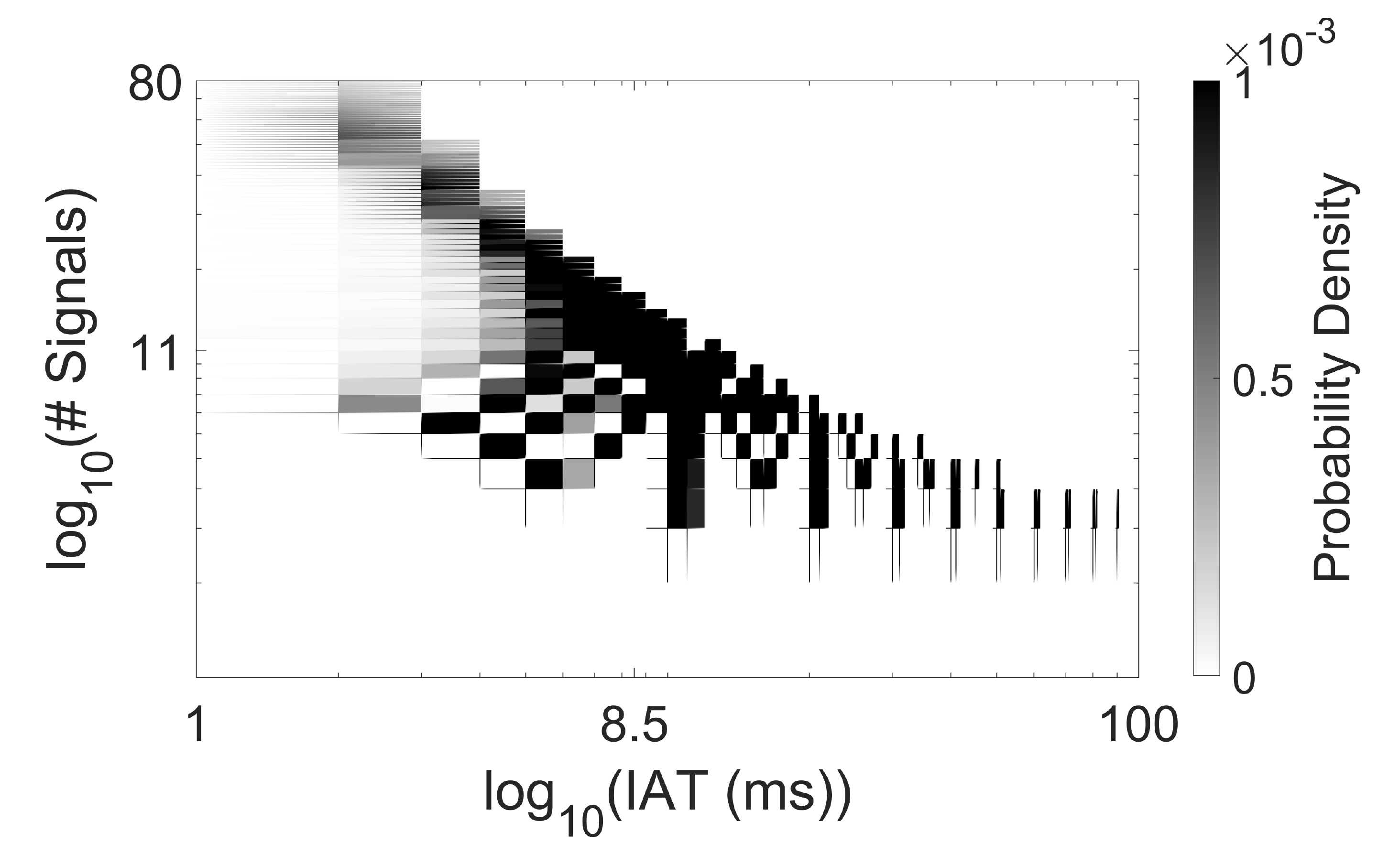}
\caption{Second week \home.} 
\label{fig:trace_home_2}	
\end{subfigure}
\caption{Probability distribution function of traces during \third from \office and \home.} 
\label{fig:traces_3weeks}
\end{figure*}

Once the acquired interference traces are characterised in terms of mean IAT and the number of interference signal arrivals per slot, we compute the two-dimensional probability distribution of the trace. Next, the hourly traffic patterns are identified by comparing the distribution with a one-hour peak-traffic distribution extracted from the trace. By comparing the trace with peak-hour traffic, we can classify peak traffic hours and off-peak traffic hours. To this end, the Normalised Cross-Likelihood Ratio (NCLR)~\cite{le:hal-00163855} was used with values of NCLR close to zero indicating highly similar distributions. Thus, peak and off-peak traffic hours can be identified from the trace, which in turn is useful for training the model. A threshold value of $0.5$ for NCLR is used to distinguish among the two. 

Next, we characterise the interference traces in two ways. The first is to compute the probability density function (PDF) of the interference traces \wrt their mean IAT and number of interference signal arrivals, as shown in Fig.~\ref{fig:traces_scenario1},~\ref{fig:traces_scenario2}, and~\ref{fig:traces_3weeks}. The other is to compute the NCLR for traces from the same campaign, from different channels (\first), locations (\second), or weeks (\third). 

\begin{figure*}
\centering
\begin{subfigure}[t]{\columnwidth}
\includegraphics[trim={1.5cm, 1.2cm, 2.0cm, 0.4cm}, clip, width=1\columnwidth, keepaspectratio]{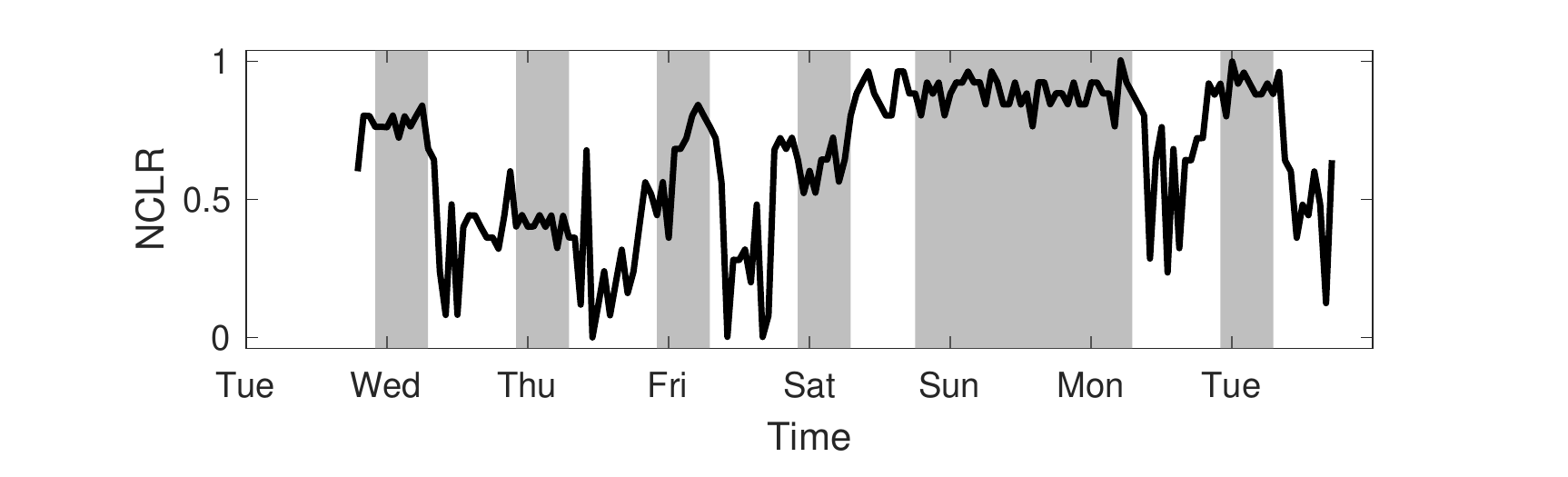}
\caption{\office week-1.} 
\label{fig:nclr_office_w1}
\end{subfigure}
\begin{subfigure}[t]{\columnwidth}
\includegraphics[trim={1.5cm, 1.2cm, 2.0cm, 0.4cm}, clip, width=1\columnwidth, keepaspectratio]{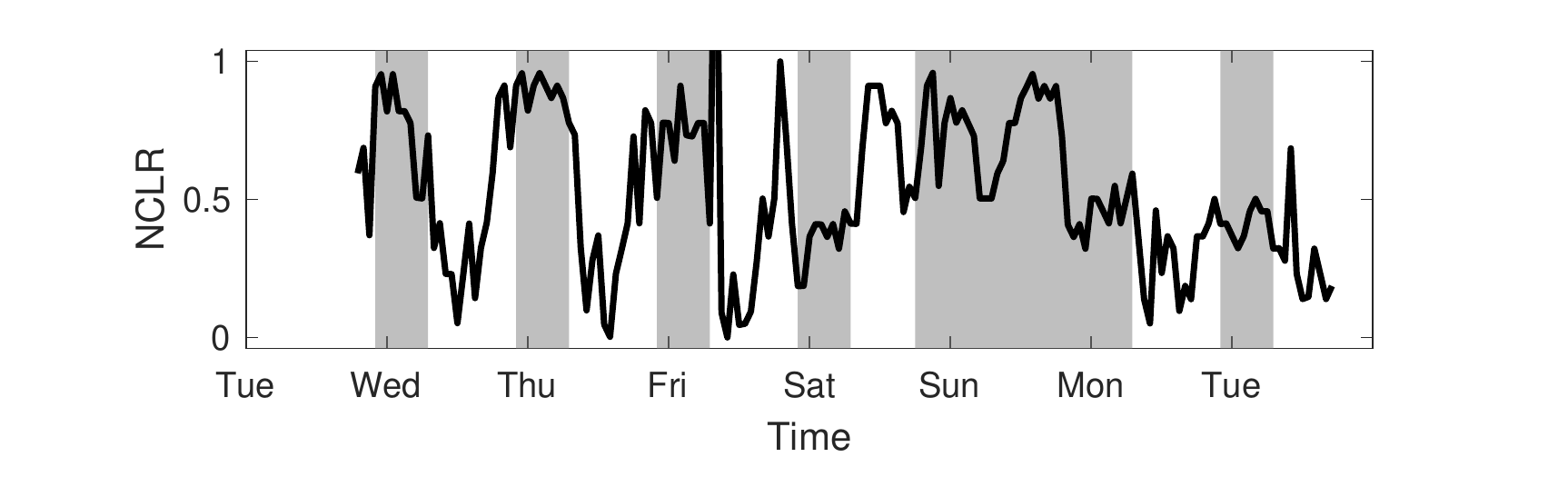}
\caption{\office week-2.} 
\label{fig:nclr_office_w2}
\end{subfigure}
\par\bigskip
\begin{subfigure}[t]{\columnwidth}
\includegraphics[trim={1.5cm, 0.2cm, 2.0cm, 0.4cm}, clip, width=1\columnwidth, keepaspectratio]{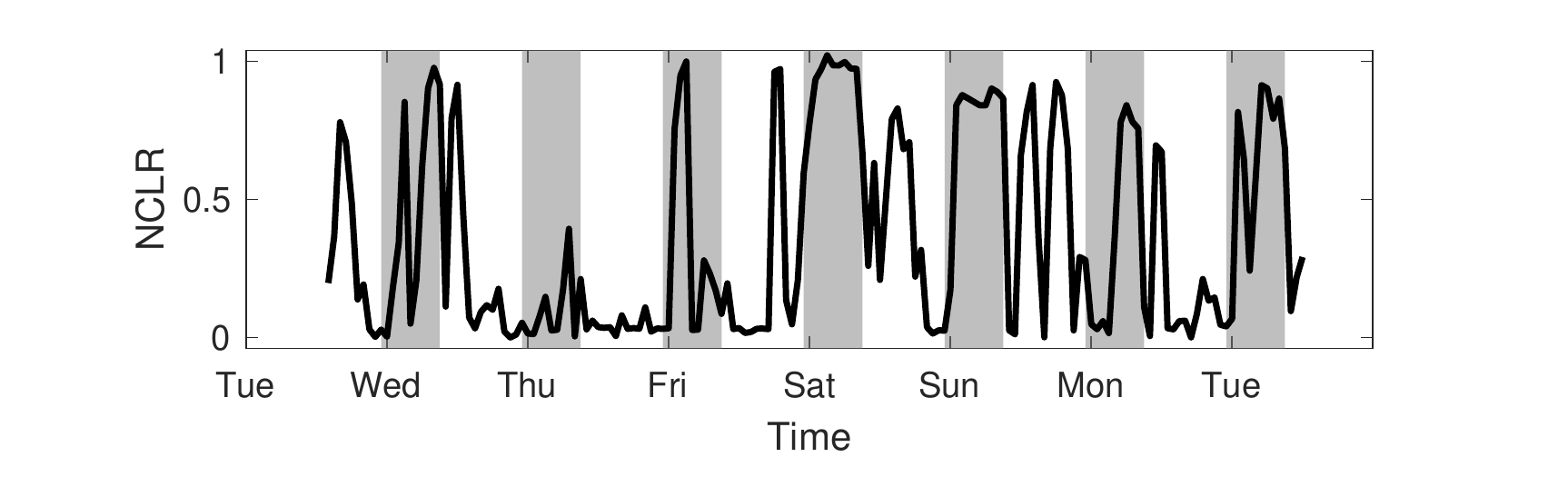}
\caption{\home week-1.} 
\label{fig:nclr_home_w1}	
\end{subfigure}
\begin{subfigure}[t]{\columnwidth}
\includegraphics[trim={1.5cm, 0.2cm, 2.0cm, 0.4cm}, clip, width=1\columnwidth, keepaspectratio]{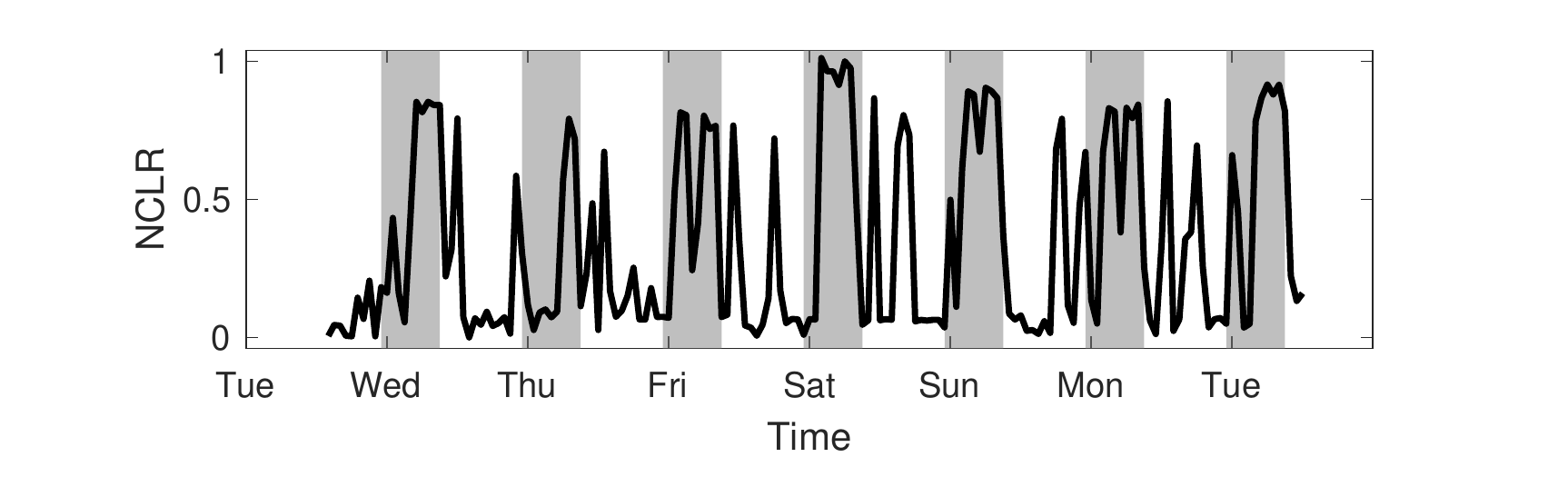}
\caption{\home week-2.} 
\label{fig:nclr_home_w2}	
\end{subfigure}
\caption{Interference patterns in \office and \home for two weeks.} 
\label{fig:nclr_2weeks}
\end{figure*}

\fakeparagraph{\first} Through the PDF lens (Fig.~\ref{fig:traces_scenario1}), it appears that the interference on channels $13$ and $23$ is similar.
This is further confirmed by the low value of $NCLR = 0.22$ in Tab.~\ref{tab:NCLR}. On the other hand, the interference on channel $18$ is different. We conjecture that this is a combined effect of the IEEE~$802.15.4$ channels overlapping with different WiFi channels and being interfered with by different APs. 

\fakeparagraph{\second} The location of the nodes induces different trends in their PDFs, in this case, location $1$ and $2$, showing similar behavior, as shown in Fig.~\ref{fig:traces_scenario2}. This can also be seen in Tab.~\ref{tab:NCLR} with $NCLR = 0.12$, and explained by different interference characteristics induced by the position of the nodes in the proximity of two different APs (\ie location $1$ and $2$ are close to AP1, while $3$ is close to AP2). 

\fakeparagraph{\third} Fig.~\ref{fig:traces_3weeks} shows the results from the \third campaign. A few trends are clearly identifiable. First, the quantity of traffic increases as one progresses from \office to \home. The trend is more marked during the second week. Secondly, the traffic in \home is more bursty than \office, the PDFs show high probabilities in the bursty zone, mostly due to the video streaming done by students in \home (\ie student dormitory). 
We now turn our attention to interference characteristics induced by day and night variations. 
Fig.~\ref{fig:nclr_2weeks} shows the NCLR obtained from the comparison of the $1$-hour peak trace with each $1$-hour interference trace for both environments. In \office the $1$-hour peak trace represents the busiest traffic period during the day and for \home during the night. In the \office, Fig.~\ref{fig:nclr_office_w1} and Fig.~\ref{fig:nclr_office_w2}, we easily identified patterns in the interference distribution over time during week days and weekends. The regions with high NCLR, match the outside of office hours ($7$:$00$--$22$:$00$) time period and the weekends ($19$:$00$ Saturday--$7$:$00$ Monday), when there is no activity in the \office building, therefore less interference. Moreover, an increase in the interference, with NCLR decreasing close to zero, can be observed during the busiest office hours, $10$:$00$--$11$:$00$ and $13$:$00$--$15$:$00$. Interestingly, Thursday night of the first week and the weekend days of the second week, show an increase in the interference, that we ascribe to a set of experiments run in the \office building.  In \home, the variations over time appear to be somewhat dependent on the night (\ie off-peak between $23$:$00$ and $9$:$00$) and day variations, but are not as clearly marked as in the \office. Also, in \home the range of variations between \free and \busy periods is more dramatic, while the \busy periods are smoother (\ie longer bursty interference periods) than in \office. These are the effects of more users and devices (WiFi/Bluetooth/microwave ovens) in the students \home than in the \office, plus no strict access time policies. 

In a nutshell, the observations from our experimental campaigns show that the environment in which the low-power wireless nodes are immersed, the location where the nodes are placed, and the channel used, have an impact on how the interference is perceived. Moreover, these observations directly inform modelling decisions, suggesting that at least two models accounting for the peak and off-peak interference patterns should be adopted. 

\subsection{Modelling Approach}
\label{sec:approach}

We built on the above analysis to exploit the set of traces to create two models:
\begin{inparaenum}[$i$)]
\item for estimating the interference, and for 
\item predicting white spaces for low-power wireless nodes in the presence of interference.
\end{inparaenum}

\subsubsection{Interference Estimation}
\label{sec:interference_estimation}

\fakeparagraph{Model} The fundamental motivation for our modelling approach for estimating the interference is that the observed traces display an arbitrary distribution, and Gaussian Mixture Models (GMMs) can produce smooth estimations of arbitrarily shaped distributions~\cite{reynolds2015gaussian}. Therefore, we use a GMM, whose defining parameters are the number of components ($\mathcal{M}$) and three matrices: mixture component weights ($\vect{W}$), component means ($\vect{\mu}$), and covariances ($\vect{\Sigma}$). The former is a stochastic matrix, which determines the weight with which each Gaussian component should model data, and $\vect{\mu}$ and $\vect{\Sigma}$ define the mean and the covariance of each component. In our approach, we use two GMM models for peak and off-peak periods. 

The choice of the number of components ($\mathcal{M}$) affects the estimation accuracy. Moreover, each component ($\mathcal{M}$) has ($\mathcal{Q}$) dimensions given by the number of features used to characterise the distributions. 

\fakeparagraph{Parameters} In our case, the components are two-dimensional ($\mathcal{Q} = 2$), as the interference traces are characterised by the mean IAT and the number of signal arrivals per slot. The number of components is estimated empirically by comparing the GMM model estimates \wrt the ground truth trace using Area Under Curve (AUC) as the metric. In Section~\ref{sec:perf_eval}, we show how this is carried out for our approach. 

\fakeparagraph{Training} Once the parameters are computed and set, the model can be trained. The matrices (\ie $\vect{W}$, $\vect{\mu}$, $\vect{\Sigma}$), of the GMM model were estimated using the expectation maximisation (EM) algorithm. A diagonal covariance matrix $\vect{\Sigma}$, the most used in the literature was adopted, requiring fewer samples for training. 

\subsubsection{White Space Prediction}
\label{sec:wsp_prediction}

\begin{figure}[b!]
\centering
\includegraphics[trim={0, 0cm, 0, 0cm}, clip, width=\columnwidth, keepaspectratio]{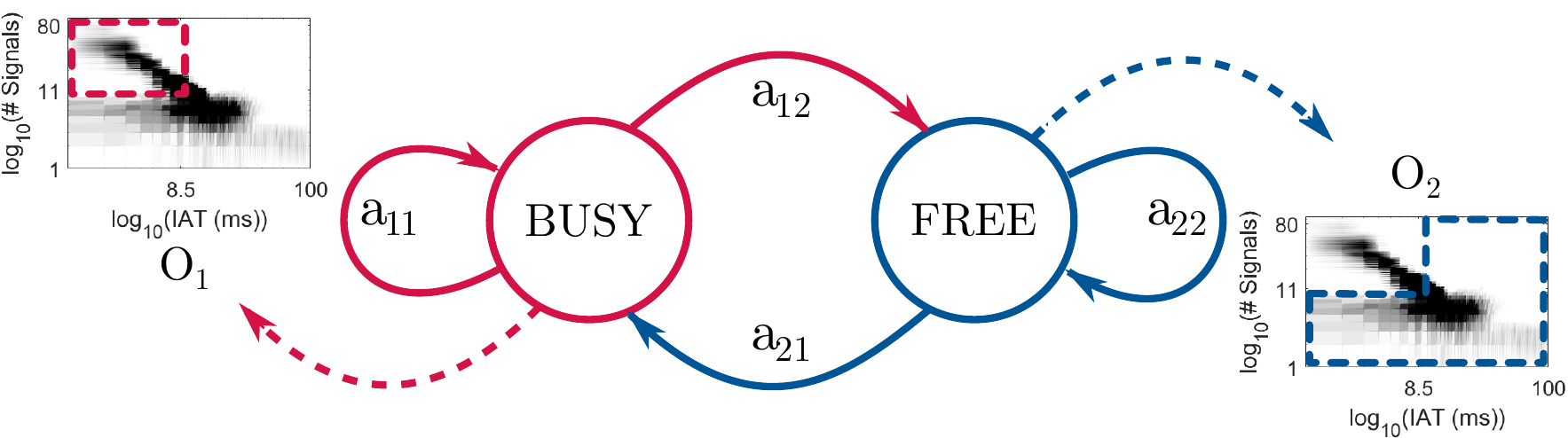}
\caption{Hidden Markov model.} 
\label{fig:n_g_components_auc}
\end{figure}

A contribution of our research is an approach for predicting transmission opportunities for low-power wireless nodes in the presence of interference. We propose to use a Hidden Markov Model (HMM) to exploit the output of the GMM model, the estimated interference, for white space prediction.

We adopt the notation from~\cite{IndikaLCN} to indicate the complete parameter set of the HMM model:
\begin{inparaenum}[$1)$]
\item hidden (unobserved) states $\vect{S} = \{\free, \busy\}$, correspond to the two different conditions of the wireless channel;
\item initial state probabilities $\vect{\pi}$;
\item observations $\vect{O} = \{o_1, o_2\}$, correspond to the two features used to characterise the interference, mean IAT and number of signal arrivals per slot;  
\item state transition probability matrix $\vect{A}$, models the evolution of the wireless channel as transitions among the set of unobserved states;
\item observation probability matrix $\vect{B}$.
\end{inparaenum}

Fig.~\ref{fig:n_g_components_auc} shows a graphical representation of the HMM model, with transition probabilities overlaid on the arrows showing the state transitions (\ie $a_{11}$, $a_{12}$, $a_{21}$, and $a_{22}$), and  the emission distributions for each state represented by the modeled interference distributions corresponding to the \busy and \free states of the channel. The model parameters $\vect{A}$ and $\vect{B}$ are initialised using uniformly distributed probability matrices, while $\vect{\pi}$ is initialised for the data set under consideration, and all are recomputed using the \textit{Baum-Welch} algorithm~\cite{Juang1986}. In addition, the training data used for the HMM is labeled as \free or \busy with the help of the two thresholds, \thiat and \thcount, introduced in Section~\ref{sec:interference_characterization}. In our approach, we use two HMM models, one for peak and one for off-peak periods.

\section{Model-based Data Communication}
\label{sec:solution}

\begin{figure*}
\centering
\includegraphics[trim={0cm, 0cm, 0cm, 0cm}, clip, width=0.85\textwidth, keepaspectratio]{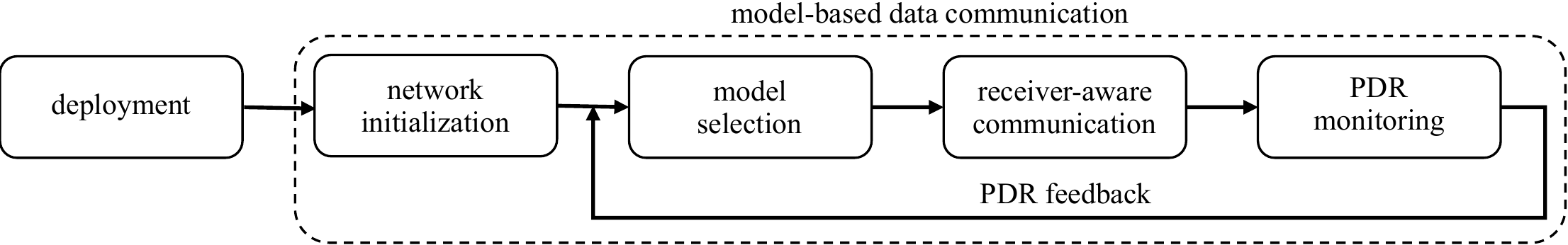}
\caption{Overview of the model-based data communication.} 
\label{fig:overview_data_collection}
\end{figure*}

In this section, we describe the design of the model-based receiver-aware MAC protocol (\proposedmac). 
Fig.~\ref{fig:overview_data_collection} depicts an overview of the operation of \proposedmac and in the following subsections, we further illustrate the design of \proposedmac including network initialisation and how we integrated receiver-aware communication into the solution. 

\subsection{Network Initialisation}
\label{sec:net_init}

The network initialisation is the first stage in the model-based data communication phase and it sets the ground for the receiver-aware communication which will be explained in Section~\ref{sec:rx_init_comm}. 

\fakeparagraph{Time synchronisation} Distributed systems, such as low-power wireless networks, often require a time synchronisation service to enable the coordination between the devices in the network. Especially for time slotted wireless networks, tight synchronisation is essential to perform transmissions at scheduled times. Because our approach also uses time-slots, a simple synchronisation mechanism with a flooding-based technique, inspired by Trident~\cite{trident}, is implemented for global time synchronisation. 

In low-power wireless data communication applications, a designated network coordinator node (often called gateway) typically collects all data packets and delivers them to the external server where the data is stored, processed, and analysed. The nodes in the wireless network communicate with the network coordinator using one of many dynamic network topologies, such as Star, Ring, Mesh, Grid, and Tree~\cite{Sohraby2007, McGrath2013}. Irrespective of the topology being used, a node sends/forwards the data packets to its next hop which follows the same procedure until the network coordinator receives the packet. As the nodes communicate with their next hop in the scheduled time-slots, it is mandatory to synchronise all the clocks of the network with that of the network coordinator so that the events in the network are harmonised. 
In IoT applications wherein smart objects are envisaged to interact with each other to achieve a particular objective, time synchronisation is essential if \proposedmac is used. 
Despite the application, the scheduling of packet transmissions is distributed and implemented locally in the nodes based on the interference models computed in the deployment phase. 

\begin{figure}
\centering
\includegraphics[trim={0cm, 0cm, 0cm, 0cm}, clip, width=\columnwidth, keepaspectratio]{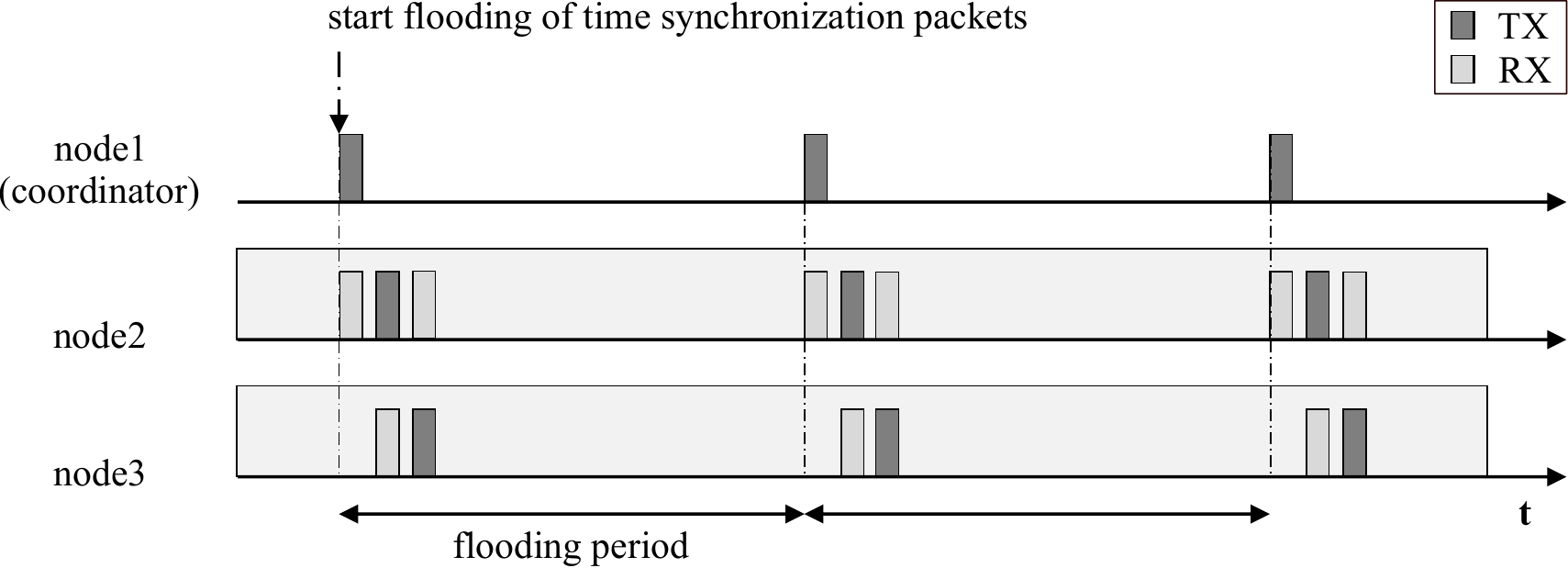}
\caption{Timeline of clock synchronisation.}
\label{fig:tsync}
\end{figure}

Our time synchronisation mechanism is initiated with the broadcast of a control packet, called a synchronisation packet, by the network coordinator, see  Fig.~\ref{fig:tsync}. This packet has three fields: 
\begin{inparaenum}[$a)$] 
\item timestamp,
\item authoritative level, and 
\item time offset with the network coordinator. 
\end{inparaenum}
When the protocol is being initialised, the network coordinator has the lowest authoritative level, \ie $0$, while the rest of the nodes in the network have $255$. After the initialisation phase is over, the network coordinator creates a time synchronisation packet by setting both the authoritative level and the time offset of the control packet to $0$ and timestamp the packet before broadcasting it. Note that a node in the wireless network updates its clock upon receiving a time synchronisation packet if and only if its authoritative level is higher than that of the received control packet. Moreover, to avoid deadlocks, when a node adjust its clock drift, the node's authoritative level is also set to one higher than that of the received control packet. 

Updating the clock drift is done as follows. Denoting a node's clock drift with its neighbour as $\Delta t_\mathrm{N}$ and the neighbour's clock drift with the network coordinator as $\Delta t^N_\mathrm{C}$, a node's clock drift with the network coordinator, $\Delta t_\mathrm{C}$, has the relationship in Equation~\ref{eq:clock_drift}. 

\begin{equation}
\Delta t_\mathrm{C} = \Delta t_\mathrm{N} + \Delta t^\mathrm{N}_\mathrm{C}
\label{eq:clock_drift}
\end{equation}

When a node receives a synchronisation packet from one of its neighbours, the node computes $\Delta t_\mathrm{N}$, which is the difference between the timestamps of the packet transmission at the neighbour and its reception at the node. The former is available in the synchronisation packet itself and the latter is available at the node as the node records the timestamp at which the packet is received. Moreover, the node receives the value of $\Delta t^\mathrm{N}_\mathrm{C}$ along with the packet. Therefore, assuming the processing time at the node is negligible, the clock drift can easily be computed by Equation~\ref{eq:clock_drift} and once done, the node updates the three fields of the packet and rebroadcasts it to ensure all the other nodes in the network are synchronised as well. 

To ensure all nodes in the network received at least one packet, the network coordinator floods the network with three synchronisation packets with a tunable time gap in between. As it is important to keep the clocks of the nodes updated to minimise time drifts, the network coordinator periodically re-initiates the synchronisation. It is worth noting that the synchronisation periodicity, $T_\mathrm{sync}$, is a trade-off between the tight synchronisation and the lifetime of the wireless network. 

\fakeparagraph{Routing and model exchange} 

We build on top of the routing information and use it for the receiver-aware communication to get the information about the next hop, as will be shown in Section~\ref{sec:rx_init_comm}. To this end, existing routing protocols, such as the Collection Tree Protocol (CTP)~\cite{Gnawali2009}, can be used on top of \proposedmac to acquire knowledge about the next hop. 

Furthermore, a node must know the interference models of its neighbours for the functioning of the proposed solution. Therefore, the nodes must share the computed models with their neighbours. This is done during the initialisation phase, which starts after the synchronisation phase. Because the models are shared on a slot basis, the nodes need to be synchronised for receiving them from their neighbours. To avoid collisions among the simultaneously transmitted packets, it is ensured that only a single node broadcasts its models in a given slot of $300$~ms that is heuristically determined. This is achieved with round-robin transmissions with the node identifier determining the turn of the node to broadcast its models using random access. 

The time at which a node broadcasts the models, $t_\mathrm{TX}^\mathrm{bc}$, is expressed in Equation~\ref{eq:share_models_rr}, wherein $t_\mathrm{start}^\mathrm{bc}$, $n$, $N_\mathrm{window}$, and $T_\mathrm{window}$ denote the start of the broadcast period, the node identifier, the number of broadcast windows, and the duration of a single window, respectively. Note that $N_\mathrm{window}$ is dependent on the maximum number of neighbours a node might have in the network, thus it is a tunable parameter and its value should be carefully chosen. 
\begin{equation}
\label{eq:share_models_rr}
t_\mathrm{TX}^\mathrm{bc} = t_\mathrm{start}^\mathrm{bc} + (n \mod N_\mathrm{window}) \times T_\mathrm{window}
\end{equation}

At the end of the model sharing period ($N_\mathrm{window} \times T_\mathrm{window}$), all nodes in the network possess the models of their neighbours. These models are important to a node for receiver-aware communication. Although only the models of the next hop are used by the node at any given time for accessing the shared medium, the rest of the models could be used when the routing topology changes due to poor link qualities. On such occasions, as the interference models of the new next hop are readily available, the node does not have to request interference models from the neighbours again. Therefore, keeping a set of interference models of its neighbours, a node saves energy significantly while minimising delays due to radio environmental changes. After sharing the interference models, the nodes start duty cycling to save energy. 

\subsection{Model Selection}
\label{sec:recalibration}

Radio interference in wireless networks is of a dynamic nature. Especially in low-power wireless networks, depending on the operating channel, deployed location, and the environment, the interference perceived by the nodes is unique, as illustrated in Section~\ref{sec:interference_characterization}. 

\proposedmac leverages the interference models to estimate interference and predict white spaces at the node's location. Due to the dynamic nature of the interference, the models should also re-calibrate their parameters to adapt to fluctuating interference patterns. This is achieved as follows. As discussed in Section~\ref{sec:interference_characterization}, generic interference perceived by wireless nodes has two distinguishing behaviours: peak and off-peak. Consequently, as advocated in Section~\ref{sec:interference_characterization}, two interference models, one for the peak and the other for the off-peak interference, should be used for the accurate estimation of interference and subsequently for predicting white spaces. Therefore, \proposedmac can choose one out of two models that is the best fit for the current interference level. 

The selection of the appropriate model is solely based on the current performance of the data communication application, which is quantified with the \pdr metric. As depicted in Fig.~\ref{fig:overview_data_collection}, the \pdr feedback loop triggers the model 
selection command. However, care should be taken when 
selecting the models as it can improve or deteriorate the performance of the wireless network. More details on how to make the decision on when to 
change the models and the impact of this decision on the performance of the proposed receiver-aware data communication mechanism are discussed in Section~\ref{sec:fb_loop}. 

After making the decision to change the model, the network coordinator disseminates the 
model selection command by flooding the wireless network with a 
model selection control packet. Fields of such 
a control packet consist of the age and the type of the model that is going to be used, \ie peak or off-peak. Similar to the authoritative field in a time synchronisation packet, see Section~\ref{sec:net_init}, a node uses the age field to avoid loops in the flooding. The coordinator initiates the flooding with the age of zero and receiving nodes increase its age by $1$ before rebroadcasting. Also, the age field is used as a means of acknowledgements; a node deems the received 
model selection packet as an acknowledgement when the packet's age is $1$ higher than the one that is saved by the node itself. To make sure all the nodes in the network receive the 
model selection command successfully, the coordinator broadcasts a maximum of five 
such control packets. 

\subsection{Receiver-aware Communication}
\label{sec:rx_init_comm}

This section presents the novel receiver-aware communication technique for low-power wireless networks. Unlike traditional receiver-aware communication with periodic beacons, the proposed approach of finding the rendezvous point between a sender and a receiver is based on the interference models that were introduced in Section~\ref{sec:approach}. 

\begin{figure}
\centering
\includegraphics[trim={0.6cm, 0.6cm, 0.6cm, 0.6cm}, clip, width=0.75\columnwidth, keepaspectratio]{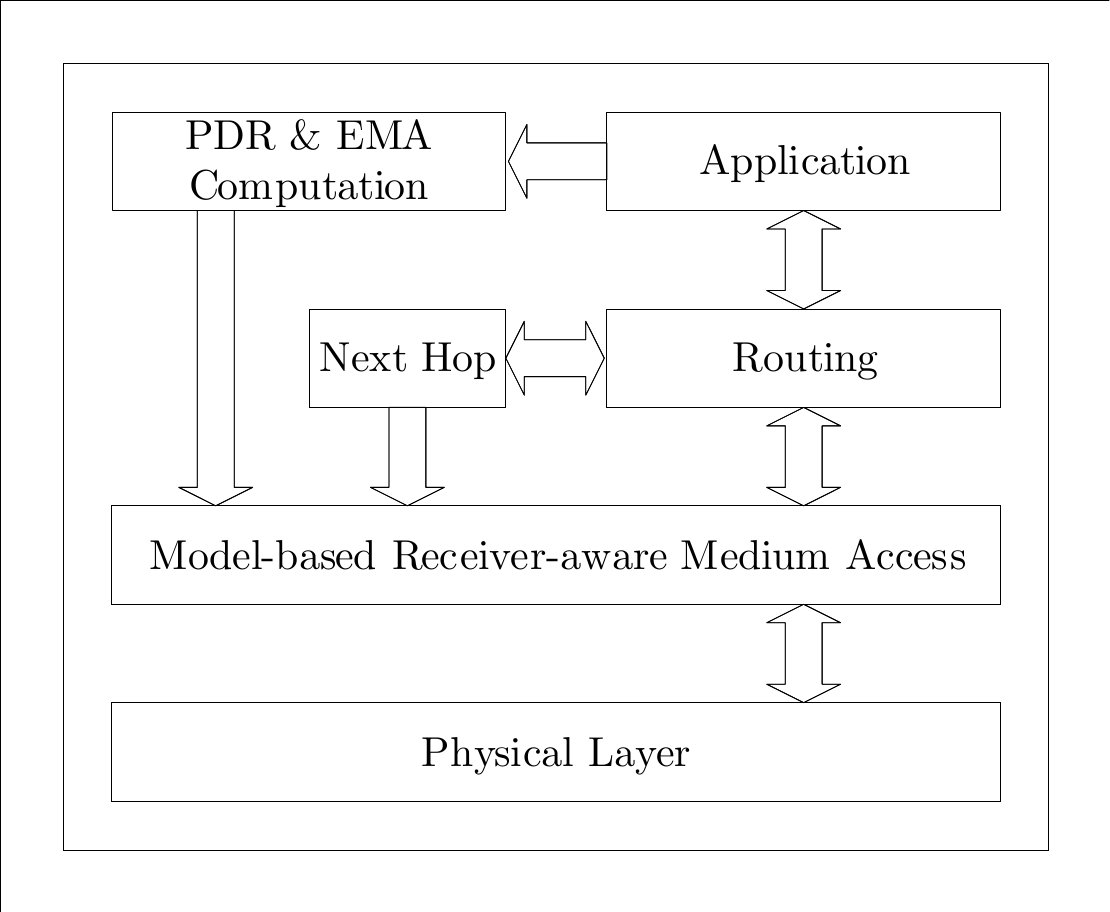}
\caption{Protocol stack of \proposedmac.}
\label{fig:protocol_stack}
\end{figure}

\proposedmac is a cross-layer approach as depicted in Fig.~\ref{fig:protocol_stack}. A big portion of our solution is based in the MAC layer, \wrt the IEEE~$802.15.4$ communication protocol stack, while the application layer and network layer provide essential inputs, which includes application performance and information on the next hop. These inputs are prerequisites for the operation of \proposedmac. 

\fakeparagraph{Operation} All nodes in the wireless network parametrise two interference models to be used during peak and off-peak interference situations, denoted as \textit{own interference models} throughout this paper. In addition to their own interference models, the nodes keep track of their \textit{next hop's interference models}. Both types of models were computed during the deployment phase as explained in Section~\ref{sec:pre_deployment}. The own and next hop's interference models can estimate generic interference and predict white spaces at the node and its next hop's location, respectively. Note that the white space prediction generates a list of \free slots in which less interference is expected, and the two lists of predictions from the two models are used for two different purposes. 

A node maintains two states: active and sleep. When the node is in the former state, it can communicate by turning the radio on, while during the latter the node switches off the radio to save energy. The nodes use the predictions from their own models for scheduling their active/sleep states. In other words, the nodes switch their radio on during these \free slots. When the interference is at its minimum, the own interference model of a node might predict a large number of \free slots within the current data period. 
Despite the ample amount of \free slots, the node utilises the first consecutive \free slots depending on the network size and goes back to sleep. 
This can drastically reduce the network-wide energy consumption while keeping the reliability of the data communication high. The number of utilisable \free slots, $N_\mathrm{slot}$, within a data period, $T_\mathrm{data}$, is dependent on the network size, thus it is a configurable parameter; the bigger the network size, the higher the number of \free slots required and the higher $N_\mathrm{slot}$ becomes. On the contrary, when the interference is high, there could be cases where the own interference model is unable to predict any \free slot. On such occasions, the node utilises the next consecutive $N_\mathrm{slot}$ slots to transmit its data irrespective of their interference conditions. 

Similar to the own interference model, the white space predictions from the next hop's interference model inform a node when its next hop is active for data communication. Note that all nodes in the network, except for the coordinator, have their next hop's interference models. Therefore, the individual node is aware of the time-slots in which the next hop is active, thus a rendezvous point can be found for the data communication. As the receiver \textit{notifies} the sender when it is listening on the channel indirectly via the interference models, the communication is considered to be receiver-aware. 

\begin{figure}
\centering
\includegraphics[trim={0cm, 0cm, 0cm, 0cm}, clip, width=\columnwidth, keepaspectratio]{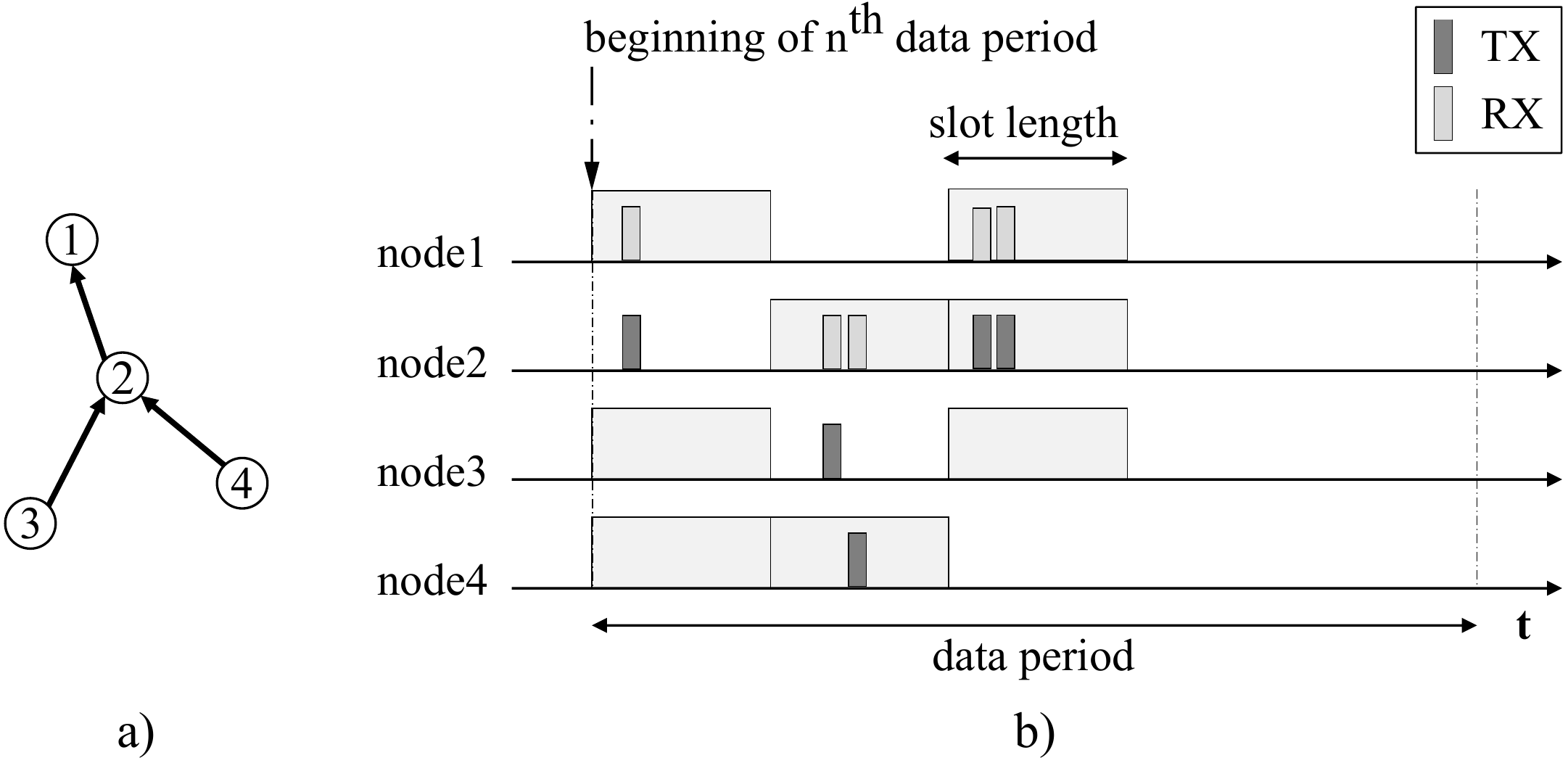}
\caption{Receiver-aware communication with interference models.}
\label{fig:rx_init}
\end{figure}

Fig.~\ref{fig:rx_init} demonstrates the proposed model-based receiver-aware communication concept with a $4$-node wireless network. According to the underlying topology, node~$1$ serves as the network coordinator, node~$2$ is a forwarder node, and node~$3$ and $4$ are end nodes. The nodes go into their active state based on the predictions by their own interference models for receiving packets. For example, node~$1$ has active states in slots $1$ and $3$, relative from the beginning of the data period. At the beginning of the $n^\mathrm{th}$ data period, as depicted in Fig.~\ref{fig:rx_init}, all nodes except the network coordinator have a data packet ready to be transmitted toward the coordinator. Therefore, to send the data packet, the individual node checks for the nearest \free slots at its next hop in which it is in the active state to receive transmissions, which is done by requesting \free slots from the next hop's interference model. Moreover, the nodes transmit/forward their packets within the next hop's \free slot in a round-robin fashion. This is for the other neighbouring nodes to refrain from simultaneous data transmissions, which otherwise leads to collisions and consequently packet losses. 

The round-robin transmissions within a \free slot are performed as follows. A \free slot is further divided into $8.512$~ms sub-slots, which is the same value used for the $TH_\mathrm{IAT}$ when characterising interference in Section~\ref{sec:interference_characterization}. The length of the sub-slot is devised from the fact that an IEEE~$802.15.4$ data frame, with its maximum size of $133$~bytes, and its acknowledgement spend $8.512$~ms on-air at a data rate of $250$~kbps for successful transmission. A node uses its identifier to choose a sub-slot within a \free slot. The transmission time of a node within a slot, $t_\mathrm{TX}^\mathrm{ss}$, is expressed in Equation~\ref{eq:subslot_rr}, wherein $t_\mathrm{start}^\mathrm{slot}$, $n$, $N_\mathrm{ss}$, and $T_\mathrm{ss}$ are the start of time of the slot, the node identifier, the number of sub-slots, and the length of a sub-slot, respectively. 
\begin{equation}
\label{eq:subslot_rr}
t_\mathrm{TX}^\mathrm{ss} = t_\mathrm{start}^\mathrm{slot} + (n \mod N_\mathrm{ss}) \times T_\mathrm{ss}
\end{equation}

It is worth noting that curtailing the slot length leads to a low number of sub-slots, which ultimately will increase the probability of concurrent transmissions by collocated nodes. This will eventually increase the number of packet losses in the network. 

\subsection{Packet Delivery Ratio Monitoring and Feedback}
\label{sec:fb_loop}

Radio interference is dynamic in a given environment and its properties could change at any time. This nature of interference significantly affects the performance of \proposedmac. To accommodate the changes in the radio environment, a Packet Delivery Ratio (\pdr) based feedback loop is used. As depicted in Fig.~\ref{fig:overview_data_collection}, the \pdr of the data communication application is continuously monitored, and when there is a decline in the value a decision is made whether or not to 
change the interference models. 

While the application is running in the network, the network coordinator computes the network-wide \pdr as expressed in Equation~\ref{eq:pdr_cal}, wherein $N_\mathrm{RX}$ and $N_\mathrm{total}$ are the number of data packets received at the network coordinator and the total number of packets supposed to be received within the data period, respectively. Because the network coordinator has the overall picture of the wireless network, it has the information on $N_\mathrm{total}$. The \pdr is computed periodically at the end of each data period with a duration of $T_\mathrm{data}$. Note that the value of $T_\mathrm{data}$ is an application layer parameter. 
\begin{equation}
\label{eq:pdr_cal}
PDR = \frac{N_\mathrm{RX}}{N_\mathrm{total}} \times 100\%
\end{equation}

To avoid unnecessary triggers to 
change the models, which will put the communication network to instability, the network coordinator also keeps track of the moving average of the \pdr. The moving average smooths the \pdr by eliminating its sudden fluctuations. There are many techniques available to compute the moving average and to quickly adapt to the changes in the radio environment, the Exponential Moving Average (\ema) was used. Equation~\ref{eq:ema} shows how the \ema is computed at the network coordinator, wherein $\alpha$ is the smoothing factor and $\ema_\mathrm{last}$ is the \ema of the previous data period. The smoothing factor is $\alpha=2/(N_\mathrm{window}+1)$, where $N_\mathrm{window}$ is the length of the moving window. \ema is initialised with the average \pdr of the first window. 
\begin{equation}
\label{eq:ema}
\ema = \alpha \times \pdr + (1 - \alpha) \times \ema_\mathrm{last}
\end{equation}

The \ema of the network-wide \pdr computed by the network coordinator is a measure of the performance of the proposed solution; when the performance drops, the interference models should be 
changed. The decision on when to trigger the 
model selection command is based on a predefined threshold, $TH_\mathrm{PDR}$, which is an end-user customised parameter. It is an indication of the degradation of the performance when the \ema goes below the threshold. However, the 
mode selection should not be initiated spontaneously when the \ema crossed the threshold. The reason is that the change in the interference could be momentary, thus the 
change of models would not help to overcome the performance loss after the temporal interference variation. Therefore, a transient period called
\textit{model selection timeout} is applied, which delays the model 
selection and allows the impulse of interference to fade away and thereby to settle the \ema down. If the interference is persistent even after the timeout, the model 
selection is triggered. This technique alleviates the instability that emerges with rapid model 
changes. 


\section{Performance Evaluation}
\label{sec:perf_eval}


We validated the interference models used in this work by conducting a statistical comparison between the interference traces (training set) collected from the two indoor environments and estimated traces from the GMM model~\cite{our_dcoss_paper_short}, a state-of-the-art Pareto model~\cite{Huang2010}, and our previously proposed MMPP(2) model~\cite{IndikaLCN}. The former is closely related to our work since it focuses on a model–based white space prediction mechanism for WSNs in the presence of WiFi interference, thus was chosen for the comparison. Moreover, we evaluated \proposedmac by comparing it with Crystal~\cite{crystal} and ContikiMAC~\cite{contikimac} with and without retransmissions. 
CRYSTAL is a synchronous data transmission protocol for low-power wireless networks that uses channel hopping and noise detection techniques to mitigate interference to deliver high dependability. On the contrary, CSMA/ContikiMAC is an asynchronous protocol available in the Contiki protocol stack (v$2.7$), which uses a power-efficient wake-up mechanism to minimise energy consumption and re-transmissions to improve data transmission reliability. These two solutions helped us to compare the performance of \proposedmac with two different channel access paradigms. Due to practical reasons, the evaluation of \proposedmac was carried out with COOJA in conjunction with MATLAB, as described in Section~\ref{sec:poc_implementation}. 

\subsection{Metrics}
\label{sec:metrics}
To assess the performance of the interference estimation, we considered two metrics: \textit{accuracy} and \textit{False Positive Rate} (\fpr). The former is a measure of the prediction performance of the model, while the latter provides an assessment of the packet loss in the network when the prediction mechanism is being used.

The performance of \proposedmac is evaluated with \pdr and \dutycycle as metrics for reliability and energy consumption, respectively. Here we consider the network-wide \pdr and \dutycycle which are computed as follows: \pdr~$ = \frac{n_\mathrm{rx}}{n_\mathrm{tot}}$, \dutycycle~$ = \frac{t_\mathrm{on}}{N*(t_\mathrm{on}+t_\mathrm{off})}$, wherein $n_\mathrm{rx}$, $n_\mathrm{tot}$ and $N$ are number of packets received at the network coordinator, total number of packets transmitted by nodes and total number of nodes in the network respectively. Moreover, $t_\mathrm{on}$ and $t_\mathrm{off}$ denote transceiver on and off durations, respectively.


\subsection{Model Parameter Selection}
\label{sec:model_parm_selection}

\fakeparagraph{GMM} The number of components was empirically identified, varying it from three to ten and computing the Area Under Curve (AUC). The results indicated that seven components ($\mathcal{M} = 7$) are enough for a satisfactory accuracy of $99.9$\% of the estimated interference. 

\fakeparagraph{HMM} We used two HMM models for peak and off-peak periods, respectively. The training traces for peak and off-peak models are obtained by computing the mean \textit{NCLR} for the traces and picking the $1$-hour trace closest to this mean. Note that a $1$-hour training set was used as it provides statistical relevance, due to self-similarity of interference~\cite{our_dcoss_paper_short}, for the channel behaviour. 

\subsection{Interference Estimation Validation}
\label{sec:model_validation}

Our evaluation of interference modelling is divided into two parts. First, we assess the performance of the GMM model with different interference characteristics. For this, traces from all campaigns were used. Second, we compare to the state-of-the-art, a Pareto model~\cite{Huang2010}, and with our previous approach based on an MMPP(2) model~\cite{IndikaLCN}. Traces from the first week of the \third campaign were used. 

We quantitatively evaluate the \textit{accuracy} and \fpr of the estimated interference trace \wrt the ground truth trace. The output of the GMM model is a trace characterised in terms of mean IAT and number of signal arrivals per slot. To perform our comparison, mean IAT and the number of signal arrivals of both traces (estimated and ground truth) were translated into a channel state, \busy and \free, using the \thiat and \thcount thresholds during each time slot. From this, the confusion matrix of two channel state sequence is derived along with the metrics. 

\begin{figure*}
\centering
\begin{subfigure}[t]{0.25\textwidth}
\includegraphics[trim={0cm, 0cm, 1.75cm, 0cm}, clip, height=0.8\textwidth, keepaspectratio, right]{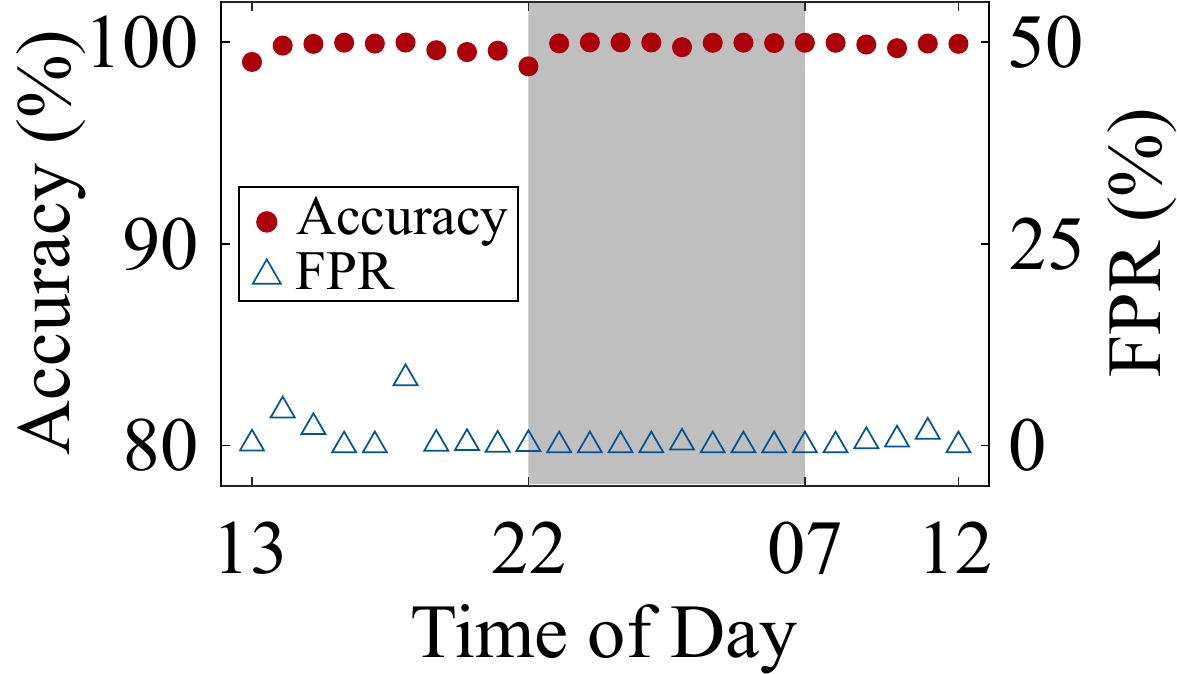}
\caption{Channel 13.} 
\label{fig:gmm_perf_first_ch13}
\end{subfigure}
\begin{subfigure}[t]{0.25\textwidth}
\includegraphics[trim={2.25cm, 0cm, 1.75cm, 0cm}, clip, height=0.8\textwidth, keepaspectratio, center]{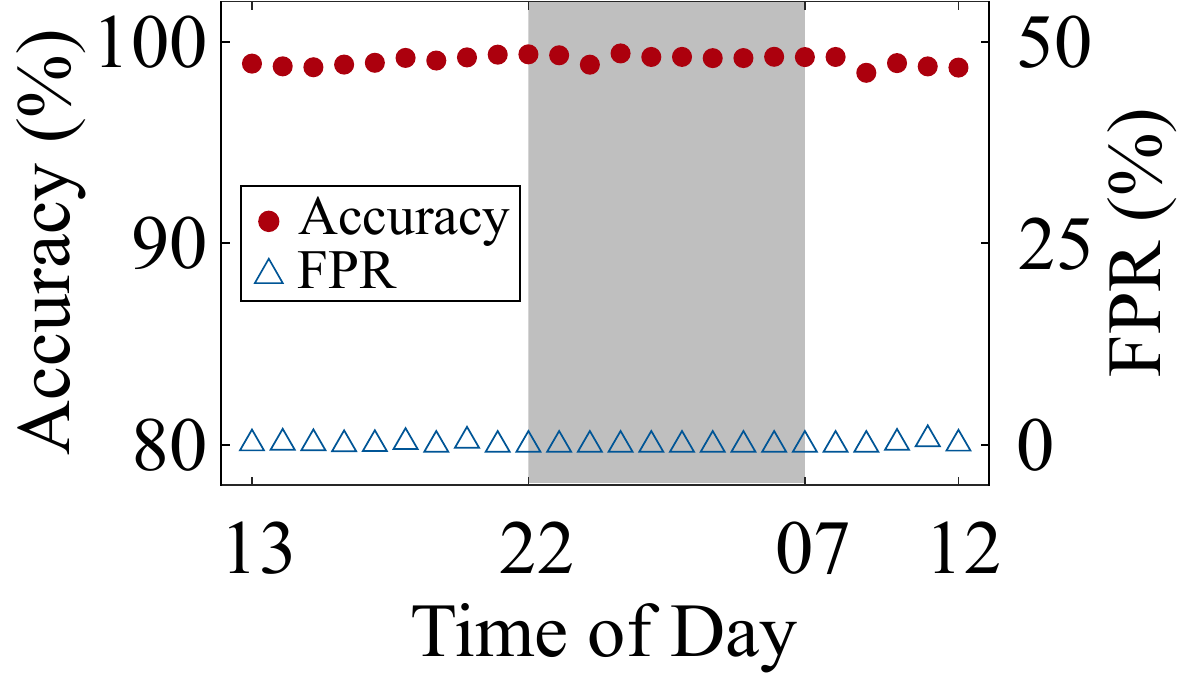}
\caption{Channel 18.} 
\label{fig:gmm_perf_first_ch18}
\end{subfigure}
\begin{subfigure}[t]{0.25\textwidth}
\includegraphics[trim={2.25cm, 0cm, 0cm, 0cm}, clip, height=0.8\textwidth, keepaspectratio, left]{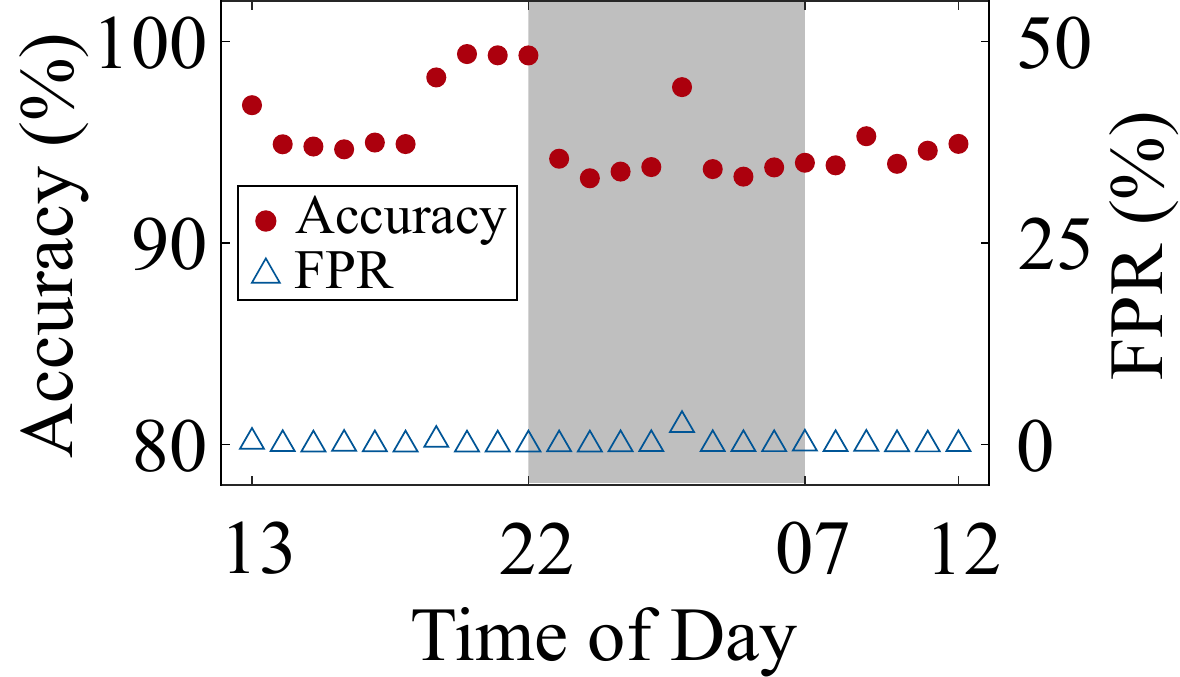}
\caption{Channel 23.} 
\label{fig:gmm_perf_first_ch23}
\end{subfigure}
\caption{Performance of GMM in \first.} 
\label{fig:gmm_perf_first}
\par\bigskip
\centering
\begin{subfigure}[t]{0.25\textwidth}
\includegraphics[trim={0cm, 0cm, 1.75cm, 0cm}, clip, height=0.8\textwidth, keepaspectratio, right]{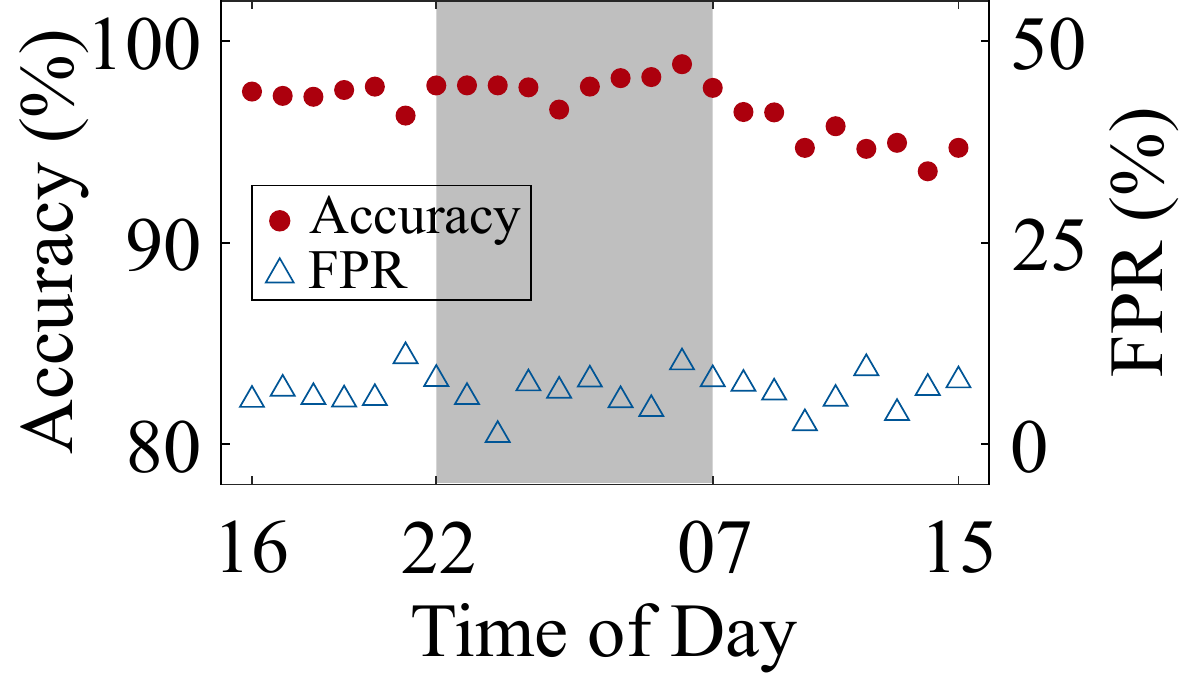}
\caption{Location 1.} 
\label{fig:gmm_perf_second_loc1}
\end{subfigure}
\begin{subfigure}[t]{0.25\textwidth}
\includegraphics[trim={2.25cm, 0cm, 1.75cm, 0cm}, clip, height=0.8\textwidth, keepaspectratio, center]{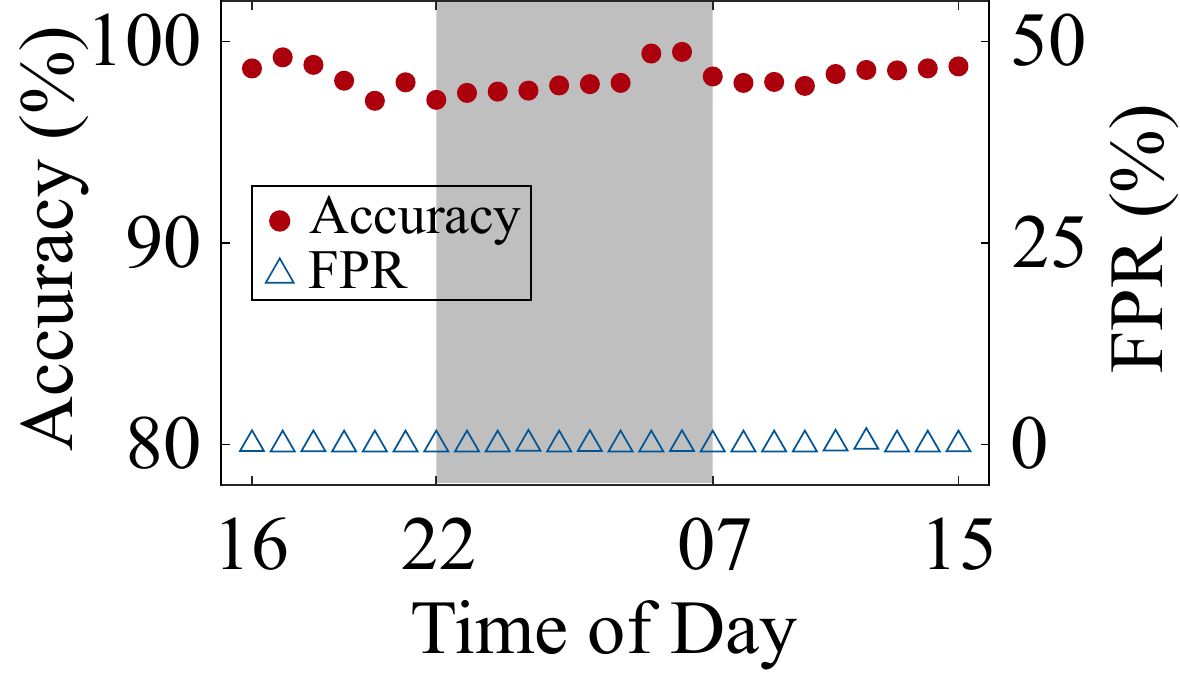}
\caption{Location 2.} 
\label{fig:gmm_perf_second_loc2}
\end{subfigure}
\begin{subfigure}[t]{0.25\textwidth}
\includegraphics[trim={2.25cm, 0cm, 0cm, 0cm}, clip, height=0.8\textwidth, keepaspectratio, left]{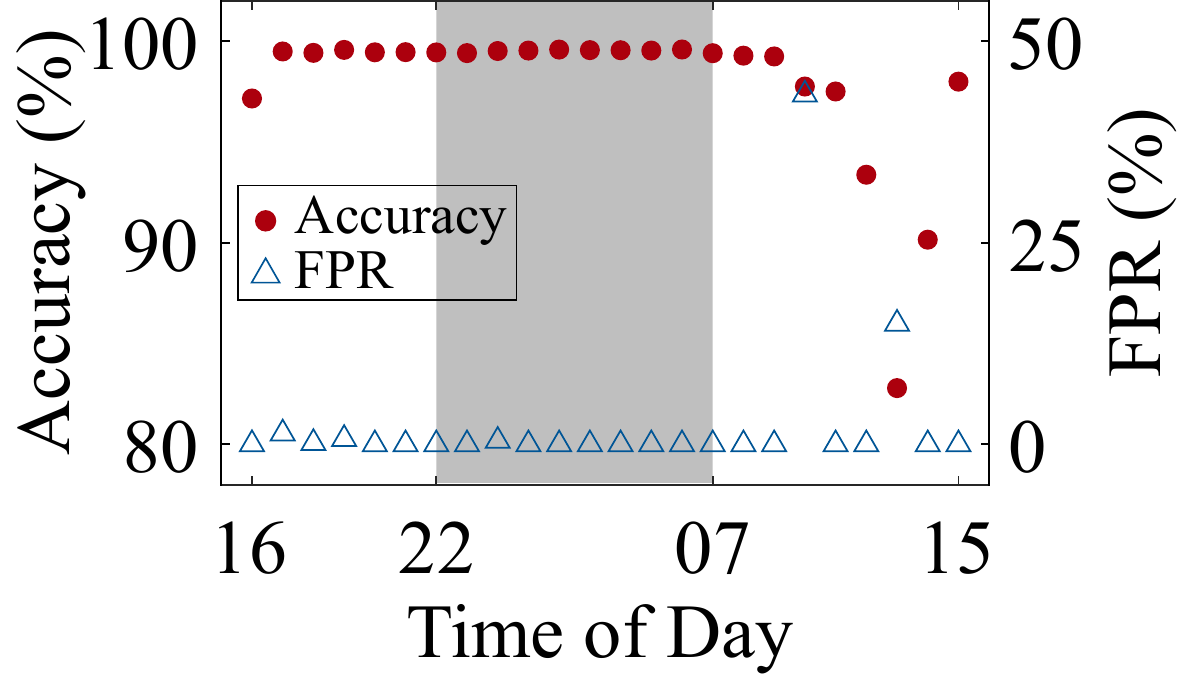}
\caption{Location 3.} 
\label{fig:gmm_perf_second_loc3}
\end{subfigure}
\caption{Performance of GMM in \second.} 
\label{fig:gmm_perf_second}
\par\bigskip
\centering
\begin{subfigure}[t]{0.225\textwidth}
\captionsetup{singlelinecheck=false, format=hang, justification=centering, labelsep=space}
\centering
\includegraphics[trim={0cm, 0cm, 1.75cm, 0cm}, clip, height=0.8\textwidth, keepaspectratio, right]{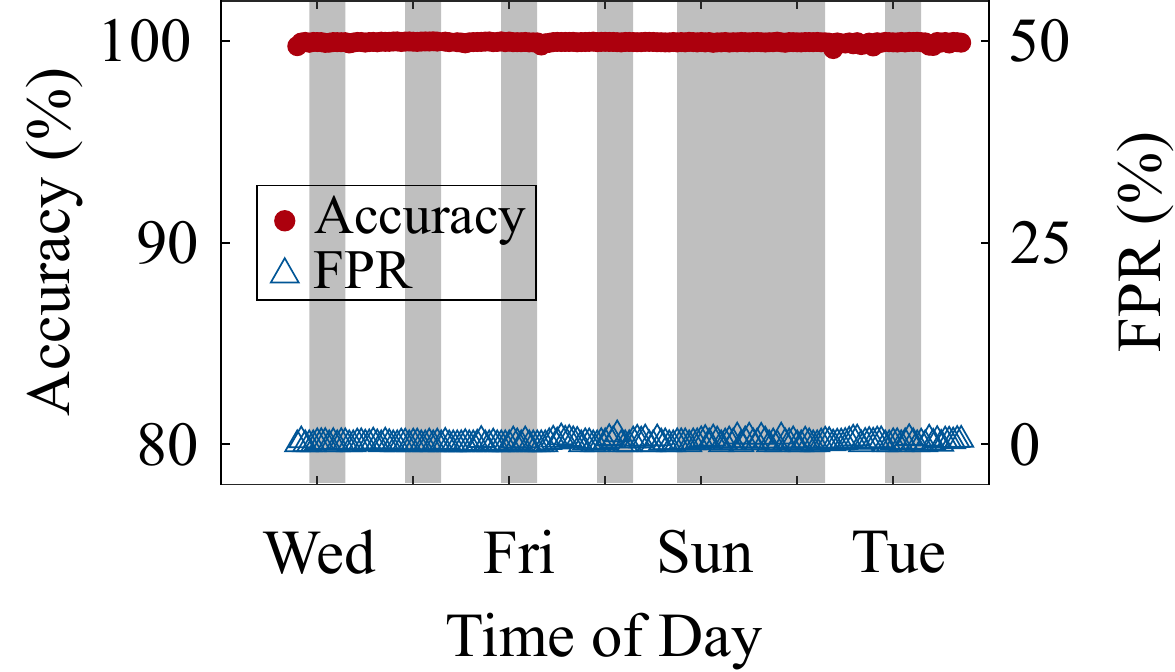}
\caption{\office: week 1.}
\label{fig:gmm_perf_third_office_w1}
\end{subfigure}
\begin{subfigure}[t]{0.225\textwidth}
\captionsetup{singlelinecheck=false, format=hang, justification=centering, labelsep=space}
\centering
\includegraphics[trim={2.25cm, 0cm, 1.75cm, 0cm}, clip, height=0.8\textwidth, keepaspectratio, right]{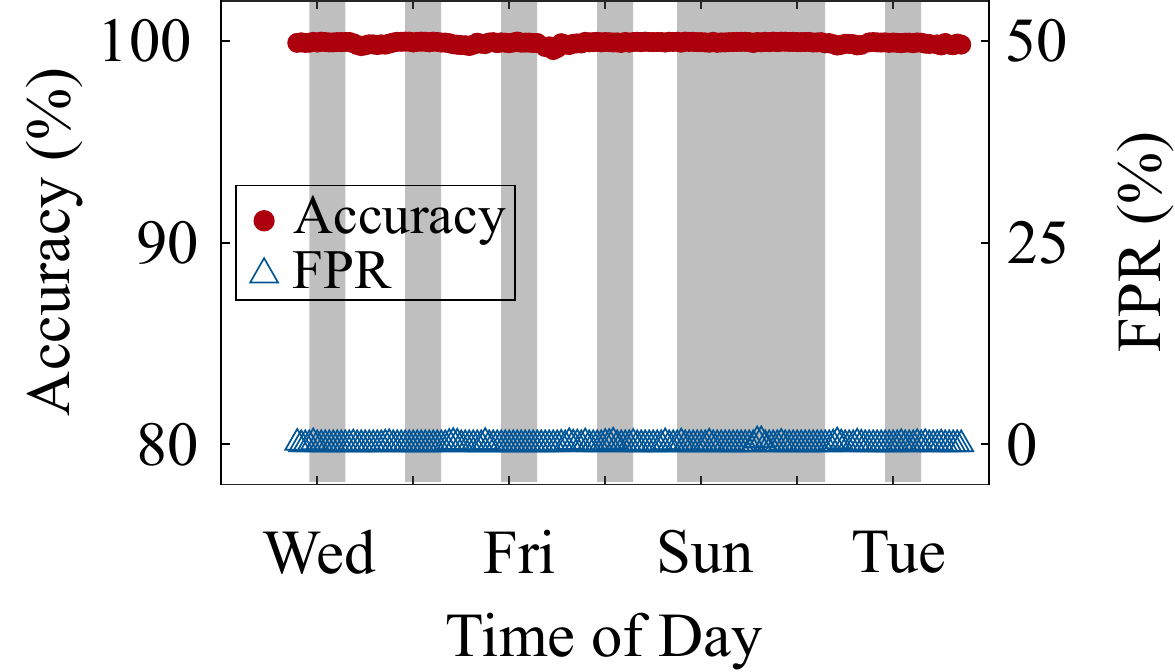}
\caption{\office: week 2.} 
\label{fig:gmm_perf_third_office_w2}
\end{subfigure}
\hspace*{1mm}
\begin{subfigure}[t]{0.225\textwidth}
\captionsetup{singlelinecheck=false, format=hang, justification=centering, labelsep=space}
\centering
\includegraphics[trim={2.25cm, 0cm, 1.75cm, 0cm}, clip, height=0.8\textwidth, keepaspectratio, left]{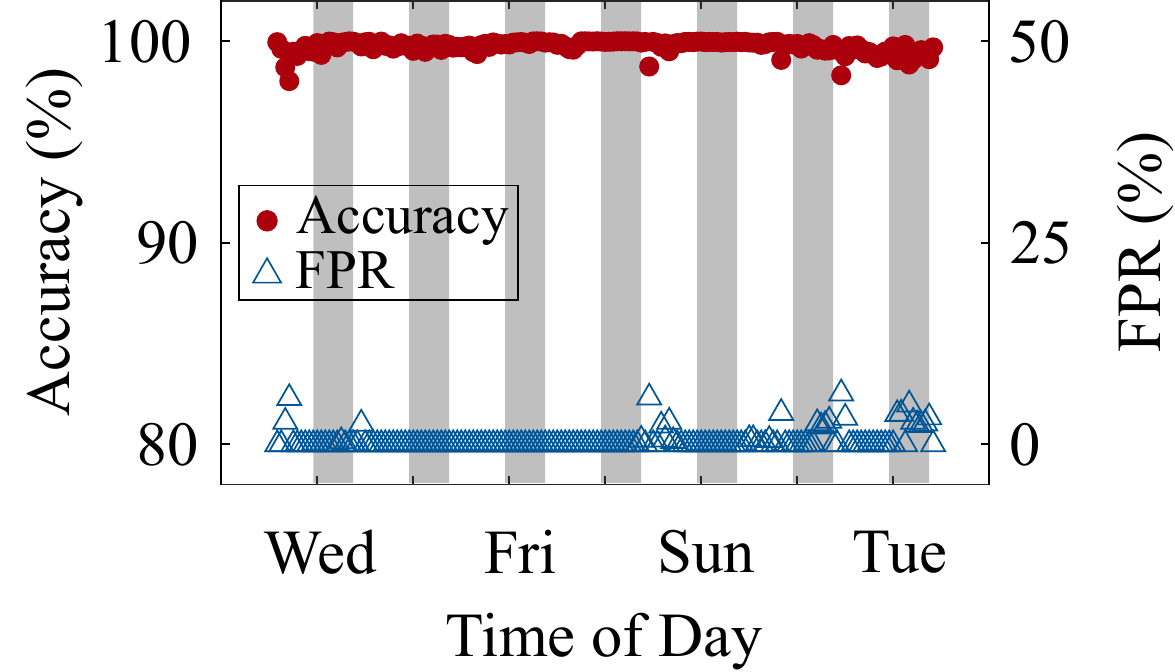}
\caption{\home: week 1.} 
\label{fig:gmm_perf_third_home_w1}
\end{subfigure}
\begin{subfigure}[t]{0.225\textwidth}
\captionsetup{singlelinecheck=false, format=hang, justification=centering, labelsep=space}
\centering
\includegraphics[trim={2.25cm, 0cm, 0cm, 0cm}, clip, height=0.8\textwidth, keepaspectratio, left]{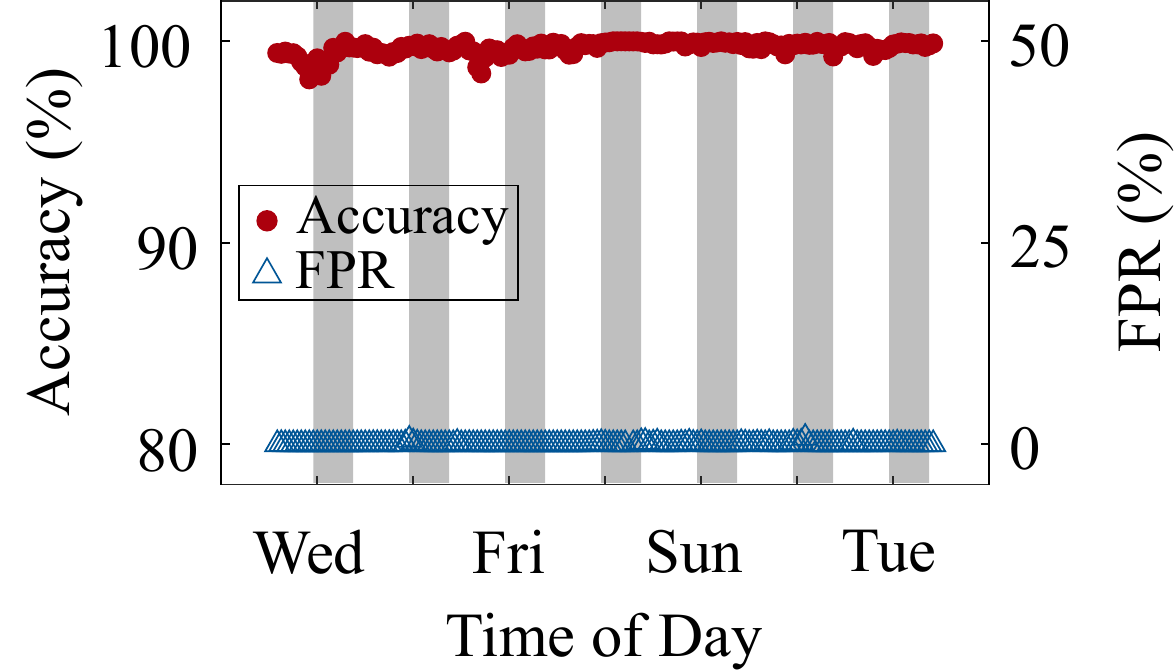}
\caption{\home: week 2.} 
\label{fig:gmm_perf_third_home_w2}
\end{subfigure}
\caption{Performance of GMM in \third.} 
\label{fig:gmm_perf_third}
\end{figure*}

The results are shown in Fig.~\ref{fig:gmm_perf_first}, Fig.~\ref{fig:gmm_perf_second}, and Fig.~\ref{fig:gmm_perf_third}. One can see that in \office, during the $24$~hours of the \first and the \second campaigns, Fig.~\ref{fig:gmm_perf_first} and Fig.~\ref{fig:gmm_perf_second}, the \textit{accuracy} of the interference estimation is high, except from $10$AM to $3$PM in location $3$ when the \textit{accuracy} decreases as low as $82.8$\% and \fpr increases up to $43.4$\%. We argue that this behavior is induced by the increase in the number of signal arrivals during those hours, as explained in our previous work~\cite{our_dcoss_paper_short}. 

During the \third measurement campaign, as shown in Fig.~\ref{fig:gmm_perf_third}, the \textit{accuracy} of the estimation over the course of two weeks is high, over $98$\%, in both environments. Moreover, it is evident that the GMM model can better estimate the behaviour of the interference in \office than \home, which can be explained with arguments similar to those for the other campaigns, wherein \home has more bursty interference~\cite{our_dcoss_paper_short}. In \office, the estimation \textit{accuracy} is stable at $100$\%, except for a few occasions; even those deviations are negligible as they are less than $1$\% during both weeks. On the contrary, \home exhibits much more frequent variations in \textit{accuracy} and \fpr, but the \textit{accuracy} does not decrease below $98$\%. 

Let's turn our attention to a comparison of the performance of our GMM modelling approach for estimating the generic interference with candidates from the state-of-the-art, e.g. Pareto and MMPP(2). Tab.~\ref{tab:cmp_modeling_performance} shows the results \wrt the \textit{accuracy} and \fpr in predicting the actual interference for both environments in \third. For each selected period in the life of the interference trace, a two-hour test trace has been chosen. The three periods in Tab.~\ref{tab:cmp_modeling_performance} correspond to peak (\textit{day}) and off-peak (\textit{night}, \textit{weekend}), the two off-peaks exhibiting different characteristics (\ie different \textit{NCLR} values). 

The Pareto-based approach relies on the self-similarity property of the interference, meaning characteristics of the interference are preserved irrespective of scaling in time. Therefore, to ensure a fair comparison with Pareto, one has to resort to at most a two-hour test trace in which the traffic exhibits self-similarity. The approach assesses the state of the channel upon the arrival of an application packet; thus, the white space prediction probability is conditioned by this state. 

Moreover, the performance of the MMPP(2) model depends on the training duration $x$ and the modelling duration factor $k$. Here, the MMPP(2) model was calibrated for maximizing the AUC value, and used $k$ = 1, $x$ = $240$, $180$, $300$~s in \office and $x$ = $420$, $240$, $540$~s in \home, for day, night and weekend. 

\begin{table*}
\centering
\caption{GMM \vs alternative solutions in \third.}
\label{tab:cmp_modeling_performance}
\setlength\extrarowheight{1pt}
\resizebox{0.65\textwidth}{!}{%
\begin{tabular}{|p{.08cm}|p{.03cm}|p{1.5cm}|C{1cm}C{1cm}C{1cm}|C{1cm}C{1cm}C{1cm}|}
\hline
\multicolumn{2}{|c|}{\multirow{2}{*}{\textbf{Period}}}     & \multirow{2}{*}{\textbf{Metric}} & \multicolumn{1}{c}{} & \textbf{\office}  & \multicolumn{1}{l|}{} & \multicolumn{1}{l}{} & \textbf{\home}    & \multicolumn{1}{l|}{} \\ \cline{4-9} 
\multicolumn{2}{|c|}{}                            &                         &  \textbf{GMM}                  & \textbf{MMPP} & \textbf{Pareto}                & \textbf{GMM}                  & \textbf{MMPP} & \textbf{Pareto}                \\ \hline
\multirow{4}{*}{\rotatebox{90}{\textbf{Weekday}}} & \multirow{2}{*}{\hspace{-.08cm}\textbf{Day}}   & \textbf{Accuracy}                     & 99.82                & 86.86   & 16.71                 & 99.42                & 68.94   & 32.41                 \\
                         &                        & \textbf{FPR}                     & 0.08                    & 98.19   & 6.51                  & 0.04                 & 97.27   & 5.99                  \\ \cline{2-9} 
                         & \multirow{2}{*}{\hspace{-.13cm}\textbf{Night}} & \textbf{Accuracy}                     & 99.98                & 87.95   & 18.26                 & 99.72                & 86.54   & 19.21                 \\
                         &                        & \textbf{FPR}                     & 0                    & 99.34   & 8.28                  & 1.16	 & 99.53   & 8.15                  \\ \hline
\multicolumn{2}{|c|}{\multirow{2}{*}{\textbf{Weekend}}}    & \textbf{Accuracy}                     & 99.96                & 97.17   & 8.59                  & 99.81                & 86.18   & 18.64                 \\
\multicolumn{2}{|c|}{}                            & \textbf{FPR}                     & 0                    & 98.86   & 8.53                  & 0.39                 & 99.32   & 7.21                  \\ \hline
\end{tabular}
}
\end{table*}

Tab.~\ref{tab:cmp_modeling_performance} shows the results \wrt the ground truth in \office and \home. The GMM approach achieves the best results, highest \textit{accuracy} and lowest \fpr, compared to the alternatives across all combinations of environments, channels, locations, and time intervals. Moreover, GMM is slightly worse in \home than \office, due to the more bursty interference. Nevertheless, the \textit{accuracy} does not decrease below $99.42$\% and the \fpr is lower than $1.16$\%. On the other hand, through the lens of both metrics, Pareto performs better in \home than \office. 
Although Pareto's \textit{accuracy} does not go over $32.41$\%, its \fpr is at $5.99$\% during the busiest traffic periods in \home. We argue that this is due to the Pareto probability distribution function which has the highest probability at the smallest IAT that it can model, result in predicting high bursty traffic, leading to high false negatives, and finally reducing its accuracy. 
Notably, MMPP(2) is better than Pareto for correctly identifying the two states of the channel, \busy and \free, translated into higher \textit{accuracy}. 
Due to the high \fpr and high \textit{accuracy}, it can be conjectured that MMPP(2) is able to better identify the \free state than the \busy state. 

\subsection{Evaluation of \proposedmac}
\label{sec:data_collection_performance}

This section presents the performance evaluation of \proposedmac for various interference conditions and compares \proposedmac with CRYSTAL~\cite{CompetitionCrystal2018, Istomin2018, Istomin2016}, and ContikiMAC~\cite{contikimac} in collaboration with CSMA protocols. 

\subsubsection{Proof of Concept Implementation}
\label{sec:poc_implementation}


To investigate its performance, \proposedmac was implemented in Contiki/COOJA~\cite{Osterlind2006, cooja}. 
In this work, Contiki~$2.7$ was used to validate the proposed proof of concept. The source code of the proof of concept implementation of \proposedmac is available in GitHub~\cite{lucid}.

\begin{figure}[b]
\centering
\includegraphics[trim={0cm, 0cm, 0cm, 0cm}, clip, width=0.9\columnwidth, keepaspectratio]{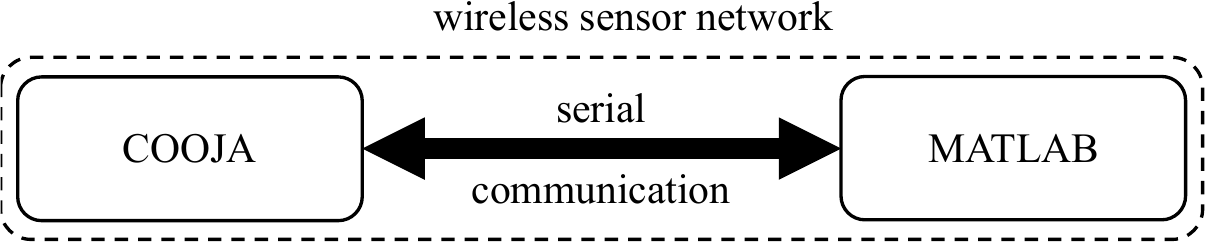}
\caption{System integration with COOJA and MATLAB.}
\label{fig:system_integration}
\end{figure}

In our proof of concept implementation of \proposedmac, the protocol needs to calculate a channel access schedule based on predictions from the interference models. While such calculations can be done on low power nodes, we only had TMoteSky nodes available, which are relatively old nodes with a slow 16 bit micro controller and limited RAM. In our implementation in COOJA, we had to use a trick to perform the schedule calculations and interference predictions in order to allow this to be performed in real-time in the node models. To get around the limitations of the TMoteSky, we "outsourced" the calculations to MATLAB~\cite{matlab}, which was running externally to COOJA and communicated with the nodes in the COOJA simulation. In this work-around, a node requests white space predictions from MATLAB through commands sent via the serial communication port. The command includes the node identifier and the interference model from which the predictions are requested. Upon receiving the command, MATLAB does prediction-related computing and conveys the result back to the node simulated in COOJA.



It is important to note that the countermeasures taken to address the computational requirements are only for the proof of concept. Newer sensor nodes with modern processors, such as the ARM Cortex-M$0$ $32$-bit processor, are powerful enough to perform the required prediction related computations on the sensor nodes and also have larger on-board RAM and FlashROM to store the interference model data. 

Furthermore, to further ameliorate the high memory requirement issue, a tree-like static topology was used throughout the evaluation. To this end, the de-facto tree-based routing protocol in WSNs, the Collection Tree Protocol (CTP)~\cite{Gnawali2009}, was used. The CTP discovers the neighbours within the communication range of the nodes and assigns a parent to a node based on the link qualities, which helps the network to create a tree topology with parents, children, and leaf nodes. The proposed solution acquires the information about the parent from the routing protocol. This piece of information is vital for the receiver-aware communication as explained in Section~\ref{sec:rx_init_comm}. 

\subsubsection{Simulation Scenario}
\label{sec:simulation_scenario}

\begin{figure}
\centering
\begin{subfigure}[t]{0.475\columnwidth}
\includegraphics[width=\textwidth, keepaspectratio]{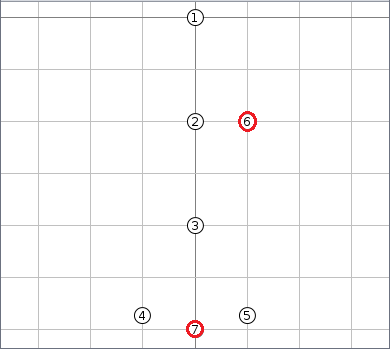}
\caption{$5$-node} 
\label{fig:scenario_n5}
\end{subfigure}
\begin{subfigure}[t]{0.475\columnwidth}
\includegraphics[width=\textwidth, keepaspectratio]{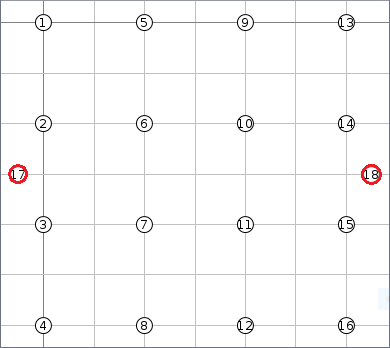}
\caption{$16$-node} 
\label{fig:scenario_n16}
\end{subfigure}
\caption{Simulation scenarios with two interferers.}
\label{fig:sim_scenarios}
\end{figure}

For the evaluation, two wireless sensor networks that consist of $5$ and $16$ nodes, respectively, were studied in varying interference conditions. Fig.~\ref{fig:sim_scenarios} depicts how the two scenarios were deployed in COOJA, with a distance between the neighbouring nodes of $20$~m. Note that the nodes in the small network (see Fig.~\ref{fig:scenario_n5}) were arranged such that it comprises of three forwarding layers to allow us to examine the data communication approaches in a multi-hop network setting. Moreover, the $16$-node network, depicted in Fig.~\ref{fig:scenario_n16}, was deployed in a grid with a $20$~m grid size. It is noteworthy that there will be bottlenecks at nodes that are closer to the network coordinator as they will end up forwarding many more packets in comparison with nodes further away from the coordinator. This is a fundamental problem for all converge-cast type multi-hop communication scenarios. The situation is exacerbated with the increasing network size, leading to declining performance of the 
protocol which 
performs transmissions. 

Furthermore, to replicate the real-life interference conditions, two jammers configured with JamLab~\cite{jamlab} were used. JamLab generates interference according to an IAT distribution. In this work, the IAT distributions extracted from the real-life generic interference measurements, presented in Section~\ref{sec:interference_traces}, were used. The two jammers use two distinct probability distributions based on the interference settings, such as peak/off-peak interference in the \office/\home. The nodes $6$ and $7$ in Fig.~\ref{fig:scenario_n5} and the nodes $17$ and $18$ in Fig.~\ref{fig:scenario_n16} resemble the interferers. Note that determined by its position, a node receives interference from a single or both jammers. 

In low-power wireless networks, Clear Channel Assessment (CCA) is used for identifying signal/interference on the operating channel. CCA in the transceiver of a node detects the presence of a signal when the energy level of the operating channel is higher than the CCA threshold, $TH_\mathrm{CCA}$. As discussed in Section~\ref{sec:interference_traces}, when measuring generic interference, the CCA threshold was set to $-77$~dBm, and the same value was used throughout the evaluation of \proposedmac. 

It is common that wireless sensor nodes generate periodic traffic. Therefore, to evaluate the performance of \proposedmac, two data generation periods were considered, \ie $T_\mathrm{data}=\{10, 60\}$~seconds. These data generation periods helped to investigate the performance of \proposedmac for both, high and low data transmission frequencies. All communication took place on IEEE~$802.15.4$ channel $18$. 

\fakeparagraph{\proposedmac} Tab.~\ref{tab:sim_config_wsp} presents a list of parameters and their values, which were used when we simulated \proposedmac in COOJA. $TH_\mathrm{IAT}$, $TH_\mathrm{count}$, and $TH_\mathrm{CCA}$ were used in the deployment phase when measuring, characterising, and deriving model parameters, while the remaining parameters were used during the execution of the data communication. 
Note that parameters, such as $T_\mathrm{sync}$, $TH_\mathrm{PDR}$, $N_\mathrm{window}$, 
model selection timeout and slot length, used in the evaluation, were empirically determined after a heuristic analysis on the subject, that we discuss in the following subsections. 

\begin{table}
\centering
\caption{Configurations of our solution.}
\label{tab:sim_config_wsp}
\resizebox{\columnwidth}{!}{%
\begin{tabular}{|lcr|}
\hline
\multicolumn{1}{|c}{\textbf{Parameter}} & \textbf{Value}    & \multicolumn{1}{c|}{\textbf{Description}} \\ \hline
$TH_\mathrm{IAT}$   & $8.512$~ms         & inter-arrival time THR            \\
$TH_\mathrm{count}$ & $11$               & THR for number of signal arrivals \\
$TH_\mathrm{CCA}$   & $-77$~dBm          & CCA threshold                           \\
$T_\mathrm{sync}$   & $5$~minutes        & time synchronisation period             \\
$TH_\mathrm{PDR}$   & $93$\%            & \pdr THR for model selection \\ 
$N_\mathrm{window}$ & $40$~samples 	& window size for EMA   \\ \hline
\end{tabular}
}
\end{table}

As mentioned in Section~\ref{sec:net_init}, the value of $T_\mathrm{sync}$ is a trade-off between the network-wide tight synchronisation and the lifetime of the low-power wireless network. Therefore, to balance both sides, we chose a value of $5$~minutes. 

$TH_\mathrm{PDR}$ is an application-dependent parameter, thus this value is specified by the end-user. However, care should be taken when selecting its value as high values of $TH_\mathrm{PDR}$ might induce instability in the performance of the wireless network, as specific values might trigger the 
model selection process more often. Therefore, in our evaluation we heuristically chose $TH_\mathrm{PDR}$ to be $93$\%. 

The length of the moving window when computing \ema was selected to be $40$~samples. This value was heuristically obtained as follows. The \ema was calculated as the window size was varied from $10$ to $100$ in steps of $10$ and the impact on the stability of the system was investigated in terms of the number of cross-overs around the threshold, $TH_\mathrm{PDR}$. Note that long windows cannot capture short-term fluctuations in the \pdr and small windows are too sensitive to short-term variations, potentially leading to an unstable system. Based on this evaluation it was found that a value of 40 for $N_\mathrm{window}$ provided the best performance. 

In addition to the parameters listed in the table, the 
model selection timeout and the length of a time-slot were varied to uncover the optimum values that deliver the best performance of \proposedmac. 

\fakeparagraph{CRYSTAL} CRYSTAL uses two unique techniques to mitigate radio interference: channel hopping and noise detection. The former tackles the interference by escaping it, \ie switching to a channel that does not experience interference, while the latter reacts to the interference by changing its termination criterion which allows to circumvent interference and renders more opportunities for communication by keeping the network awake until the interference fades away~\cite{CompetitionCrystal2018}. However, for a fair comparison of the channel access techniques under investigation, CRYSTAL was run without channel hopping enabled, thus only IEEE~$802.15.4$ channel $18$ was permitted. 

\fakeparagraph{ContikiMAC} CSMA is a well-known MAC protocol that has been widely used in wireless communications. Nonetheless, CSMA does not control the duty-cycling which is crucial for low-power wireless networks to save energy. Contiki provides different Radio Duty-Cycling (RDC) protocols, such as Low-Power Probing (LPP)~\cite{lpp}, X-MAC~\cite{xmac}, and ContikiMAC~\cite{contikimac}. The latter is the default RDC protocol in Contiki which delivers the highest performance among them in terms of energy savings. Thus, in the performance comparison, ContikiMAC in collaboration with CSMA was used. Moreover, the CSMA protocol exploits re-transmissions as a countermeasure to tackle interference. Therefore, the comparison of the performance was done with/without re-transmissions. When re-transmissions were enabled, we intuitively set the maximum number of transmissions to $3$. 

\fakeparagraph{Simulation Execution} The simulation was conducted based on the scenarios depicted in Fig.~\ref{fig:sim_scenarios}. In both scenarios the interference was injected into the simulation environment with $4$ distinct configurations, as shown in Tab.~\ref{tab:simulation_execution}. These interference settings reproduce the environments where the generic interference measurements were taken, see Section~\ref{sec:interference_traces}, and their interference patterns, as illustrated in Section~\ref{sec:interference_characterization}. Moreover, the data collection application was run for $2$ hours in all interference configurations for the two data periods. 

\begin{table}
\centering
\caption{Simulation execution.}
\label{tab:simulation_execution}
\resizebox{\columnwidth}{!}{%
\begin{tabular}{|cccc|}
\hline
\begin{tabular}[c]{@{}c@{}} \textbf{Envir-} 		\\ \textbf{onment} 		\end{tabular} &
\begin{tabular}[c]{@{}c@{}} \textbf{Interference} 	\\ \textbf{Type}		\end{tabular} &
\begin{tabular}[c]{@{}c@{}} \textbf{Network Size} 	\\ \textbf{(\#nodes)} 	\end{tabular} &
\begin{tabular}[c]{@{}c@{}} \textbf{Data Periods} 	\\ \textbf{(seconds)} 	\end{tabular} \\ \hline
\multirow{2}{*}{office} & peak     & \multirow{4}{*}{\{5, 16\}} & \multirow{4}{*}{\{10, 60\}} \\
                        & off-peak &                            &                             \\
\multirow{2}{*}{home}   & peak     &                            &                             \\
                        & off-peak &                            &                             \\ \hline
\end{tabular}
}
\end{table}

Because the interference was injected with the probability distributions of the previously collected interference traces, and the interference models were already parametrised, the simulation did not execute the deployment phase but interference models were implemented in MATLAB where the outsourced computation was done. 

\begin{figure}
\centering
\begin{subfigure}[t]{0.48\columnwidth}
\captionsetup{singlelinecheck=false, format=hang, justification=raggedright, labelsep=space}
\includegraphics[trim={1.25cm, 1.25cm, 1.25cm, 1.25cm}, clip, height=0.75\columnwidth, keepaspectratio, left]{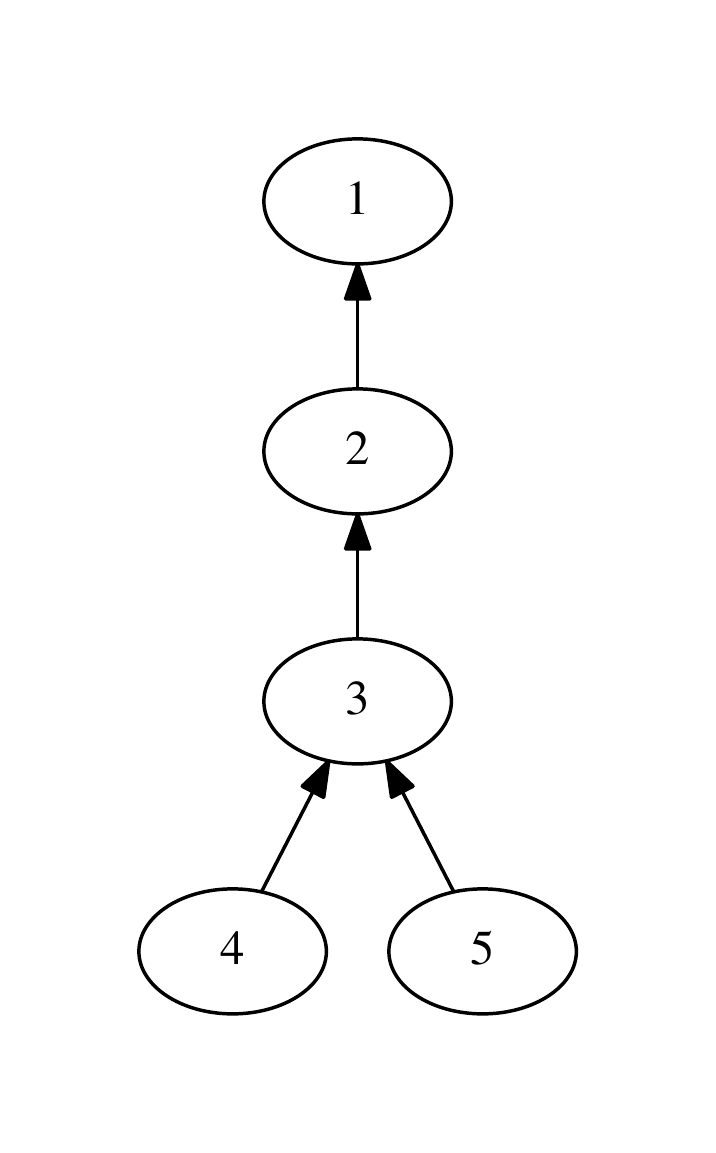}
\caption{$5$-node} 
\label{fig:topo_5}
\end{subfigure}
\begin{subfigure}[t]{0.48\columnwidth}
\captionsetup{singlelinecheck=false, format=hang, justification=raggedright, labelsep=space}
\includegraphics[trim={1.25cm, 1.25cm, 1.25cm, 1.25cm}, clip, height=0.75\columnwidth, keepaspectratio, right]{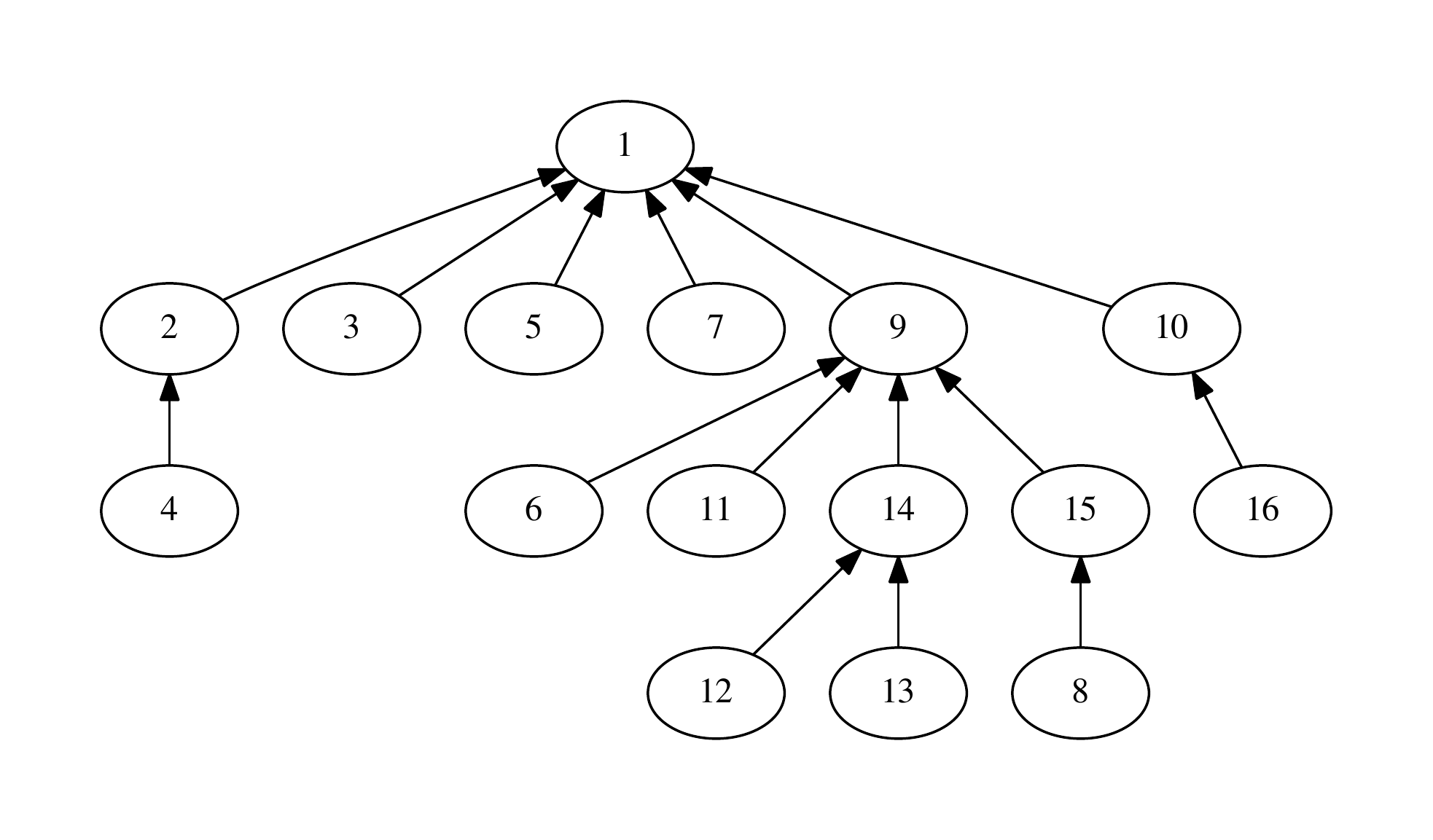}
\caption{$16$-node} 
\label{fig:topo_16}
\end{subfigure}
\caption{Network topologies.}
\label{fig:topologies}
\end{figure}

During the network initialisation phase, the nodes are time-synchronised and the routing topologies are formed. Fig.~\ref{fig:topologies} depicts the snapshots of the tree topologies that the two sensor networks, shown in Fig.~\ref{fig:sim_scenarios}, created. Both routing topologies comprise of $3$ forwarding levels, which is beneficial for evaluating the performance of \proposedmac in a multi-hop setting. 

While running, the nodes dump logs that consist of all the information related to the operation of the communication. This includes \pdr, \ema, and \dutycycle related data, which are extracted and processed after the simulations have finished to evaluate the protocols' performance. 

Furthermore, to have statistical significance in the results, each simulation was conducted $3$ times with varying seeds and the average results were used in the evaluation. 

\subsubsection{Performance of \proposedmac}
\label{sec:performance_data_collection_solution}

Evaluation of \proposedmac is split into three parts. First, the importance of model 
selection and its timeout are investigated. In this regard, the $5$-node network with two interferers, shown in Fig.~\ref{fig:scenario_n5}, was used with varying 
model selection timeouts. Second, the impact of the slot-length on the performance of \proposedmac was studied. For this, the 
model selection enabled $5$-node network was used with different slot-lengths. Third, \proposedmac was compared with 
CRYSTAL and ContikiMAC. The latter was compared with/without re-transmissions. 

The performance metrics used in the evaluation of the communication reliability are the \pdr and its exponential moving average, \ema. Moreover, the energy efficiency of the network was analysed with the average \dutycycle of the nodes. 

\fakeparagraph{Model Selection} Section~\ref{sec:fb_loop} advocated the use of a feedback loop for tackling sudden interference variations in the radio environment where the low-power wireless network has been deployed. Here we investigate and demonstrate the necessity of a feedback loop. 

\begin{table*}
\centering
\caption{EMA statistics without model selection.}
\label{tab:ema_stats_wo_model_switching}
\resizebox{0.8\textwidth}{!}{%
\begin{tabular}{|c|cc|cc|cc|cc|}
\hline
\multirow{3}{*}{\textbf{EMA Statistic}} & \multicolumn{4}{c|}{\textbf{10s}} & \multicolumn{4}{c|}{\textbf{60s}} \\ \cline{2-9} 
 & \multicolumn{2}{c|}{\textbf{HOME}} & \multicolumn{2}{c|}{\textbf{OFFICE}} & \multicolumn{2}{c|}{\textbf{HOME}} & \multicolumn{2}{c|}{\textbf{OFFICE}} \\ \cline{2-9} 
 & \textbf{Peak} & \textbf{Off-peak} & \textbf{Peak} & \textbf{Off-peak} & \textbf{Peak} & \textbf{Off-peak} & \textbf{Peak} & \textbf{Off-peak} \\ \hline
\textbf{min. (\%)} & 82.20 & 88.88 & 90.48 & 93.57 & 90.50 & 94.18 & 94.13 & 94.89 \\
\textbf{max. (\%)} & 92.48 & 96.60 & 96.62 & 98.63 & 95.33 & 97.56 & 97.64 & 98.80 \\
\textbf{avg. (\%)} & 88.04 & 93.47 & 94.27 & 96.54 & 93.16 & 96.19 & 95.94 & 97.08 \\
\textbf{std. (\%)} & 1.82 & 1.36 & 1.19 & 0.87 & 1.24 & 0.82 & 0.95 & 1.10 \\ \hline
\end{tabular}%
}
\end{table*}

\begin{table*}
\centering
\caption{EMA statistics with model selection.}
\label{tab:ema_stats_w_model_switching}
\resizebox{\textwidth}{!}{%
\begin{tabular}{|c|c|cc|cc|cc|cc|}
\hline
\multirow{3}{*}{\begin{tabular}[c]{@{}c@{}} \textbf{Timeout} \\ \textbf{(data periods)} \end{tabular}} & \multirow{3}{*}{\textbf{EMA Statistic}} & \multicolumn{4}{c|}{\textbf{10s}} & \multicolumn{4}{c|}{\textbf{60s}} \\ \cline{3-10} 
 &  & \multicolumn{2}{c|}{\textbf{HOME}} & \multicolumn{2}{c|}{\textbf{OFFICE}} & \multicolumn{2}{c|}{\textbf{HOME}} & \multicolumn{2}{c|}{\textbf{OFFICE}} \\ \cline{3-10} 
 &  & \textbf{Peak} & \textbf{Off-peak} & \textbf{Peak} & \textbf{Off-peak} & \textbf{Peak} & \textbf{Off-peak} & \textbf{Peak} & \textbf{Off-peak} \\ \hline
\multirow{4}{*}{\textbf{3}} & \textbf{min. (\%)} & 84.42 & 91.55 & 88.25 & 93.57 & 91.26 & 94.18 & 91.98 & 94.07 \\
 & \textbf{max. (\%)} & 95.58 & 96.99 & 97.36 & 98.63 & 95.15 & 97.56 & 96.86 & 98.80 \\
 & \textbf{avg. (\%)} & 90.74 & 94.72 & 93.92 & 96.45 & 93.44 & 96.20 & 94.45 & 96.79 \\
 & \textbf{std. (\%)} & 2.26 & 1.02 & 1.65 & 0.92 & 0.92 & 0.82 & 1.57 & 1.42 \\ \hline
\multirow{4}{*}{\textbf{4}} & \textbf{min. (\%)} & 85.73 & 91.11 & 92.17 & 93.57 & 91.26 & 94.18 & 92.99 & 94.25 \\
 & \textbf{max. (\%)} & 96.76 & 97.01 & 98.03 & 98.63 & 95.49 & 97.56 & 97.07 & 98.80 \\
 & \textbf{avg. (\%)} & 93.02 & 94.81 & 95.22 & 96.48 & 93.60 & 96.20 & 94.96 & 96.84 \\
 & \textbf{std. (\%)} & 1.71 & 1.08 & 1.09 & 0.90 & 0.92 & 0.82 & 1.25 & 1.39 \\ \hline
\multirow{4}{*}{\textbf{5}} & \textbf{min. (\%)} & 85.73 & 91.90 & 92.79 & 93.57 & 91.26 & 94.18 & 92.99 & 94.89 \\
 & \textbf{max. (\%)} & 96.87 & 97.42 & 98.53 & 98.63 & 95.79 & 97.56 & 97.49 & 98.80 \\
 & \textbf{avg. (\%)} & 93.76 & 94.72 & 95.70 & 96.44 & 93.73 & 96.20 & 95.43 & 97.08 \\
 & \textbf{std. (\%)} & 1.49 & 1.02 & 0.98 & 0.92 & 1.00 & 0.82 & 1.28 & 1.10 \\ \hline
\multirow{4}{*}{\textbf{6}} & \textbf{min. (\%)} & 85.73 & 91.26 & 92.79 & 93.57 & 91.26 & 94.18 & 93.44 & 94.89 \\
 & \textbf{max. (\%)} & 97.82 & 98.02 & 99.06 & 98.63 & 95.83 & 97.56 & 97.24 & 98.80 \\
 & \textbf{avg. (\%)} & 94.50 & 94.93 & 95.87 & 96.43 & 94.04 & 96.19 & 95.46 & 97.08 \\
 & \textbf{std. (\%)} & 1.52 & 1.39 & 1.01 & 0.92 & 1.00 & 0.82 & 0.96 & 1.10 \\ \hline
\end{tabular}%
}
\end{table*}

Tab.~\ref{tab:ema_stats_wo_model_switching} presents the performance of \proposedmac without using the \pdr feedback loop. With this setting, when there is a short-term change in the radio environment, the functionality of the interference model 
selection is disabled. As can be seen in the first and second-order \ema statistics of \pdr, when the data period $T_\mathrm{data}$ increases, all the \ema statistics tend to increase their values for all the interference settings in both environments. The long $T_\mathrm{data}$ allows sufficient time for the quick burst of interference in the radio environment to fade away, which is the reason for the higher performance in comparison with small $T_\mathrm{data}$. 

Moreover, \proposedmac delivers higher performance in the \office in comparison with the \home in terms of average \ema. Also, the prediction performance of the interference models is always higher during off-peak interference compared to peak interference periods. The \ema is always above the $TH_\mathrm{PDR}$ of $93$\% in all the environments and interference type combinations except during the peak interference in the \home, where it is reduced to $88.04$\% average \ema with the highest standard deviation of $1.82$\%. 

From the results in Tab.~\ref{tab:ema_stats_wo_model_switching}, one can conclude that higher data rate applications running in heavily bursty interference settings deliver poor performance with the solution due to a higher frequency of sudden changes in the radio environment compared to low data rate applications and mild interference. The interference models should be able to adapt to the dynamic nature of the interference, which is the motivation behind the model 
selection with the \pdr feedback loop. 

As elaborated in Section~\ref{sec:fb_loop}, the model 
selection is not triggered as soon as the \ema falls below the $TH_\mathrm{PDR}$ by the network coordinator. Instead, the network coordinator allows a transition period, 
model selection timeout, which is measured in terms of data periods, $T_\mathrm{data}$. Because the small timeouts introduced more instability to the communication system causing rapid model 
changes, the 
model selection timeout was varied from $3$ to $6$ data periods in steps of $1$ to evaluate the influence of the timeout and 
model selection.  

The results are shown in Tab.~\ref{tab:ema_stats_w_model_switching}. The inclusion of the model 
selection helped the average \ema to exceed the $TH_\mathrm{PDR}$ for all interference and data period combinations after introducing $4$ as the 
model selection timeout. With this timeout, the \ema has increased by $5$\% and $1.3$\% in the \home with $10$~seconds data period, respectively in peak and off-peak interference settings. Moreover, the increment of \ema in the \office with peak interference and $10$~seconds data period is $1$\%. Because of the non-bursty interference or low data rate, all the other combinations showcase negligible changes in average \ema. 

Note that the average \ema further increases with the increasing 
model selection timeout, and its standard deviation also decreases and starts to settle down when the timeout is at $5$ for all combinations. Therefore, this value of the 
model selection timeout was selected and used in the rest of the analysis, which allows nodes sufficient time until the sudden burst of interference weakens. 

\fakeparagraph{Impact of Slot-length} The number of data transmissions that can occur within a data period depends on the size of the wireless network. Consequently, the load of the forwarding nodes which forward data packets to their parents builds up. To cope with this, the \dutycycle of the nodes should be augmented. Therefore, the slot length has a significant impact on the dependability of a wireless network. 

The number of data packets, assuming a maximum size of $133$~bytes, that can be transmitted within a single slot is determined by its length. The longer the length, the more the number of packets that can be transmitted within a slot. On the contrary, increasing slot length also escalates the energy consumption of the resource-constrained wireless network. Thus, in the heuristic analysis we performed, we found that going above $150$~ms slot length leads to a much higher duty cycle than a resource-constrained node can sustain. 

Furthermore, decreasing the slot length also reduces the number of available sub-slots in which data transmission takes place, as discussed in Section~\ref{sec:rx_init_comm}, diminishing the medium access opportunities for the nodes in the network. This has a significant negative impact on forwarding nodes wherein multiple data packets need to be transmitted within the slot, leading to packet losses. With this regard, $40$~ms was identified as the critical slot length below which significant packet losses are inevitable even for a small wireless network with $5$ nodes. 

\begin{figure}
\centering
\includegraphics[trim={0cm, 0cm, 0cm, 0cm}, clip, width=\columnwidth, keepaspectratio]{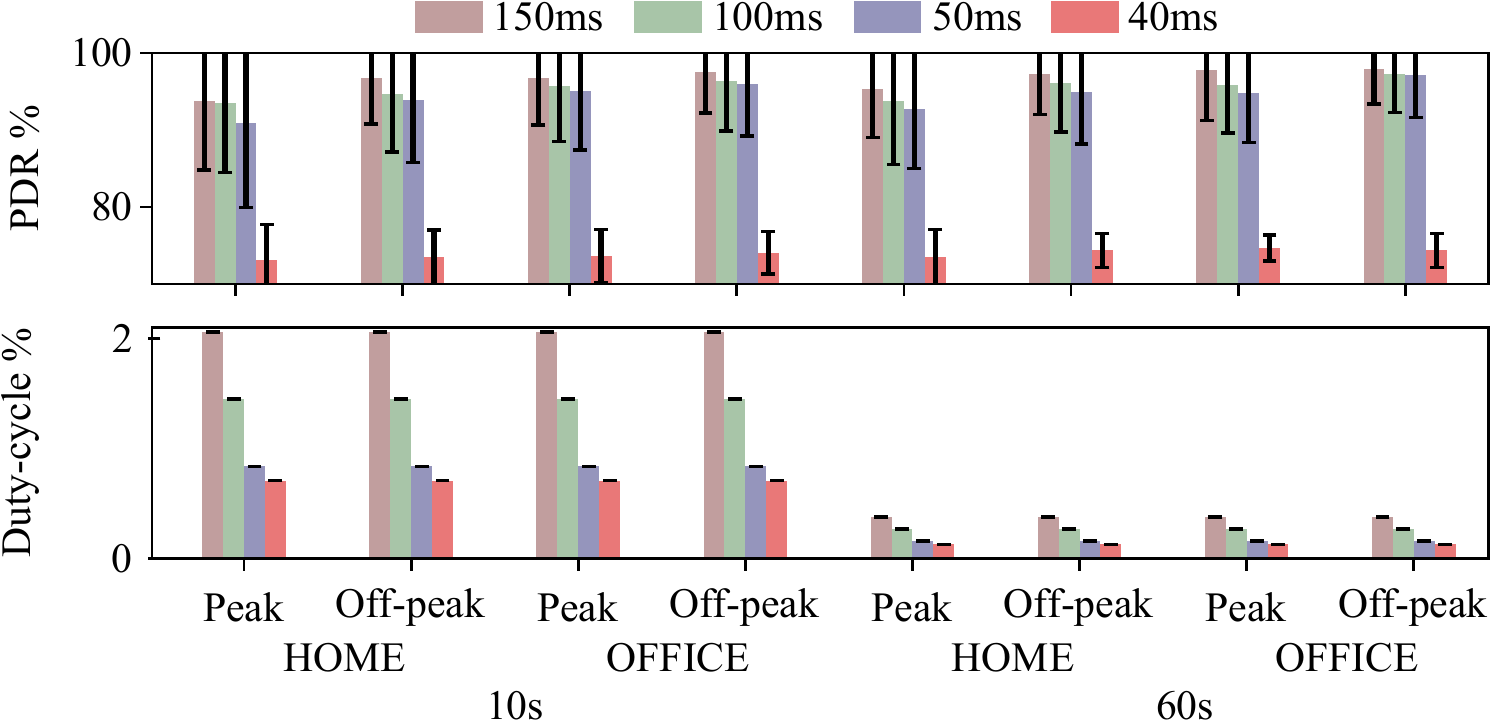}
\caption{Change in \pdr (top) and \dutycycle (bottom) of \proposedmac with varying slot lengths.}
\label{fig:wsp_slotlen}
\end{figure}

Therefore, to investigate its effect on the performance, the slot length was varied with the following values \{$150, 100, 50, 40$\}~ms, while running \proposedmac in the $5$-node network. These values were chosen based on the heuristic analysis of the slot length. Fig.~\ref{fig:wsp_slotlen} presents the change in the network-wide \pdr and \dutycycle with varying slot lengths. 

As Fig.~\ref{fig:wsp_slotlen} demonstrates, \proposedmac with $40$~ms slot length delivers less than $75$\% network-wide \pdr in all the data periods, environments, and interference types combinations. In comparison with the performance of \proposedmac with $50$~ms slot length, the difference in \pdr is more than $17.8$\% in all the cases. Therefore, $40$~ms slot length is not acceptable as it degrades the reliability of the communication network significantly. 

With $50, 100$, and $150$~ms slot lengths, the wireless network delivers similar performance, and the differences \wrt the $50$~ms slot length are less than $3$\%. Moreover, when the system applies the $50$~ms slot length, the \dutycycle is at its minimum, \ie $0.84$\%, and $0.16$\%, respectively in $10$ and $60$~seconds data periods. This is valid for all the combination of settings. 

By considering the above results, the $50$~ms slot length is best suited as the solution to obtain high dependability of the wireless network. Therefore, in the following analysis of \proposedmac, $50$~ms will be used as the slot length. 

\fakeparagraph{Comparison with CRYSTAL and ContikiMAC} Thus far, the performance of \proposedmac was studied with distinct data periods in varying interference settings. In the following, \proposedmac's performance is studied and compared to two relevant state-of-the-art protocols. The comparison focuses on CRYSTAL~\cite{CompetitionCrystal2018} and ContikiMAC~\cite{contikimac} with the configurations specified in Section~\ref{sec:simulation_scenario}. 

\begin{figure}
\centering
\includegraphics[trim={0cm, 0cm, 0cm, 0cm}, clip, width=\columnwidth, keepaspectratio]{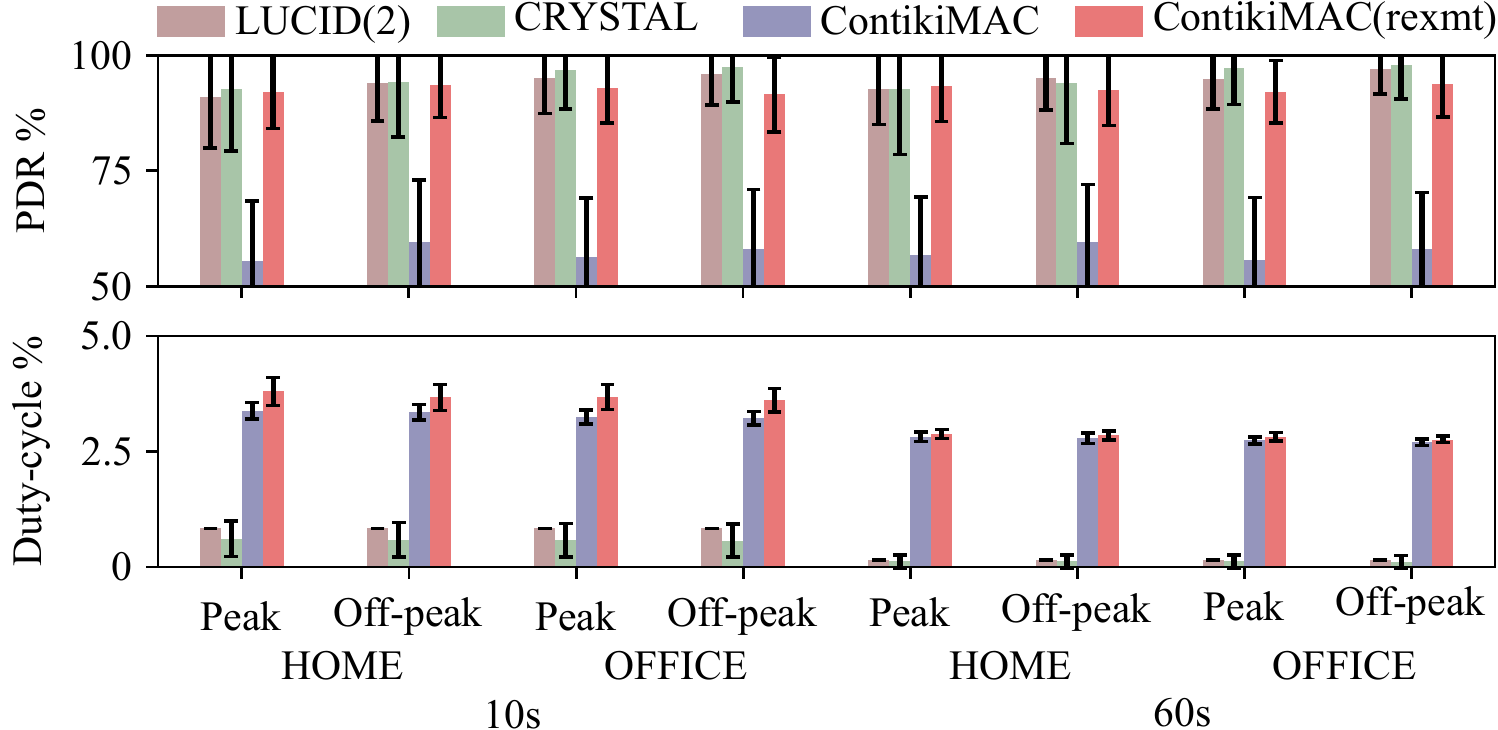}
\caption{Comparison of \pdr (top) and \dutycycle (bottom) in the $5$-node network with the SoA.}
\label{fig:cmp_wsp_soa_n5}
\end{figure}

Fig.~\ref{fig:cmp_wsp_soa_n5} depicts the comparison of network-wide \pdr and \dutycycle of the $5$-node wireless network for varying interference conditions. It is evident from the figure that ContikiMAC without re-transmissions performs the worst in terms of reliability as it achieves less than $60$\% \pdr. This is true for all combinations of data periods, environments, and interference conditions. CSMA is a random channel access mechanism, thus it can be expected that ContikiMAC in collaboration with CSMA without re-transmissions has poor performance. However, when  re-transmissions are enabled, ContikiMAC improves its \pdr by more than $30$\%. This proves the efficiency of re-transmissions implemented in CSMA for interference mitigation. 

\proposedmac, CRYSTAL, and ContikiMAC with re-transmissions produce more than $90$\% \pdr in all cases. Nonetheless, \proposedmac outperforms ContikiMAC in all cases except in the \home peak interference 
with $1.2$\% and $0.6\%$ decrease in \pdr, respectively for $10$ and $60$~seconds data periods \wrt re-transmissions enabled ContikiMAC. In the collected dataset the \home peak exhibits highly bursty interference, during which \proposedmac detects performance degradation through \pdr monitoring, leading to model 
selection. As illustrated in Section~\ref{sec:recalibration}, to avoid frequent 
model selections, a transition period is introduced. However, the transition period contributes to a slight decline in performance \wrt ContikiMAC, especially in very bursty interference settings. However, re-transmissions in ContikiMAC perform well in combating bursty interference but at the cost of enormous energy consumption. 

The \pdr of \proposedmac, in comparison with CRYSTAL, shows mixed performances, and the differences are negligible. However, \proposedmac shows the lowest standard deviation of the \pdr. Notably, CRYSTAL reports the highest standard deviation in the \home peak interference settings, $13.4$\% and $14.1$\%, while \proposedmac exhibits $10.9$\% and $7.7$\%, respectively for $10$ and $60$~seconds data periods. Here, the difference in the standard deviations of \proposedmac is more marked than that of the CRYSTAL with the two data periods. The high variation in the \pdr of \proposedmac arises when the $10$~seconds data period was used, wherein the interference model 
selections are more common than that of the $60$~seconds data period due to the interference dynamics. Because CRYSTAL tackles such sudden interference changes with noise detection techniques, keeping the radio on for a longer period than usual until the interference burst fades away, the difference of standard deviations in the \pdr for both data periods is insignificant. This is evident through the lens of the duty-cycle, which shows higher standard deviations for $10$~seconds data period than $60$~seconds period for CRYSTAL. This demonstrates CRYSTAL's compromise between reliability and energy consumption when using noise detection mechanisms. 

In terms of \dutycycle in the $5$-node network, irrespective of re-transmissions, ContikiMAC shows the worst results exhibiting at least $3.2$\% and $2.7$\% of an average \dutycycle, respectively for $10$ and $60$~seconds data periods. The re-transmissions are an additional overhead, which increases the energy consumption of the wireless network, and is reflected in the results by showing a higher \dutycycle for ContikiMAC with re-transmissions. 

CRYSTAL surpasses \proposedmac in all interference settings, with a maximum difference of $0.3$\% and $0.1$\% in \dutycycle, respectively for $10$ and $60$~seconds data periods in the $5$-node network. Nonetheless, the standard deviation of the \dutycycle is at least $0.4$\% and $0.1$\% higher in CRYSTAL than \proposedmac, respectively for $10$ and $60$~seconds data periods. The high standard deviation of CRYSTAL is due to the nature of its noise detection mechanism, where it keeps the radio on until the interference fades away. This additional radio on duration depends on the length of the burst, which leads to a high standard deviation of the \dutycycle. 

\begin{figure}
\centering
\includegraphics[trim={0cm, 0cm, 0cm, 0cm}, clip, width=\columnwidth, keepaspectratio]{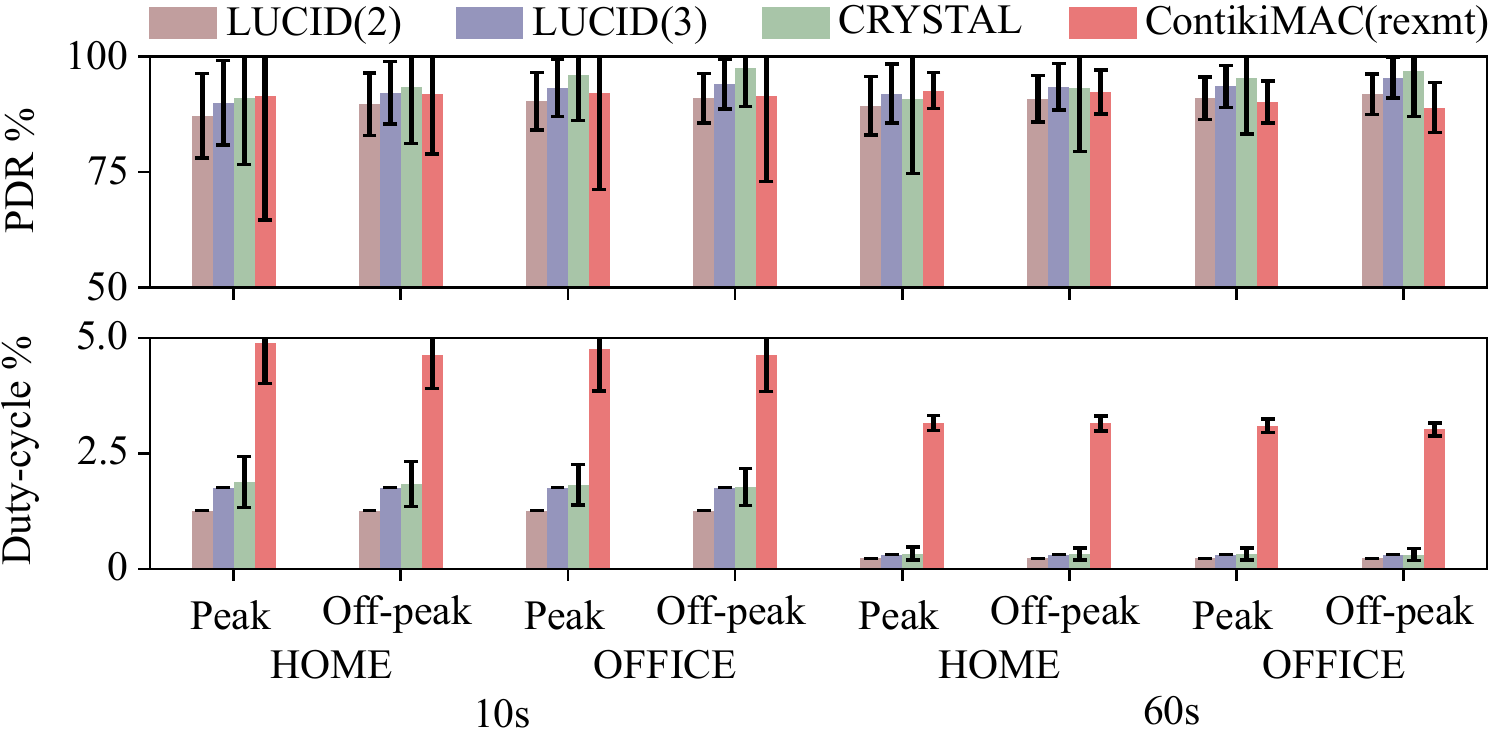}
\caption{Comparison of \pdr (top) and \dutycycle (bottom) in the $16$-node network with the SoA.}
\label{fig:cmp_wsp_soa_n16_2n3slots}
\end{figure}

Now, let's turn our attention to the performance of the $16$-node low-power wireless network which contains more load and data forwarders than the simple $5$-node network. This investigation was performed with two settings in addition to the configurations illustrated in Section~\ref{sec:simulation_scenario}. Because the $16$-node network consists of a high number of data forwarders, these might become congested with a large number of data packets to be forwarded, which makes them bottlenecks and which can lead to packet losses. Therefore, to allow the nodes to forward as many packets as possible, the number of \free slots used by a node within the duration of a data period, \ie a session, is varied. 

Fig.~\ref{fig:cmp_wsp_soa_n16_2n3slots} compares the results with $2$ and $3$ \free slots per session alongside CRYSTAL and re-transmission enabled ContikiMAC. The increase in the number of \free slots per session boosts the network-wide \pdr in all  combinations of the interference settings. This gain is prominent in the \office off-peak interference with a gain of $3.1$\% and $3.6$\% in \pdr, respectively for both $10$ and $60$~seconds data periods. However, as the transceiver is on for longer, the proposed solution with $3$ \free slots per session requires more energy than that with $2$ \free slots. This is evident in the \dutycycle, wherein a $0.5\%$ and $0.08$\% increase in the \dutycycle is visible. Furthermore, in comparison with the simple $5$-node network, the $16$-node sensor network with $2$ \free slots per session exhibits more than $3.4$\% degradation in  \pdr. This is due to the congestion especially in node $9$, as depicted in Fig.~\ref{fig:topo_16}. With the adjustment of $3$ \free slots, it was possible to ease the congestion, and the network managed to improve the difference in \pdr of the two networks by $78$\%. 

\proposedmac outperforms CRYSTAL in the \home with a $60$~seconds data period, $1.2$\% increase in \pdr and $0.02$\% decrease in \dutycycle, and ContikiMAC in all the cases except for the \home peak interference. Because the interference is more bursty in the \home than the \office as discussed in Section~\ref{sec:model_validation}, neither CRYSTAL's noise detection technique nor the random access method was able to mitigate the radio interference better than \proposedmac in a relatively large multi-hop network. However, CRYSTAL is capable of mitigating light interference slightly better than \proposedmac, as it delivers a \pdr increase of $3.4$\% in the \office off-peak interference conditions \wrt \proposedmac. Furthermore, the \dutycycle of \proposedmac is always less than that of CRYSTAL and ContikiMAC even with the increase in the number of \free slots per session.

\section{Discussion and Conclusions}
\label{sec:discussion}

This paper aims at investigating cross-technology interference in IEEE~$802.15.4$ based low-power wireless networks and designing and developing solutions that increase their dependability. In particular, the focus was on indoor environments that exhibit varying interference conditions. 

We proposed a novel proactive model-based receiver-aware MAC protocol, \proposedmac. Contrary to the traditional receiver-initiation concept, interference models, trained with the collected interference traces, are used to decide on the rendezvous point between a sender and a receiver for data communication. The accurate interference models are key to \proposedmac. Moreover, to adapt to the dynamic interference conditions, the performance of the network is continuously monitored to trigger the interference model selection, wherein the models are substituted by new ones when the radio medium changes its properties. This feedback loop is pivotal to keep the performance of the solution above the desired level of dependability for the wireless network. 

The performance of \proposedmac was evaluated in COOJA with different data periods that represent high and low data rate applications under varying interference settings. \proposedmac was compared with CRYSTAL and ContikiMAC, which are solutions in two different paradigms to address the CTI problem. The results demonstrated that \proposedmac tackles high bursty interference in dense wireless networks well while consuming less energy for communication than CRYSTAL and ContikiMAC, increasing the dependability of the low-power wireless networks. 

Although \proposedmac was designed for IEEE~$802.15.4$ based low-power wireless networks, the principles are fundamental to other wireless networks as well. Therefore, the work presented in this paper opens opportunities to research the feasibility of using the white space prediction mechanism and the MAC protocol in other wireless communication networks as well.

It is confirmed through this research that the slot based collaborative use of the IAT and the number of signals provided an efficient way to characterise generic interference rather than using IAT alone. The NCLR is a useful metric to identify patterns in the interference perceived by the nodes. 

The white space prediction mechanism was designed as a technique to tackle CTI. However, later, it turns out that the same mechanism can be easily adopted as a MAC protocol for finding rendezvous points between a sender and a receiver.

Although \proposedmac delivers productive and encouraging results, there are a number of limitations that come along with it. Those limitations and their potential improvements are discussed here, as they will open further research directions. 

First, the proposed medium access technique is not a plug and play solution for improving the dependability of low-power wireless networks. \proposedmac requires to assess and understand the radio medium sufficiently enough to accurately parametrise the interference models. This additional phase in which interference is measured and characterised is an overhead in terms of time and energy consumption.

In the design, two interference models, each for peak and off-peak interference conditions were used. However, in environments wherein sudden and rapid interference variations occur, two interference models might not be able to adequately capture interference characteristics. This can be addressed by adding more than two interference models, depending on the interference patterns. 
However, the use of multiple interference models increases the memory requirement for nodes. Note that individual nodes must have access to the interference models of their neighbours as well. Managing a large number of interference models while generating predictions from them increases the computational complexity of nodes. Therefore, more research is needed to minimise the memory footprint of interference models, cost of computation and associated energy consumption. 

There could be occasions when the wireless networks experience heavily bursty interference on the operating channel, leading to low network reliability. Channel hopping could overcome such circumstances and is extensively used in many protocols as a countermeasure to tackle high interference. Even though \proposedmac does not use channel hopping, it can be adopted in the solution if the interference models for all the channels that are in the hopping sequence are available in the nodes, including that of their neighbours. The same procedure that was introduced in this work can be used to check the availability of white spaces at the receiver and the channel with the nearest white space could be utilised by the nodes to accomplish the communication.
Nonetheless, this will increase the memory requirement, computational complexity, and energy consumption of nodes, potentially leading to low dependable networks. Therefore, further research is needed to optimise the dependability of low-power wireless networks as channel hopping will incur additional energy requirements. 

Finally, a pragmatic application of \proposedmac would be its integration with a CSMA-based solution where transmission decisions are taken by leveraging the white space prediction. Moreover, the GMM interference estimation model can be used for emulating radio interference in testbeds which are used for testing/benchmarking wireless communication networks. 


\section*{Acknowledgments}
\label{sec:acks}


The authors would like to thank S. Palipana, and P. Agrawal for their help in carrying out this research.

\bibliographystyle{unsrt}
\bibliography{references}

\begin{IEEEbiography}[{\includegraphics[width=1in,height=1.25in,clip,keepaspectratio]{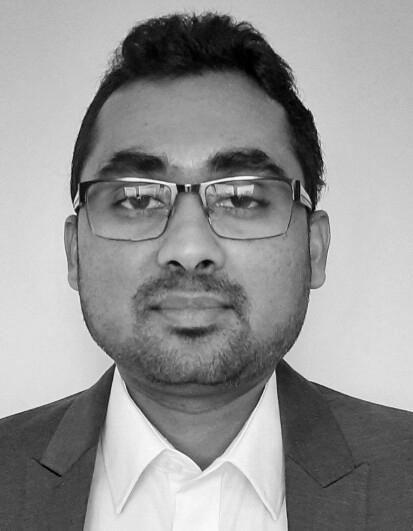}}]{Indika S. A. Dhanapala} received the B.Sc. degree (Hons.) in Electrical \& Electronic Engineering from the University of Peradeniya, Sri Lanka, in 2010, and the master's degree from the Department of Communication \& Information Technology, University of Bremen, Germany, in 2014. He is currently a doctoral candidate in the Nimbus Research Centre, Munster Technological University, Ireland. His research interests are in low-power wireless networks, MAC protocols, interference modelling, and interference mitigation using machine learning. 
\end{IEEEbiography}

\begin{IEEEbiography}[{\includegraphics[width=1in,height=1.25in,clip,keepaspectratio]{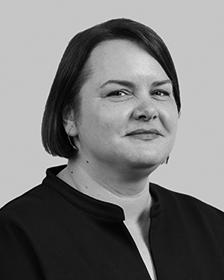}}]{Ramona Marfievici} is a Lead Engineer at Digital Catapult, London, UK. Before her appointment she was a lecturer at the Department of Computer Science at Technical University of Cluj-Napoca, Romania, and a senior researcher at Nimbus Research Centre, Cork, Ireland. She received her Ph.D. in information and communication technology from the University of Trento, Italy, in 2015, and the M.Sc. degree in information systems from the Technical University of Cluj-Napoca, Romania, in 2003. Her research interests are in the areas of internet of things and cyber-physical systems. 
\end{IEEEbiography}

\begin{IEEEbiography}[{\includegraphics[width=1in,height=1.25in,clip,keepaspectratio]{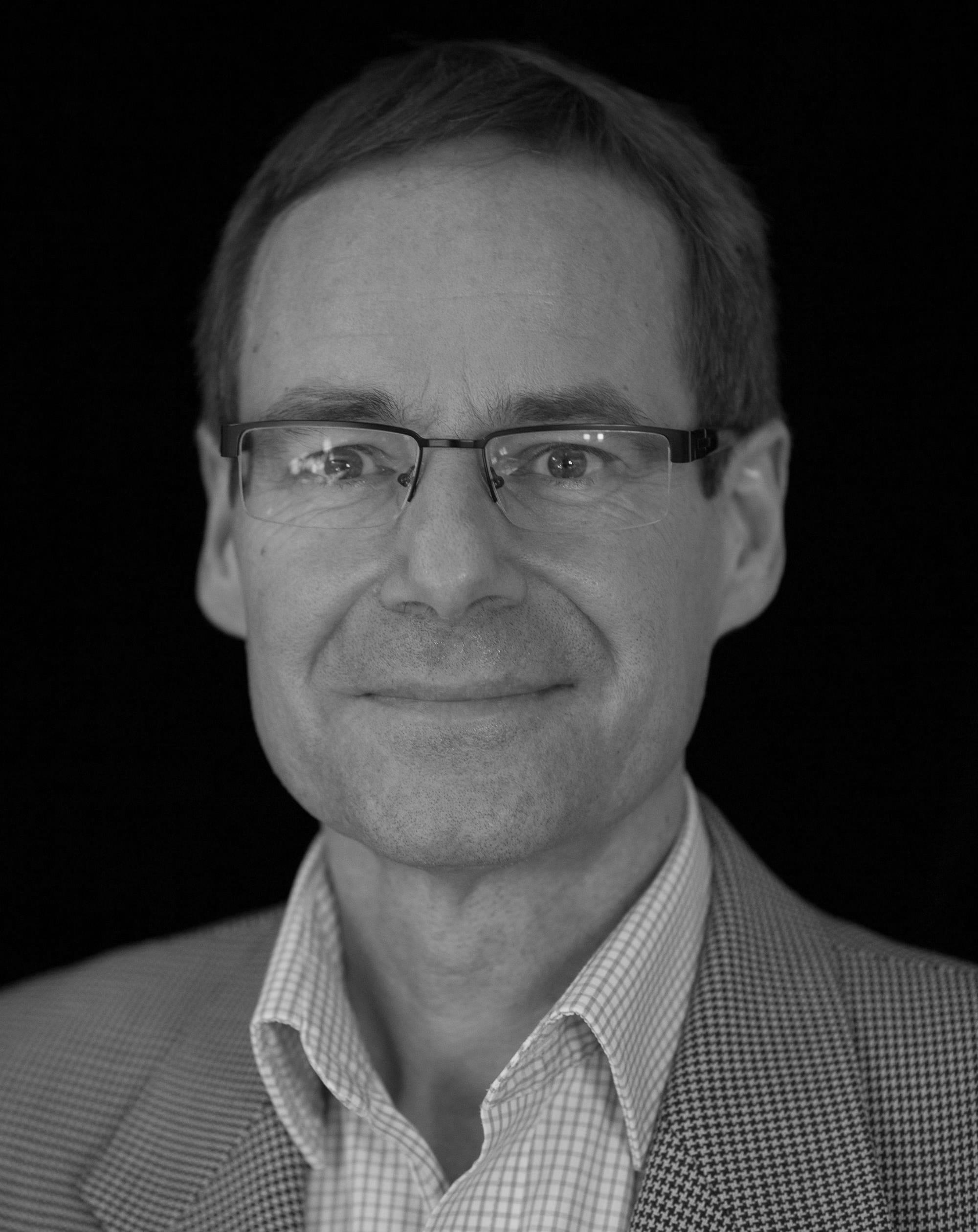}}]{Dirk Pesch} (S'96-M'00-SM'17) is a Professor in the School of Computer Science and IT at University College Cork, Ireland. His research interests include architecture, design, algorithms, and performance evaluation of low power, dense and moving wireless/mobile networks and services for Internet of Things (IoT) and cyber-physical system's applications and interoperability issues associated with IoT applications. He has over $25$ years research and development experience in both industry and academia and has coauthored over $200$ scientific articles. He is a principle investigator in the Science Foundation Ireland funded CONNECT Centre for Future Networks and the CONFIRM Centre for Smart Manufacturing and is the director of the SFI Centre for Research Training in Advanced Networks for Sustainable Societies. He is on the editorial board for a number of international journals, and contributes to international conference organization in his research area. 
He has also been active in many EU funded research projects, including as coordinator, and was involved with two startups. Prior to joining UCC, he was a Professor and the founding Head of Centre of the Nimbus Research Centre at Munster Technological University, Cork, Ireland and a design engineer with Nokia in Germany and the UK. Dirk received a Dipl.Ing. degree from RWTH Aachen University, Aachen, Germany, and a Ph.D. degree from the University of Strathclyde, Glasgow, U.K., both in electrical and electronic engineering.  
\end{IEEEbiography}


\EOD
\end{document}